\documentclass[a4paper,fleqn,usenatbib]{mnras}

\usepackage{mathptmx}

\usepackage[T1]{fontenc}
\usepackage{ae,aecompl}

\usepackage{graphicx}	
\usepackage{amsmath}	
\usepackage{amssymb}	

\title[Instabilities and mass-loss in OB-type stars]{Instability, finite amplitude pulsation and mass-loss in models of massive 
OB-type stars}

\author[A. P. Yadav and W. Glatzel]{
Abhay Pratap Yadav\thanks{E-mail: yadav@astro.physik.uni-goettingen.de}
and Wolfgang Glatzel
\\
Institut f\"ur Astrophysik (IAG), Georg-August-Universit\"at G\"ottingen, 
              Friedrich-Hund-Platz 1, D-37077 G\"ottingen, Germany
}

\date{Accepted 2017 July 14. Received 2017 July 14; in original form 2017 June 8}

\pubyear{2017}

\begin{document}
\label{firstpage}
\pagerange{\pageref{firstpage}--\pageref{lastpage}}
\maketitle

\begin{abstract}
Variability and mass-loss are common phenomena in massive OB-type stars.
It is argued that they are caused by violent strange mode instabilities identified in corresponding
stellar models. We present a systematic linear stability analysis with respect to radial perturbations 
of massive OB-type stars with solar chemical composition and masses between 23 to 100 M$_{\sun}$.
For selected unstable stellar models, we perform non-linear simulations of the evolution of the instabilities 
into the non-linear regime. Finite amplitude pulsations with periods in the range between hours and 100 days 
are found to be the final result of the instabilities. The pulsations are associated with a mean acoustic 
luminosity which can be the origin of a pulsationally driven wind. Corresponding mass-loss rates lie in the 
range between  10$^{-9}$ and 10$^{-4}$ M$_{\sun}$ yr$^{-1}$ and may thus affect the evolution of massive stars.

\end{abstract}





\section{Introduction}



Spectroscopic and 
photometric variability is a common phenomenon in OB-type stars 
\citep[see e.g.,][]{lucy_1976, genderen_1985, balona_1992, cat_2007}. 
Recently, \citet{laur_2017} have investigated stellar variability within 
young open star clusters and identified $\alpha$ Cyg, $\beta$ Cep and slowly pulsating B (SPB) variables.  
Concerning low metallicity environments, \citet{kourniotis_2014} have studied photometric variability
of 4646 massive 
stars in the Small Magellanic Cloud and reported the presence of regular and irregular variabilities 
in many OB-type stars considered in their sample. A similar study for the Large Magellanic
 Cloud has been performed by 
\citet{szczy_2010}.
The origin as well as the mechanism responsible
for the variability are still a matter of debate. Among other possible explanations,
pulsations have been suggested as the cause of variability in massive 
stars \citep[see e.g.,][]{lucy_1976, fullerton_1996}.

Similar to variability, mass-loss is a common phenomenon in massive stars from the main sequence to 
advanced evolutionary stages. Mass-loss is 
an important still not sufficiently understood ingredient for stellar evolution with consequences even for 
the galactic environment. Accordingly, it has been  
 the subject of a number of studies \citep[see, e.g., the reviews on 
mass-loss by][]{puls_2008,smith_2014}.
\citet{lucy_1970} proposed line driven winds (where the photon momentum is transferred to the gas in the envelope) 
as the mechanism responsible for mass-loss in OB-type stars. 
However, episodes of enhanced, strong mass-loss and eruptions
observed, e.g., in luminous blue variables (LBVs) are not satisfactorily explained on the basis of line driven winds. 
Meanwhile, there is growing evidence for a connection between variability, pulsations and mass-loss 
\citep[see e.g.,][]{glatzel_1999, townsend_2007, kraus_2015, yadav_2016, mcdonald_2016}.

With respect to an explanation and understanding of the observed variability,
linear stability analyses have been performed for models of massive stars. 
They cover  
main sequence models \citep{glatzel_1993} and models in advanced evolutionary stages 
\citep{wood_2014, jeffery_2016}
and consider 
stability with respect to both radial 
and non-radial perturbations \citep{osaki_1975, saio_1980, glatzel_1996}.  
As a result, violent instabilities with growth rates in the dynamical range have been identified which 
are associated with the occurrence of strange modes. This type of instability has previously been found 
in models of hydrogen deficient carbon (HdC) \citep[see, e.g.,][]{glatzel_1992, gautschy_1990b, saio_1984, wood_1976}
 and Wolf-Rayet stars \citep{glatzel_1993b}. 
For the mechanism of strange mode instabilities, we refer to \citet{glatzel_1994, papaloizou_1997} and \citet{saio_1998}.

 Concerning radial perturbations, numerical simulations of the 
 evolution of strange mode instabilities into the non-linear regime 
 show \citep{glatzel_1999, yadav_2016} that they can drive episodes of enhanced mass-loss and eruptions. 
In a recent study \citep{yadav_2017}, we have performed a stability analysis of models for massive zero 
age main sequence (ZAMS) stars having solar chemical 
composition in the mass range between 50 and 150 M$_{\sun}$. 
The linear analysis reveals that models having masses above 
58 M$_{\sun}$ are violently unstable. Non-linear simulations show that the instabilities lead to finite amplitude
pulsations with periods in the 
range between  3 and 24 h. The pulsations are associated with a mean acoustic luminosity which might be 
the origin of pulsationally driven mass-loss and allows for an estimate of the mass-loss rate. 
For the models considered, 
the maximum mass-loss rate estimated is found to be of the order of 
10$^{-7}$ M$_{\sun}$ yr$^{-1}$.

In this paper, we intend to extend the previous study of \citet{yadav_2017} to post main sequence
models, i.e., to models of massive OB-type stars for the same, solar, chemical composition. 
After construction of stellar models, we shall test them for linear stability to identify unstable models. 
For selected unstable models, we then perform  numerical simulations of the instabilities into the non-linear regime
to determine their final result. 
The paper falls into five sections:
stellar models are discussed in section \ref{models}, the linear stability analysis is described in section \ref{lsa}
and non-linear simulations are presented in
section \ref{ns}. Our conclusions follow (section \ref{dac}).


\section{Stellar models}
\label{models}
\begin{figure}
    \centering $
 \Large
 \begin{array}{c}
   \scalebox{0.65}{ \input{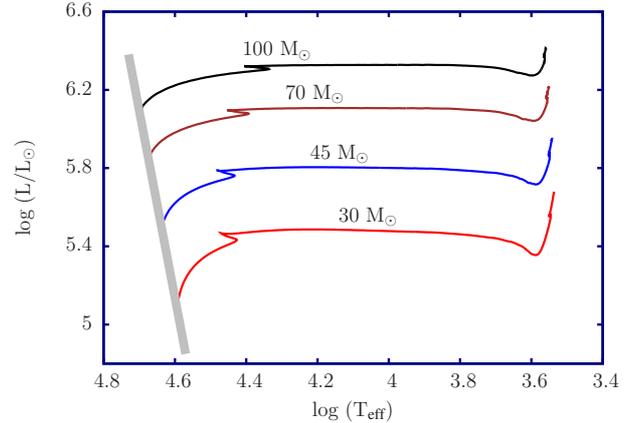} } \\
 \end{array}$
 
 \caption{ HR diagram of evolutionary tracks with solar chemical composition and the initial masses indicated.}
 \normalsize
 \label{hrd}
 \end{figure}

\begin{figure}
\centering $
\Large
\begin{array}{c}
  \scalebox{0.62}{ \input{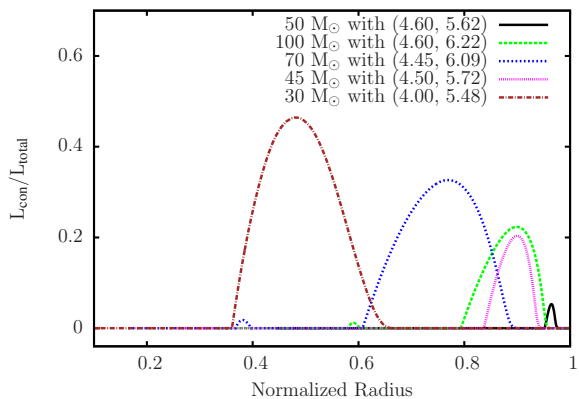} } \\
 \end{array}$
 \caption{ Fraction of convective luminosity as a function of normalized radius for selected models along the evolutionary 
 tracks shown in Fig. \ref{hrd}.
 The values of $ \log T_{\rm{eff}}$ and $ \log L/L_{\sun}$ are indicated in brackets.}
 \normalsize
 \label{ratio}
 \end{figure}

To study instabilities and pulsationally driven mass-loss, we have 
considered a wide range of stellar models having properties close to that of OB-type stars. 
For illustration, the evolutionary tracks from the main sequence to the Hayashi line are shown in Fig. \ref{hrd} 
for initial masses of 30, 45, 70 and 100 M$_{\sun}$, respectively and  solar
chemical composition. 
These tracks have been created using the `mad star EZ-Web' code\footnote{\url{http://www.astro.wisc.edu/~townsend/static.php?ref=ez-web}}.
Effective temperatures, luminosities and masses obtained in this way have been used later on to construct 
high resolution envelope models.
Since the minimum effective temperature for 
B-type stars lies around 10000 K,  this study is restricted to models 
having effective temperatures above this value. 
Energy transport by convection becomes significant in stellar models having effective temperatures below 10000 K. In order to avoid 
difficulties with the treatment of convection, in particular with the still poorly understood coupling of pulsation and 
convection we will not consider models with effective temperature below 10000 K. We can then assume that energy transport is 
dominated by radiation diffusion. 
For illustration, the ratio of convective and total luminosities is displayed in Fig. \ref{ratio} for selected models. 
We deduce that the contribution of the convective luminosity to the total luminosity can
become as high as 45 per cent for  models having effective temperatures close to 10000 K.

Since the dynamical instabilities in models of OB-type stars found by \citet{glatzel_1993} 
are restricted to the envelopes of these objects, the stellar cores may be disregarded for their investigation.  
Therefore, in this study,
we have restricted ourselves to considering the envelopes of OB-type stellar models only.
Using effective temperatures, luminosities and masses from the evolutionary calculations for 30, 45, 70 
and 100 M$_{\sun}$, highly resolved envelope 
models have been constructed as the basis for the subsequent linear stability analysis.
These sets of stellar models will be referred to as `Evolutionary models' hereafter.

In several studies \citep[see e.g.,][]{glatzel_1994, saio_1998, saio_2009}, it has been shown that a high luminosity to 
mass ratio favors the presence of strong instabilities in the envelopes 
of massive stars. In order to study the effect of the luminosity to mass ratio 
we have chosen a representative point in the Hertzsprung-Russell diagram (HRD) having log (L/L$_{\sun}$) = 5.62 and log T$_{\rm{eff}}$ = 4.6. 
For these two parameters fixed, envelope models 
in the mass range between 23 and 80 M$_{\sun}$ have been created and will be investigated for stability too.
This set of envelope models will be referred to as `Models with distinct L/M values' in our analysis.

For the construction of envelope models the 
Stefan-Boltzmann law together with the photospheric pressure as given by 
\citet{kippenhahn_2012} are used as the boundary conditions for the integration of the initial value problem
posed by the equations of stellar structure. 
The inner boundary is defined by a maximum cutoff temperature
of the order of 10$^{7}$ K. 
Magnetic fields and rotation are disregarded. For the 
onset of convection, Schwarzschild's criterion has been used. 
Convection is treated according to standard mixing length theory \citep{bohm_1958} 
with 1.5 pressure scale heights for the mixing length. Opacities are taken from the 
OPAL tables \citep{rogers_1992, rogers_1996, iglesias_1996}.

\section{Linear stability analysis}
\label{lsa}
In this study, we restrict ourselves to considering radial perturbations only. 
The equations governing stellar stability and pulsations in the linear approximation are then
taken in the form given by \cite{gautschy_1990b}.
They form a fourth order boundary eigenvalue 
problem with complex eigenfrequencies ($\sigma_{r}$ + i$\sigma_{i}$) and eigenfunctions. The real part ($\sigma_{r}$) 
of the eigenfrequency provides the
pulsation period while the imaginary part ($\sigma_{i}$) indicates damping (for $\sigma_{i}$ > 0) or excitation 
(for $\sigma_{i}$ < 0). 
In the present study, the eigenfrequencies are normalized by the global free fall time
$\tau_{\rm{ff}}$ = $\sqrt{R^{3}/3\,GM}$, where $R$, $G$ and $M$ 
denote the radius of the stellar model, 
the gravitational constant and the stellar mass, respectively. 
The system of equations is solved using the Riccati method as described by \cite{gautschy_1990a}.

For the treatment of convection, the 
frozen in approximation \citep{baker_1965} has been used here. 
This approximation consists of neglecting the Lagrangian perturbation of the convective luminosity 
which
holds for models where energy transport is dominated by radiation and the pulsation time scale is significantly 
shorter than the convective turnover time. 
In fact these conditions are met in the models considered here (see Fig. \ref{ratio}, where the fraction of 
the convective luminosity is plotted as a function of relative radius for representative models).
For the dependence on the treatment of convection of stellar (strange mode) instabilities in massive stars we refer to 
\citet{sonoi_2014}.

\begin{figure*}
\centering $
\Large
\begin{array}{cc}
  \scalebox{0.68}{ \input{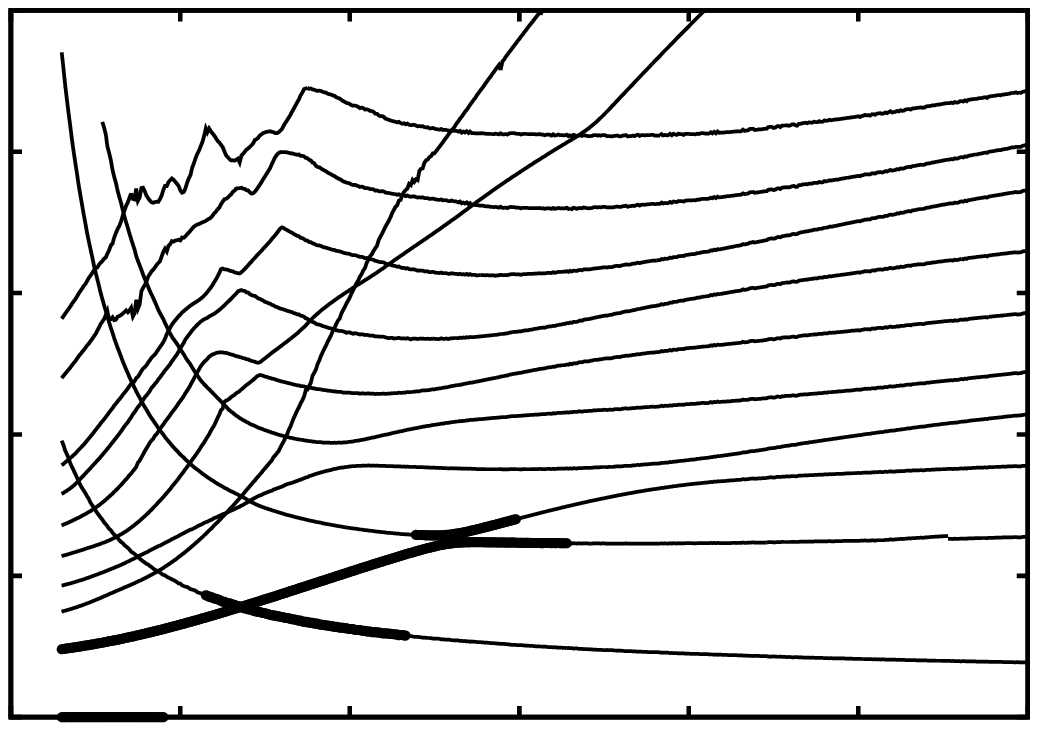} } 
   \scalebox{0.68}{ \input{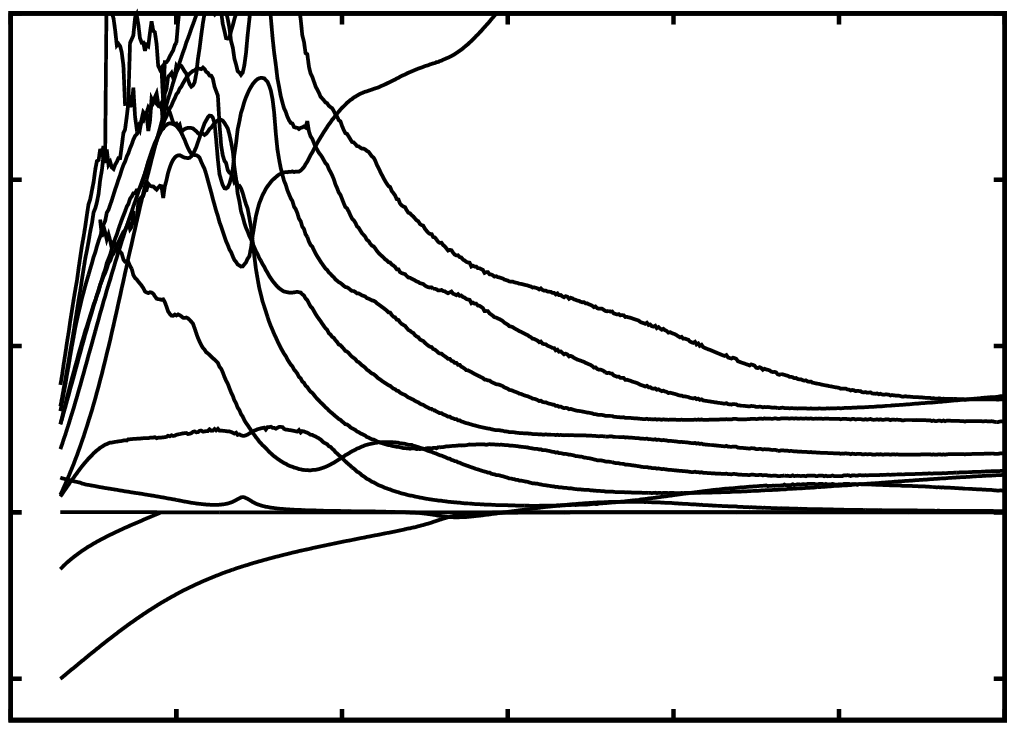} } \\
 \end{array}$
 \caption{ Modal diagram for envelope models having log L/L$_{\sun}$ = 5.62, 
 log T$_{\rm{eff}}$ = 4.6 and solar chemical composition. Real (a) and imaginary (b) parts of the eigenfrequencies 
 normalized by the global free fall time are given as a function of mass. Unstable modes are indicated by thick dots in 
 (a) and negative imaginary parts in (b). }
 \normalsize
 \label{23_80_solar_modal}
 \end{figure*}

 \subsection{Models with distinct L/M values}
In order to study the dependence on the luminosity to mass ratio L/M of dynamical instabilities in massive stars, 
 we have selected a location in the HR diagram 
having log (L/L$_{\sun}$) = 5.62 and log T$_{\rm{eff}}$ = 4.6. 
Keeping these parameters fixed, a linear stability analysis has been performed 
for models having masses between 23 and 80 M$_{\sun}$. 
The result of the linear stability analysis is displayed as a modal diagram in Fig. \ref{23_80_solar_modal}. 
A modal diagram represents eigenfrequencies and modes as a function of a control parameter, which is the mass in the case of 
Fig. \ref{23_80_solar_modal} \cite[for a general discussion of modal diagrams see][]{saio_1998}.

Real and imaginary parts of the eigenfrequencies normalized by the global free fall time are given in 
Fig. \ref{23_80_solar_modal}(a) and Fig. \ref{23_80_solar_modal}(b), respectively. Thick dots in Fig. \ref{23_80_solar_modal}(a)
and negative imaginary parts in Fig. \ref{23_80_solar_modal}(b) correspond to unstable modes. 
In general, the damping of modes increases with their frequency and order. Above a certain order (or sufficiently high
frequency) which 
depends on stellar models, no unstable 
modes are found. Therefore, we have restricted our investigation to low order modes. 
 From Fig. \ref{23_80_solar_modal} we deduce that models having masses below 
53 M$_{\sun}$ are unstable. The selected luminosity and effective temperature would correspond to an evolutionary track with
an initial 
mass of 45 M$_{\sun}$ (see Fig. \ref{hrd}). 
With decreasing mass the modal diagram becomes more complicated and exhibits a variety of mode crossings and interactions.  
Apart from oscillatory unstable modes, we also note the presence of a non-oscillatory (monotonically) unstable mode 
 in models having masses below 30 M$_{\sun}$. 
  The existence of radial monotonically unstable modes with dynamical growth rates 
 has been reported previously \citep[e.g.,][]{saio_2011, yadav_2016}. However, their physical origin and 
 observational consequence
 is still an open question. 
 In general, the strength of instabilities (growth rate) increases with the luminosity to mass ratio
(see Fig. \ref{23_80_solar_modal}).

\subsection{Evolutionary models}

\subsubsection{Models with initial masses of 100 and 70 M$_{\sun}$}

A modal diagram along the evolutionary track from the ZAMS to log T$_{\rm{eff}}$ = 4.0 for an initial mass of 
100 M$_{\sun}$ (see Fig. \ref{hrd}) 
with the effective temperature as a parameter is shown in Fig. \ref{100m_solar_modal}, where
real (a) and imaginary (b) parts of the eigenfrequencies normalized by the global free fall time are displayed.
All models considered are unstable with growth rates in the dynamical regime. 
Similar to Fig. \ref{100m_solar_modal}, a modal diagram for an initial mass of 70 M$_{\sun}$ is presented 
in Fig. \ref{70m_solar_modal}. 
Compared to the 100 M$_{\sun}$ mass models, the growth rates, the number of unstable modes and the range of instability 
have decreased slightly.

\subsubsection{Models with initial masses of 45 and 30 M$_{\sun}$}
Modal diagrams along the evolutionary tracks with initial masses of 45 and 30 M$_{\sun}$ are given in 
Figs. \ref{45m_solar_modal} and \ref{30m_solar_modal}, respectively. 
In both cases, 
models in the vicinity of ZAMS are linearly stable, which is in agreement with the previous study by \cite{yadav_2017} on the stability of ZAMS models
with solar chemical composition. This investigation has revealed that only models having masses above 58 M$_{\sun}$ are unstable. 
For the 45 M$_{\sun}$ track, instability sets in around log T$_{\rm{eff}}$ = 4.62 and persists to 
log T$_{\rm{eff}}$ = 4.40. 
Another strange mode is then unstable in the range between log T$_{\rm{eff}}$ = 4.33 and 4.20. 
For the 30 M$_{\sun}$ track, instability is found only in a narrow range of effective temperatures 
for 4.42 < log T$_{\rm{eff}}$ < 4.56 (see Fig. \ref{30m_solar_modal}).

\subsubsection{Comparison with previous investigations}
\begin{figure*}
\centering $
\Large
\begin{array}{cc}
  \scalebox{0.68}{ \input{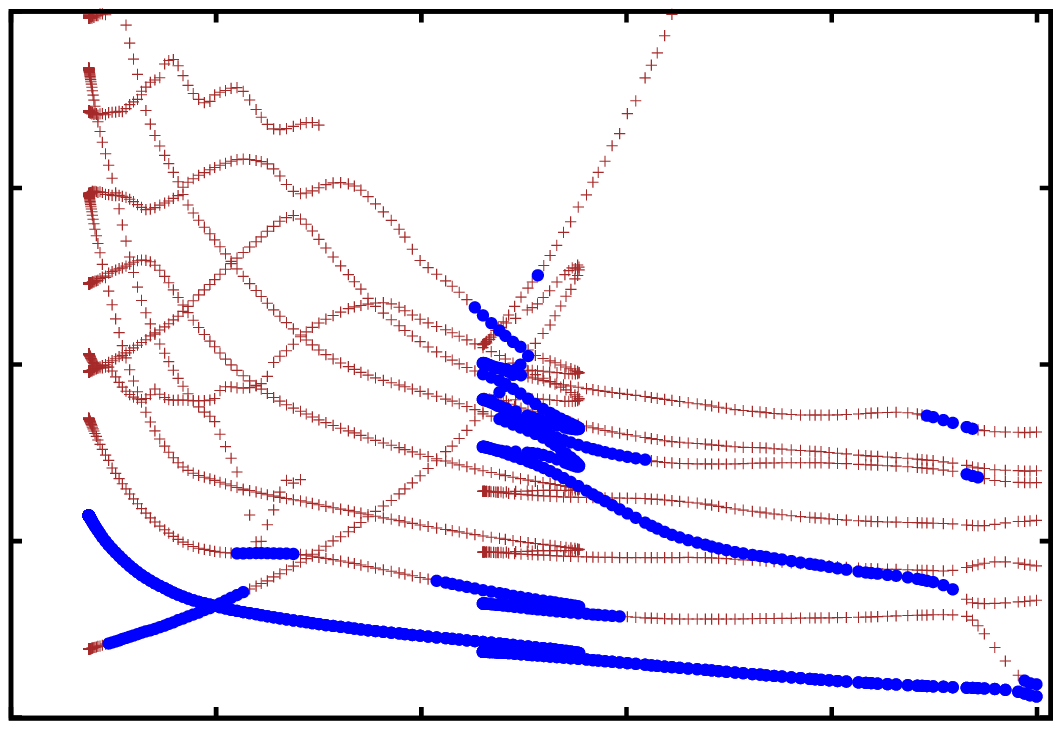} } 
   \scalebox{0.68}{ \input{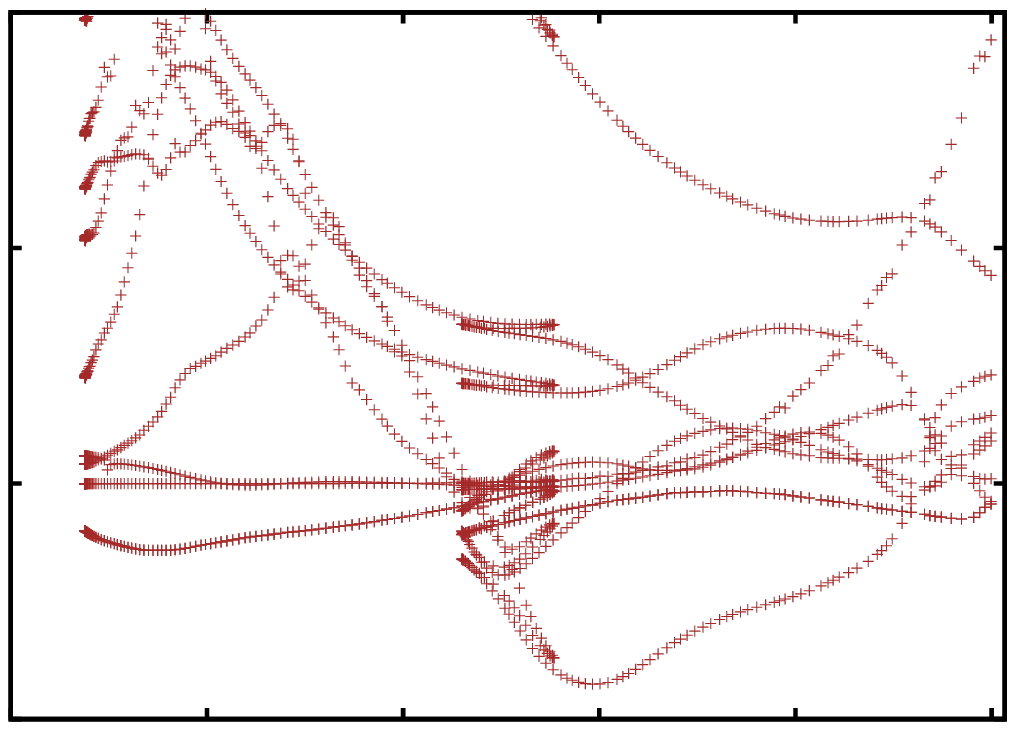} } \\
 \end{array}$
 \caption{ Modal diagram of envelope models along the evolutionary track from the ZAMS to an effective temperature of 10000 K 
 of a star with solar chemical composition and an initial mass of 100 M$_{\sun}$. Real (a) and imaginary (b) parts of the eigenfrequencies 
 normalized by the global free fall time are given as a function of effective temperature. Unstable modes are indicated by thick dots in 
 (a) and negative imaginary parts in (b).}
 \normalsize
 \label{100m_solar_modal}
 \end{figure*}

\begin{figure*}
\centering $
\Large
\begin{array}{cc}
  \scalebox{0.68}{ \input{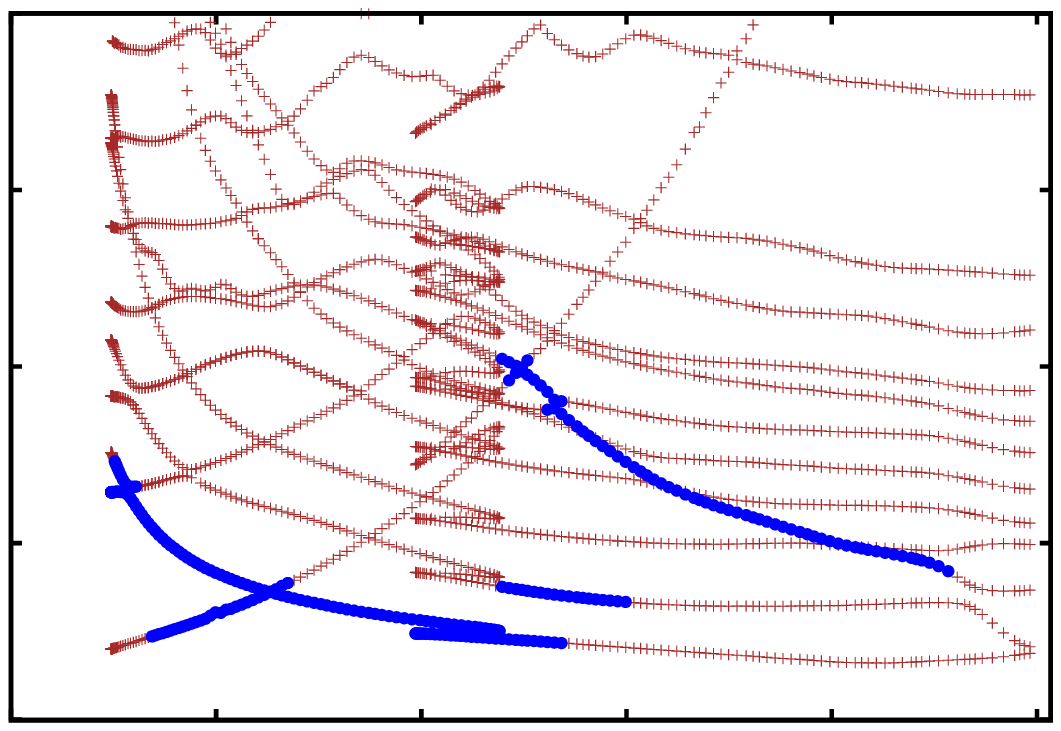} } 
   \scalebox{0.68}{ \input{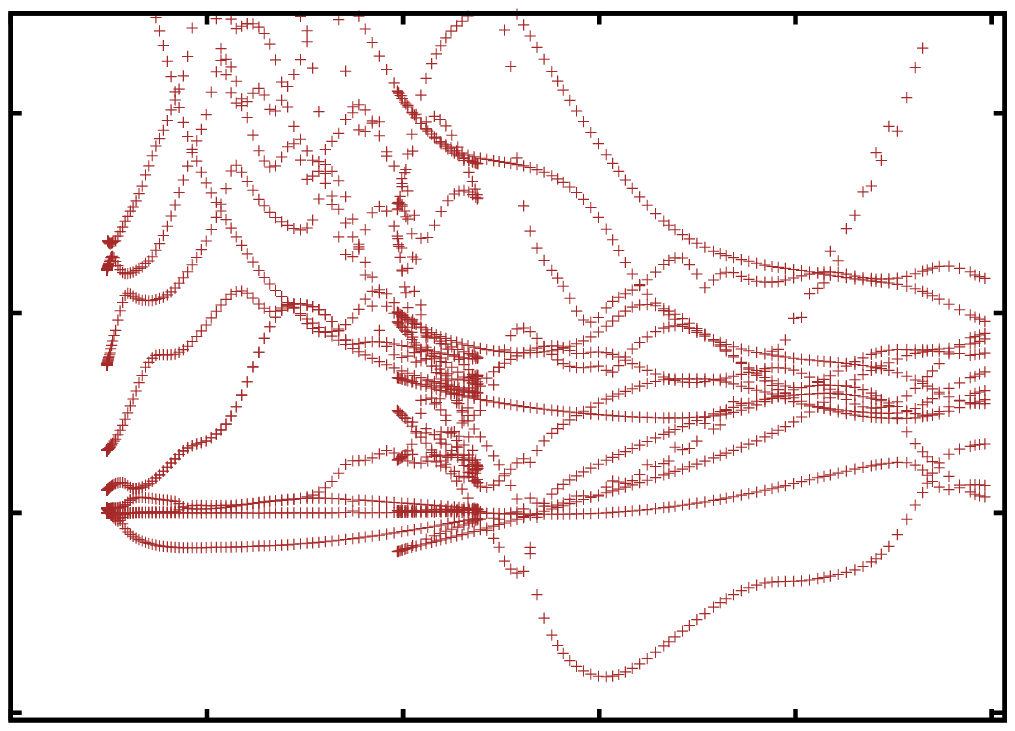} } \\
 \end{array}$
 \caption{ Same as Fig. \ref{100m_solar_modal} but for an initial mass of 70 M$_{\sun}$.}
 \normalsize
 \label{70m_solar_modal}
 \end{figure*}

\begin{figure*}
\centering $
\Large
\begin{array}{cc}
  \scalebox{0.68}{ \input{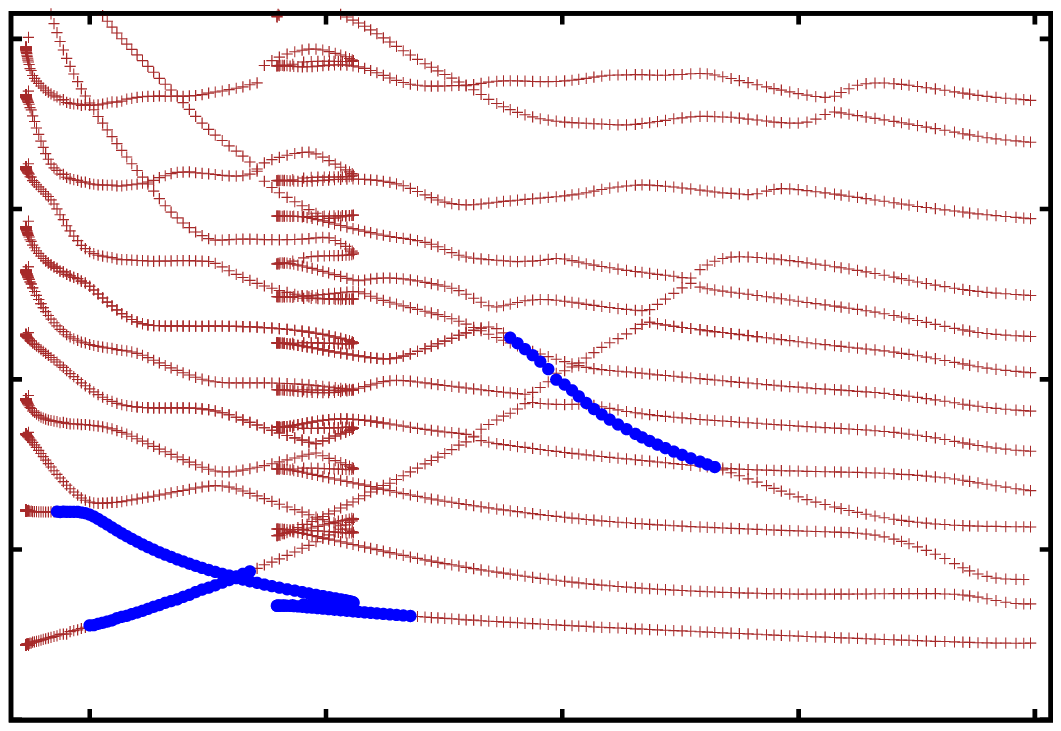} } 
   \scalebox{0.68}{ \input{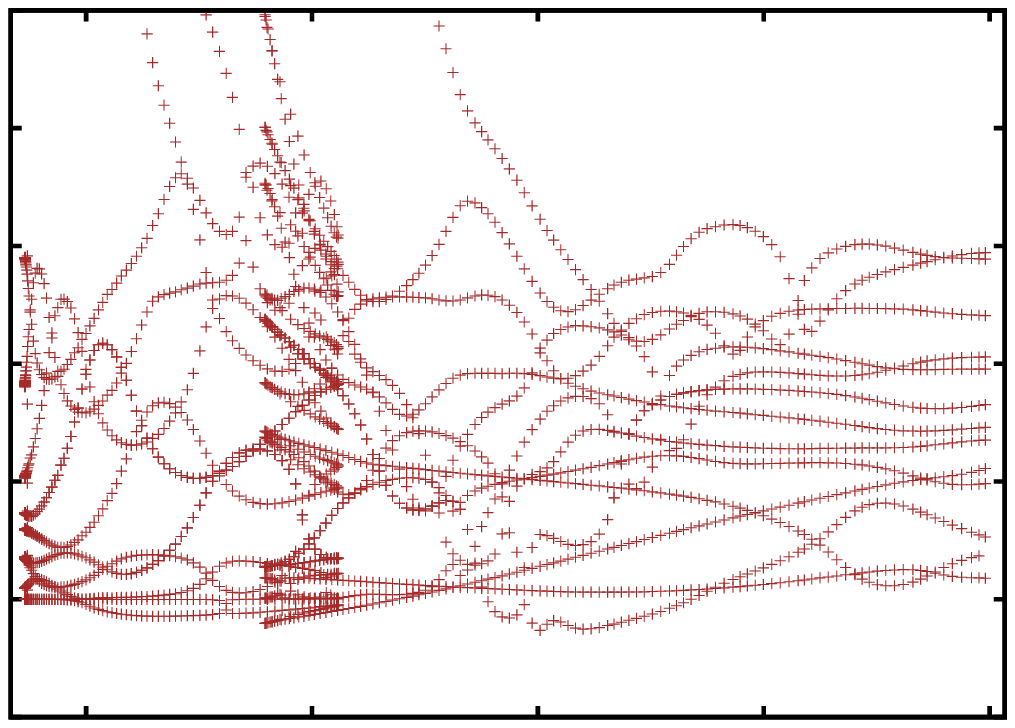} } \\
 \end{array}$
 \caption{  Same as Fig. \ref{100m_solar_modal} but for an initial mass of 45 M$_{\sun}$.}
 \normalsize
 \label{45m_solar_modal}
 \end{figure*}

\begin{figure*}
\centering $
\Large
\begin{array}{cc}
  \scalebox{0.68}{ \input{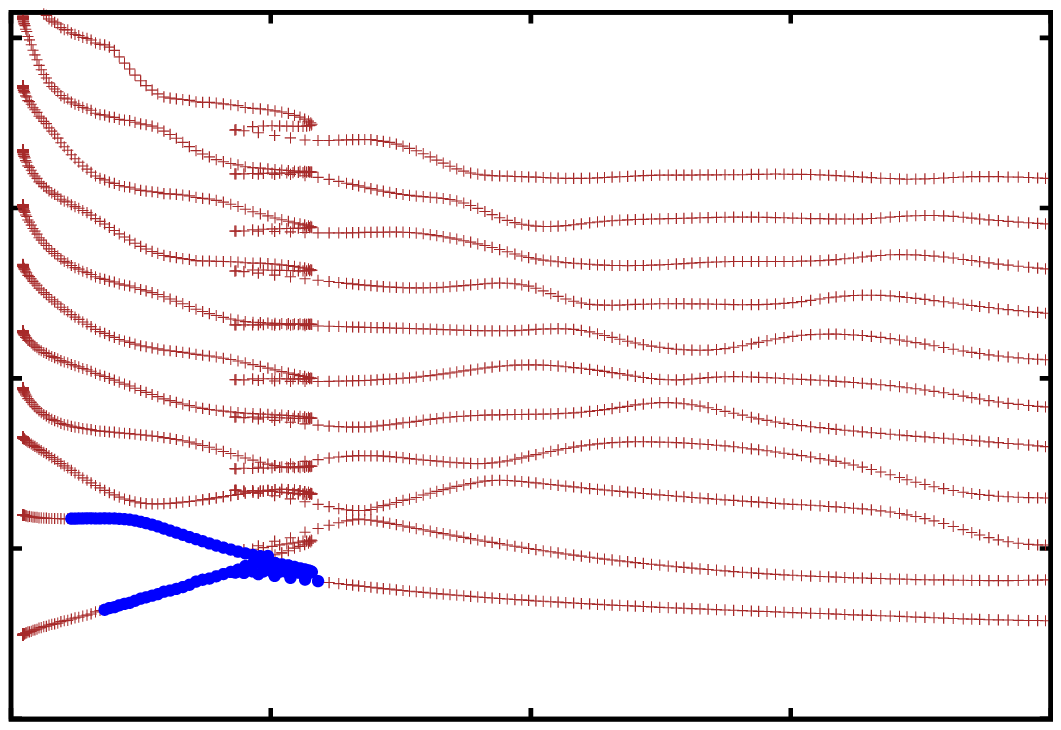} } 
   \scalebox{0.68}{ \input{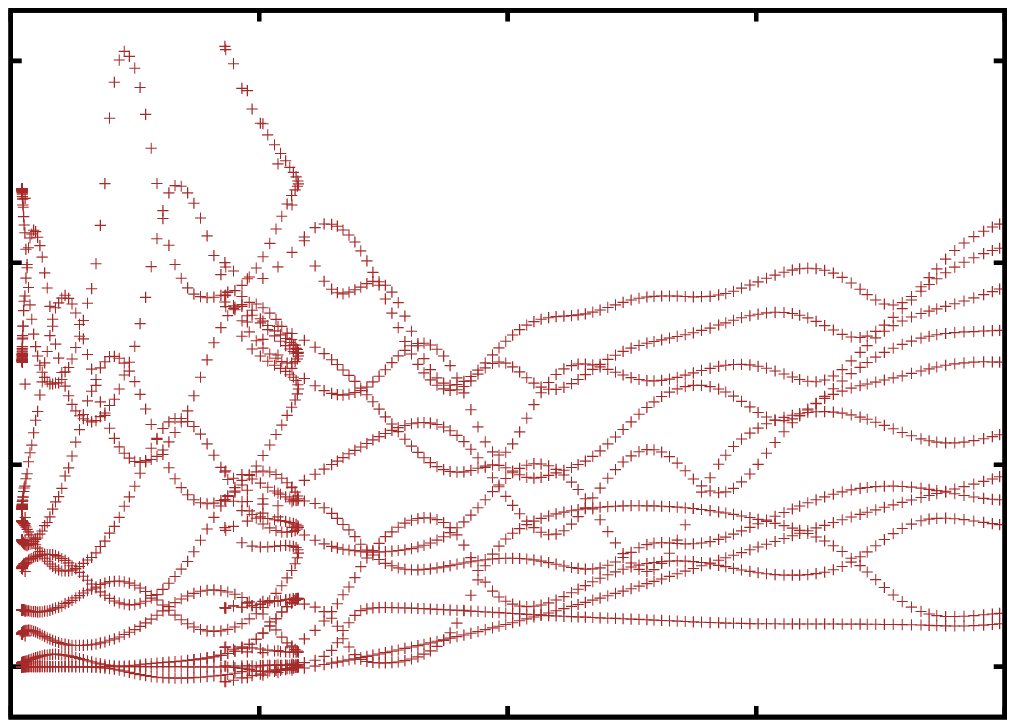} } \\
 \end{array}$
 \caption{  Same as Fig. \ref{100m_solar_modal} but for an initial mass of 30 M$_{\sun}$.}
 \normalsize
 \label{30m_solar_modal}
 \end{figure*}

For low initial masses (see Fig. \ref{30m_solar_modal}), modes are well separated and regularly spaced. If present, mode interactions occur 
via avoided crossings. With increasing initial mass (see Figs. \ref{45m_solar_modal}, \ref{70m_solar_modal}, \ref{100m_solar_modal} ), 
mode interactions become more pronounced and the modal diagrams exhibit a bewildering complexity.  Additional modes, some of them associated with 
instabilities in the dynamical range, appear in addition to the ordinary acoustic modes. In previous investigations 
\citep[see, e.g.,][]{glatzel_1993c, kiriakidis_1993}, they have been addressed as 
strange modes and strange mode instabilities, respectively. The number of unstable modes and the strength of the instabilities increase with the initial 
mass or the luminosity to mass ratio, respectively. 
Comparing our results with those of \citet{kiriakidis_1993}, we find an overall agreement even quantitatively: the strange modes and associated 
instabilities fall into two groups related to the opacity bumps due to heavy elements and helium ionization, respectively. 
 This becomes particularly obvious for the 45 M$_{\sun}$ track, where the two groups are well separated and lead to a local maximum growth rate at 
 log T$_{\rm{eff}}$ $\approx$ 4.5 for the `heavy element strange modes' and a local maximum growth rate at log T$_{\rm{eff}}$ $\approx$ 4.3 for the 
 `helium strange modes'. 
 With respect to a recent publication by \citet{D_D_2017} claiming the discovery of avoided crossings for radial 
 main sequence stellar pulsations, we note that 
 the phenomenon of mode coupling either through instability bands or avoided
 crossings is common in physics and astrophysics. For the spectrum of radial stellar pulsations, it was described by 
 \citet{gautschy_1990b,glatzel_1993,glatzel_1993b,kiriakidis_1993} for many types of stars including the ZAMS. 
 For recent studies involving mode coupling phenomena in massive stars, we refer to \citet{yadav_2016, yadav_2017}.

 The boundary conditions for the perturbation equations at the photosphere are ambiguous. To account for this ambiguity we have performed 
 linear stability analyses for a variety of photospheric boundary conditions. Concerning the eigenfrequencies obtained for the models 
 considered here the maximum relative differences for different boundary conditions amount to less than 15 per cent.

\section{Non-linear simulations}
\label{ns}
In case of instability, the linear theory can predict neither the amplitude of an unstable perturbation nor the final fate of 
the system. In order to obtain the final fate 
of the unstable models which have been identified by linear stability analyses 
(see section \ref{lsa}), instabilities have to be followed into the non-linear regime by numeral simulation. 
For the stellar models considered, instabilities may imply the following consequences:
\begin{enumerate}
 \item Finite amplitude periodic or non-periodic pulsations, as found, e.g., for unstable models of massive 
 stars by \citet{grott_2005, yadav_2017}.
 \item Eruption of the outer layers of a star, as noticed by \citet{glatzel_1999} in models of evolved massive stars 
 and by \citet{yadav_2016} in models of 55 Cygni. In these cases, the photospheric velocity exceeded the escape velocity in the course 
 of the evolution of the instabilities.   
 \item Re-arrangement of the internal structure of the star, as pointed out by \citet{yadav_2016} for one of the models for the 
 supergiant 55 Cygni. 
  
\end{enumerate}

In order to determine the final fate of instabilities, we have followed them into the non-linear regime for 
selected unstable stellar models using the numerical scheme described by \citet{grott_2005}. 
During pulsations, the acoustic energy fluxes and kinetic energies, being of interest here, are smaller than 
the gravitational and internal energies by several orders of magnitude. This requires the energy balance to be satisfied 
with an extremely high precision, which is achieved by an, with respect to energy, intrinsically conservative scheme.    
A consequence of the requirement of the conservativity is that the scheme has to be implicit with respect to time. 
The importance of conservativity in connection with the construction of numerical schemes and codes was repeatedly emphasized 
by \citet{grott_2003, grott_2005, glatzel_2016}. 
Using the procedure by \citet{grott_2005}, for details we refer the reader to this publication,
we can explicitly prove the energy balance to be satisfied intrinsically and thus show 
the results of our simulations to be reliable. Due to the occurrence of shock waves, artificial numerical viscosity has to be 
introduced to represent them. The choice of the artificial viscosity will be discussed in connection with the results 
in the subsequent sections.

Among various tests, the scheme used here had been applied during its development to classical Cepheids. Their instability 
caused by the $\kappa$ - mechanism  with imaginary parts 
of the eigenfrequencies corresponding to $\sigma_{i} \, \propto \, - 10^{-2}$ is significantly weaker than the strange mode instabilities 
studied here. In these tests, the simulations have shown the instabilities 
to lead to finite amplitude pulsations with periods close to the linearly determined values and velocity amplitudes of the order of 
10 km s$^{-1}$. Likewise, we expect corresponding simulations for RR Lyrae models with stability properties similar to Cepheids to exhibit finite 
amplitude pulsations with periods close to the linearly obtained values. Non-linear studies for RR Lyrae models date back to 
\cite{stellingwerf_1975}. We note, however, that his approach prescribing a periodic behaviour is different from our study without any 
assumption on the evolution in the non-linear regime.

\subsection{Models with distinct L/M values}
The linear stability analysis of models having a fixed luminosity and effective temperature 
( log L/L$_{\sun}$ = 5.62 and log T$_{\rm{eff}}$ = 4.6) has revealed 
that models with masses below 53 M$_{\sun}$ are unstable. To determine their final fate,
we have followed the instabilities
into the non-linear regime for selected unstable models with masses of 35, 30, 27 and 23 M$_{\sun}$. 
\begin{figure*}
\centering $
\LARGE
\begin{array}{ccccccccc}
  \scalebox{0.455}{ \input{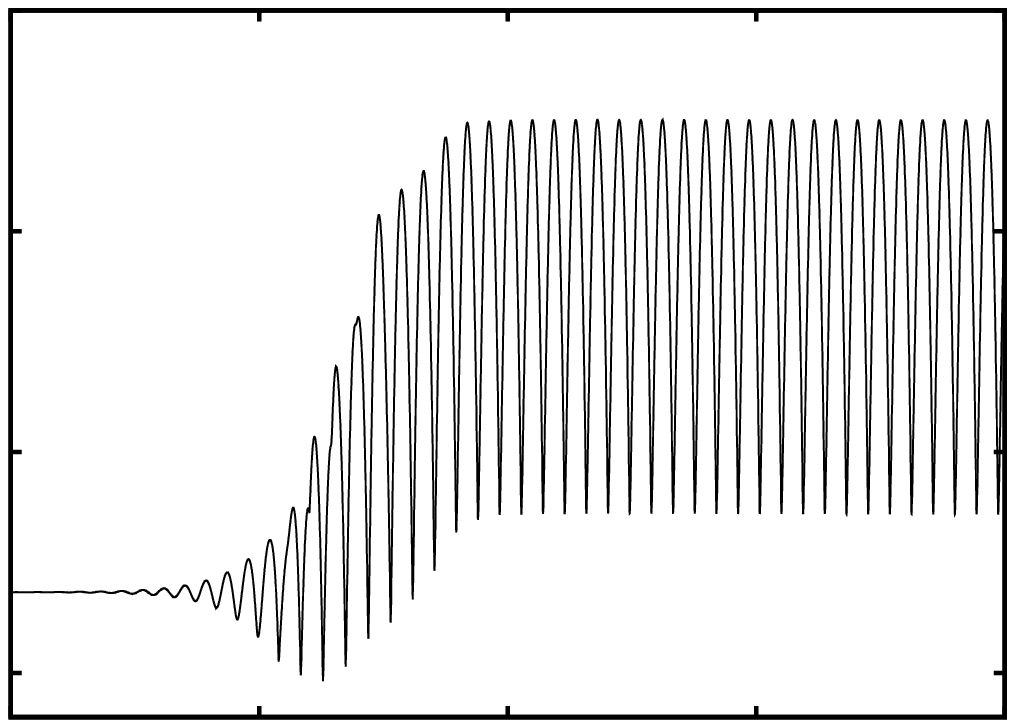} } 
  \scalebox{0.455}{ \input{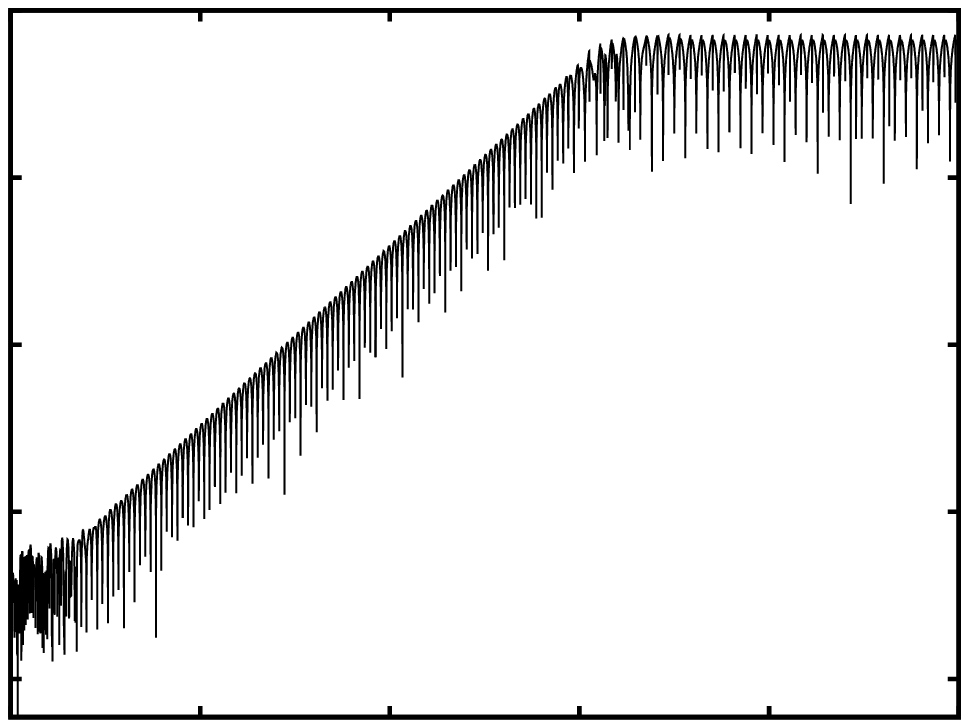} }
   \scalebox{0.455}{ \input{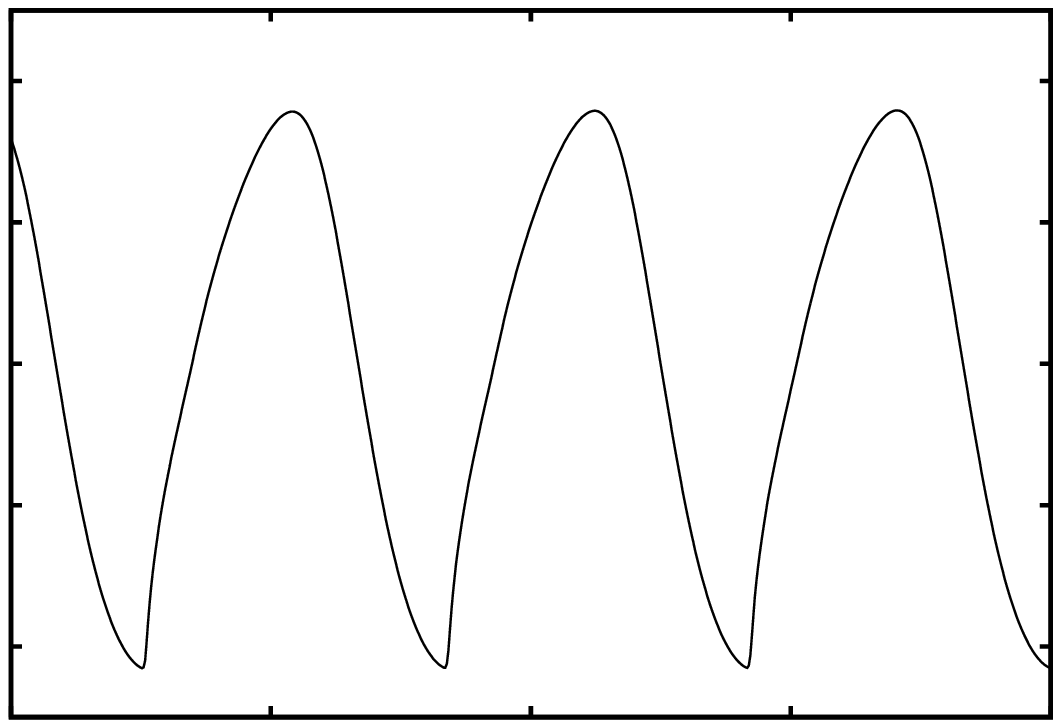} } \\

    \scalebox{0.455}{ \input{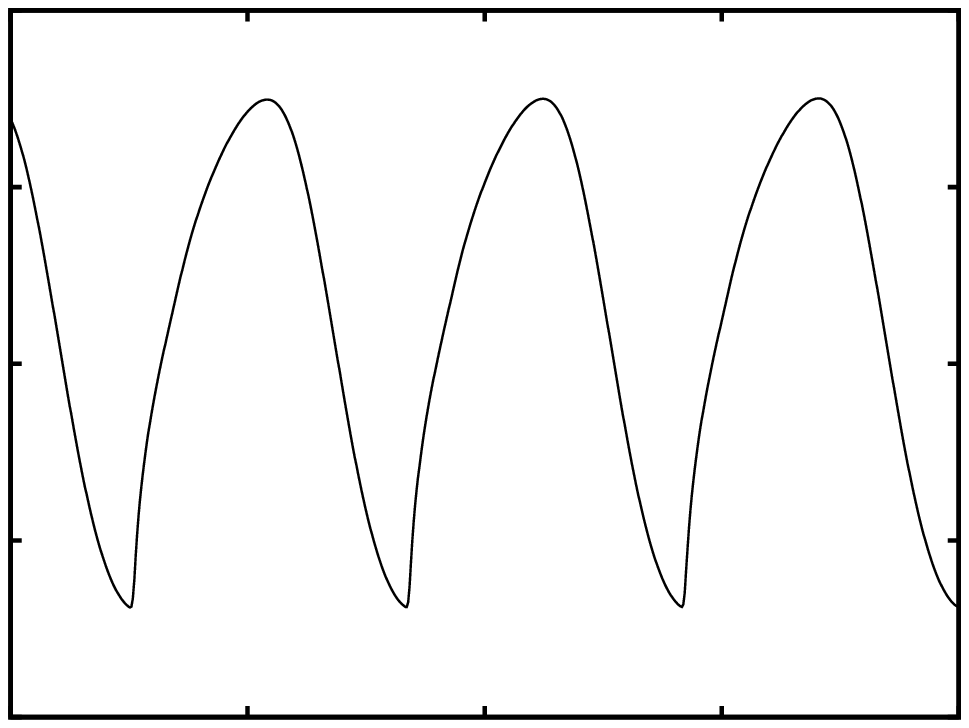} } 
  \scalebox{0.455}{ \input{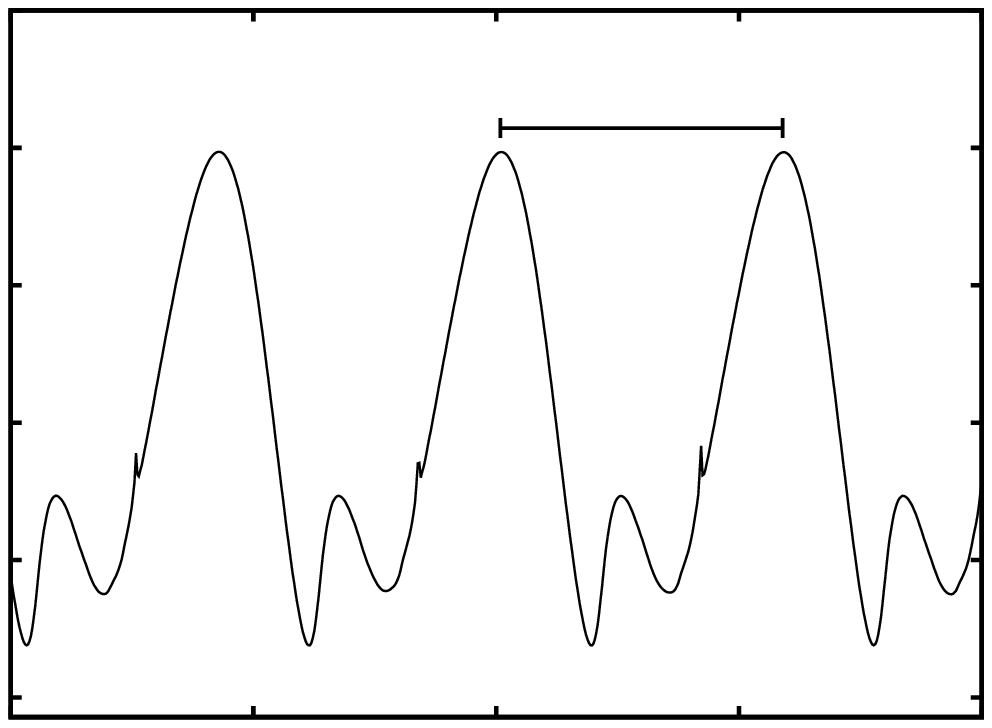} }
   \scalebox{0.455}{ \input{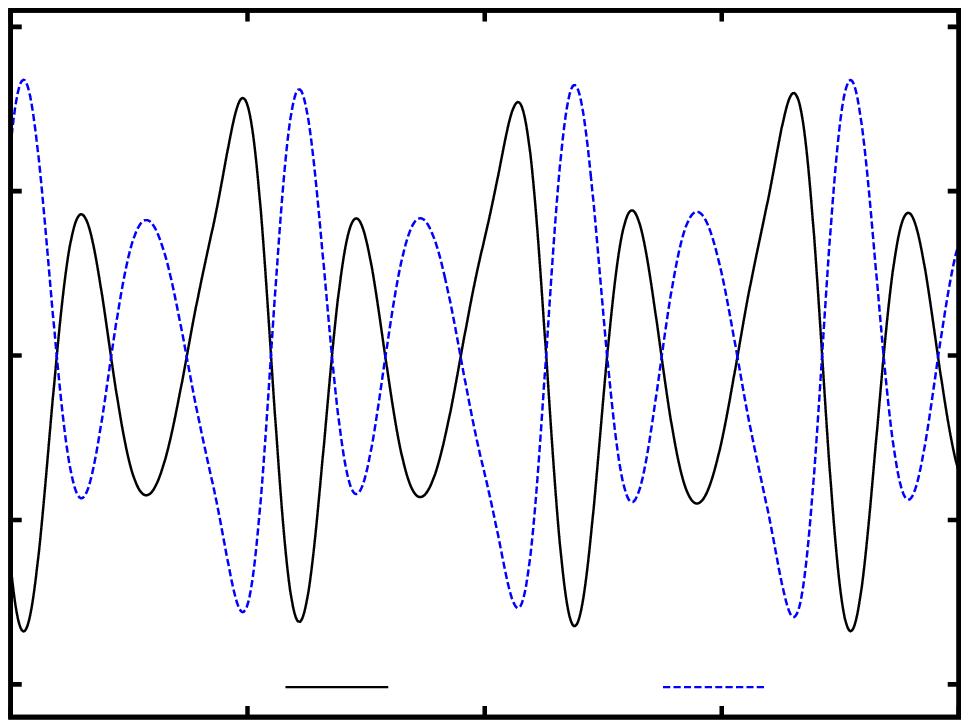} } \\
   
   \scalebox{0.455}{ \input{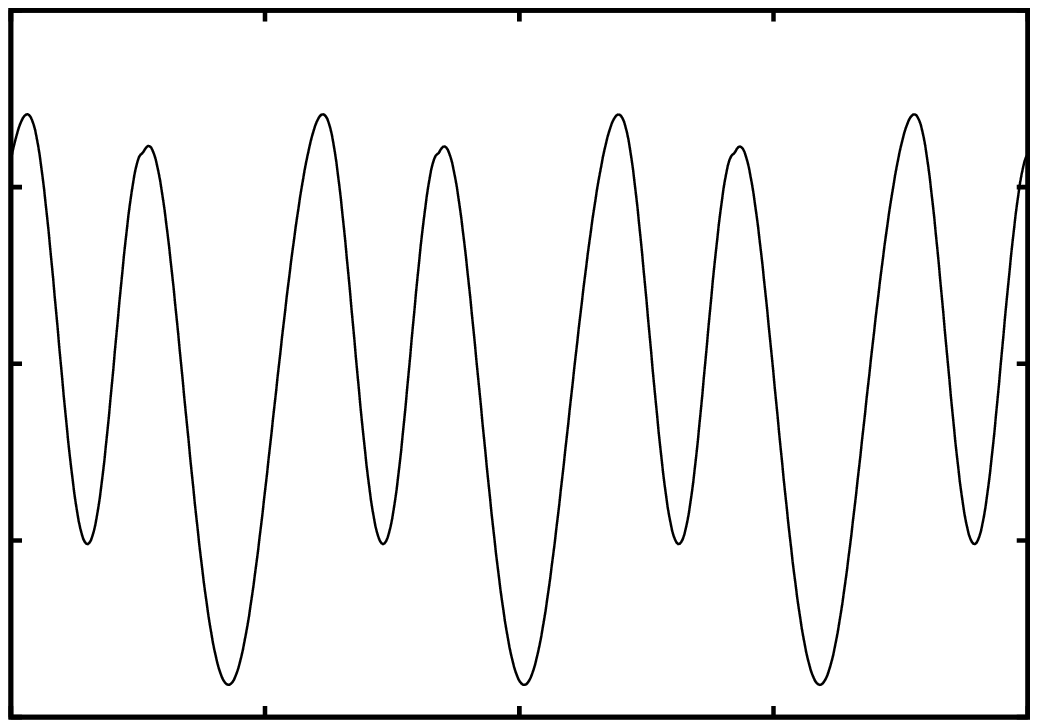} } 
  \scalebox{0.455}{ \input{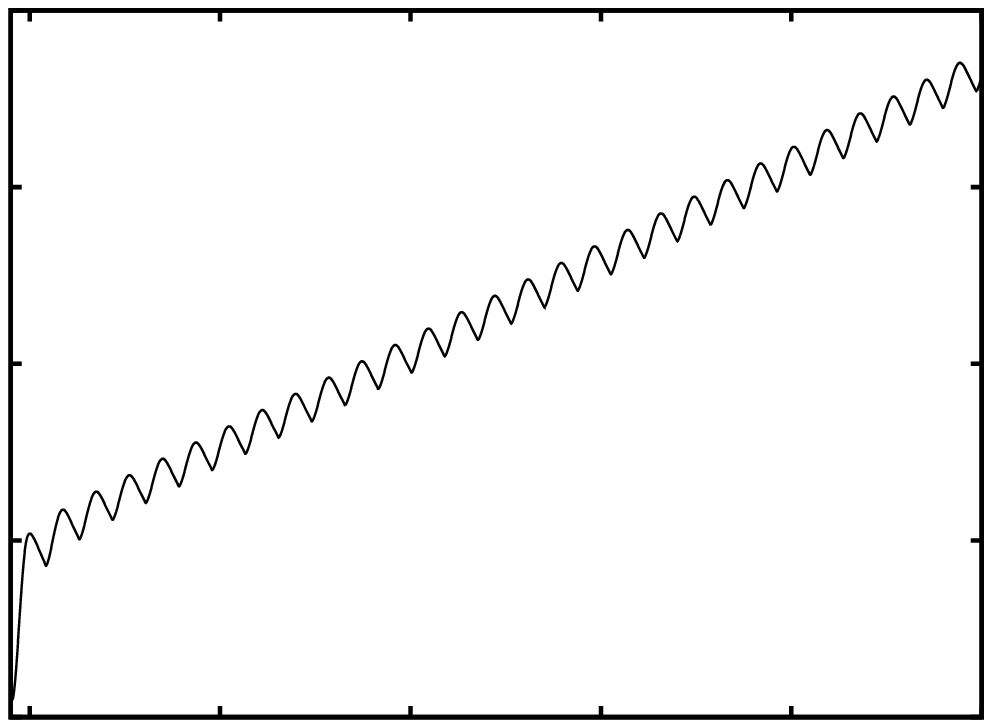} }
   \scalebox{0.455}{ \input{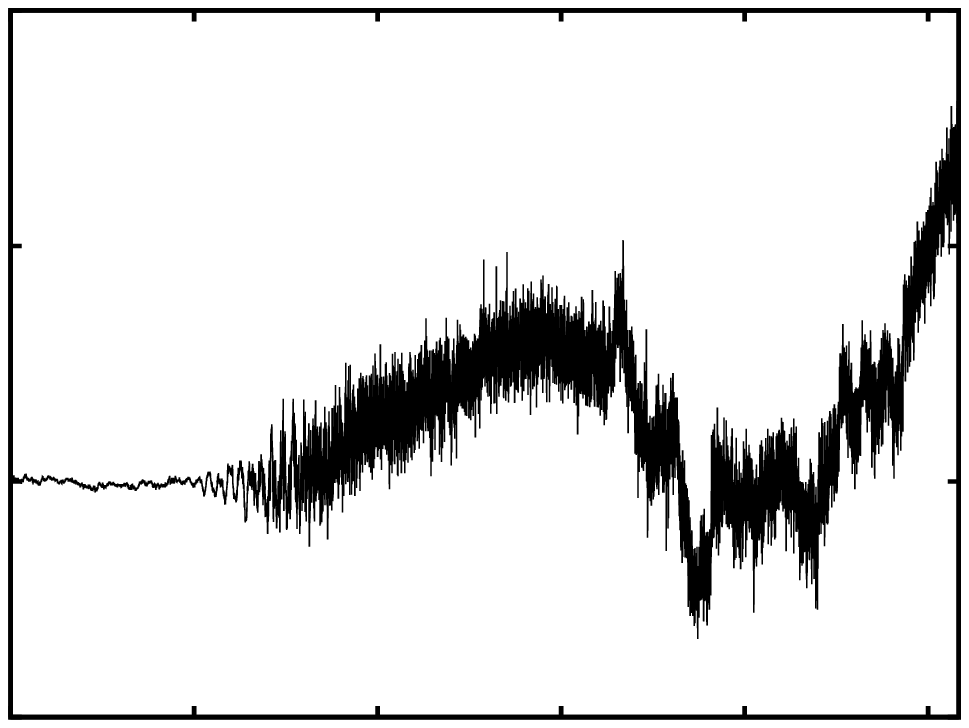} } \\
   \end{array}$
 \caption{ Evolution of the instability into the non-linear regime and finite amplitude pulsations 
 for a model having M = 35 M$_{\sun}$, log L/L$_{\sun}$ = 5.62 and 
 T$_{\rm{eff}}$ = 40000 K. As a function of time, the stellar radius, the velocity, density and temperature at the 
 outermost grid point are shown in (a) - (d) respectively, the variation of the bolometric magnitude is given in (e).
 From the velocity diagram (b), we deduce that the evolution of the instability starts from hydrostatic equilibrium
 with velocity perturbations of the order of 10$^{-6}$ cm s$^{-1}$, undergoes the linear phase of exponential growth 
 and saturates in the non-linear regime with a velocity amplitude of 190 km s$^{-1}$. Compared to the hydrostatic value, the mean 
 radius is increased by approximately 10 per cent in the non-linear regime. The various terms appearing 
 in the energy balance are shown in (f) - (h), where the hydrostatic values have been subtracted. 
 Potential and internal energy with almost identical modulus have opposite sign. They are by three and four 
 orders of magnitude
 bigger than the kinetic and the time integrated acoustic energy respectively. The sum of all energy terms is given in 
 (i) indicating the error in the energy balance. Note that this error is smaller 
 than the smallest term in the energy balance by at least four orders of
 magnitude.}
 \normalsize
 \label{35m_nonlin}
 \end{figure*}

  \begin{figure*}
\centering $
\LARGE
\begin{array}{ccc}
  \scalebox{0.455}{ \input{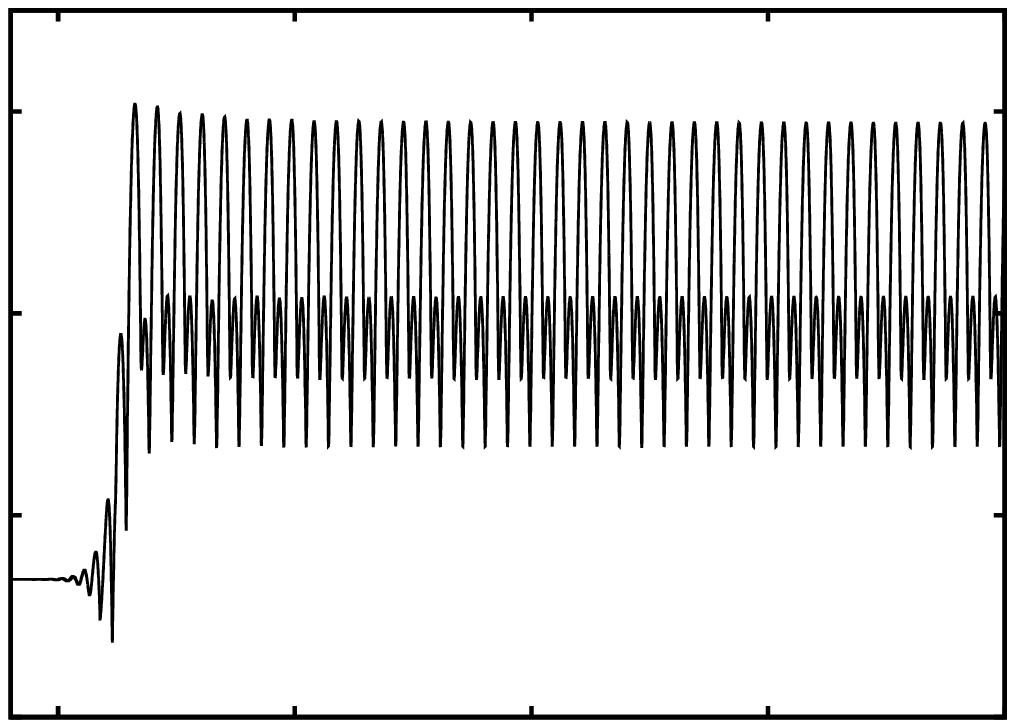} } 
  \scalebox{0.455}{ \input{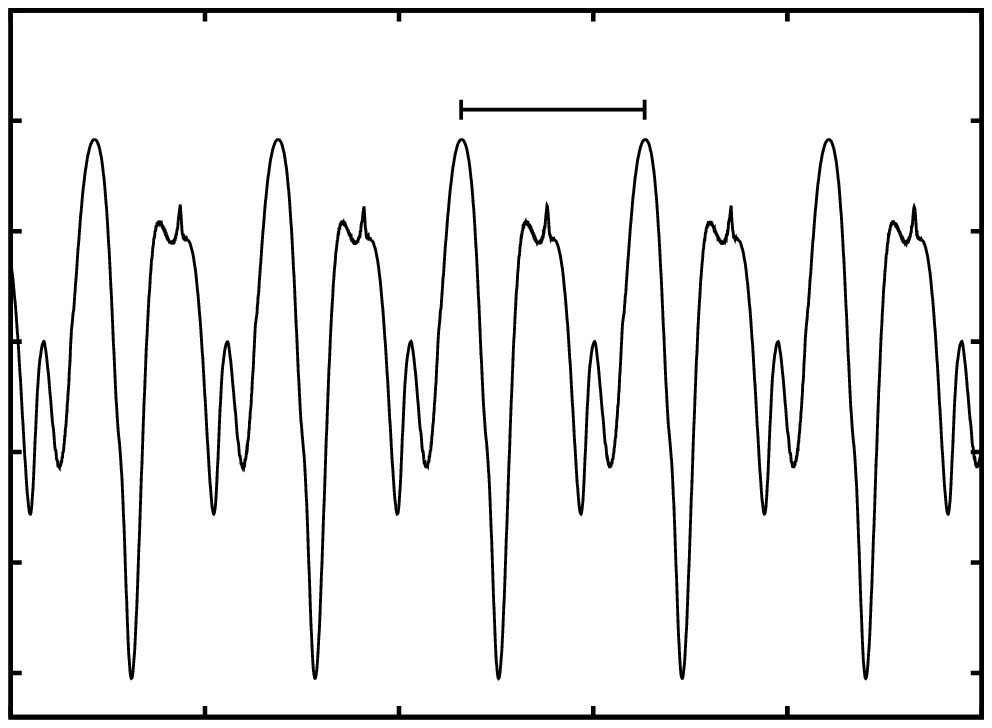} }
   \scalebox{0.455}{ \input{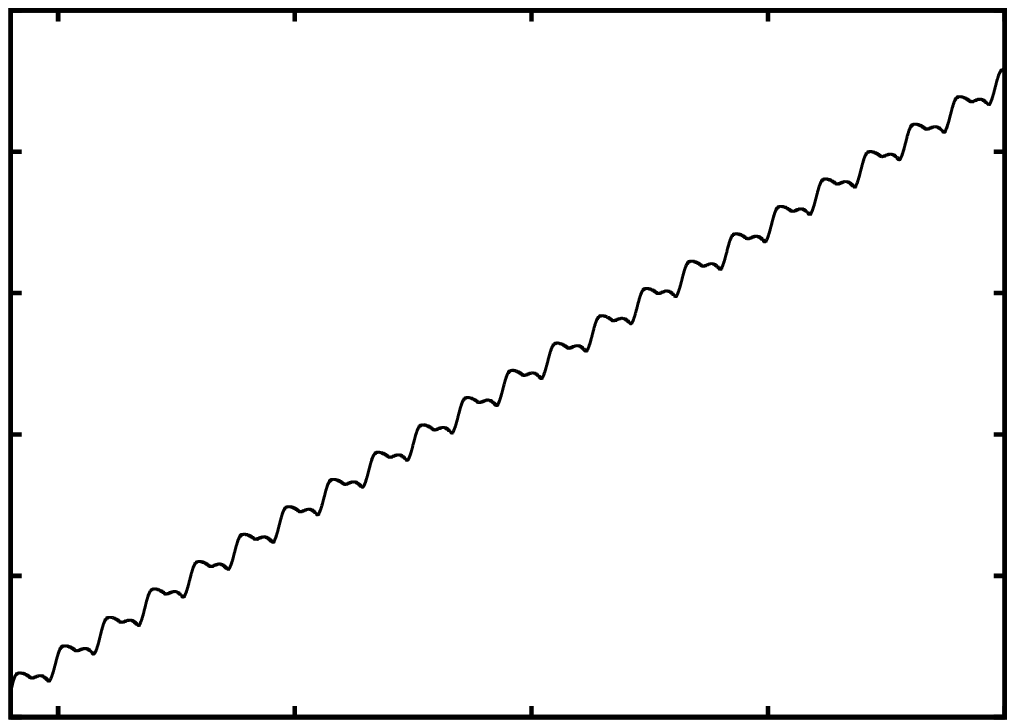} } \\

   \end{array}$
 \caption{Same as Fig. \ref{35m_nonlin} but for M = 30 M$_{\sun}$. }
 \normalsize
 \label{30m_fix_nonlin}
 \end{figure*}
 
 \begin{figure*}
\centering $
\LARGE
\begin{array}{ccc}
  \scalebox{0.455}{ \input{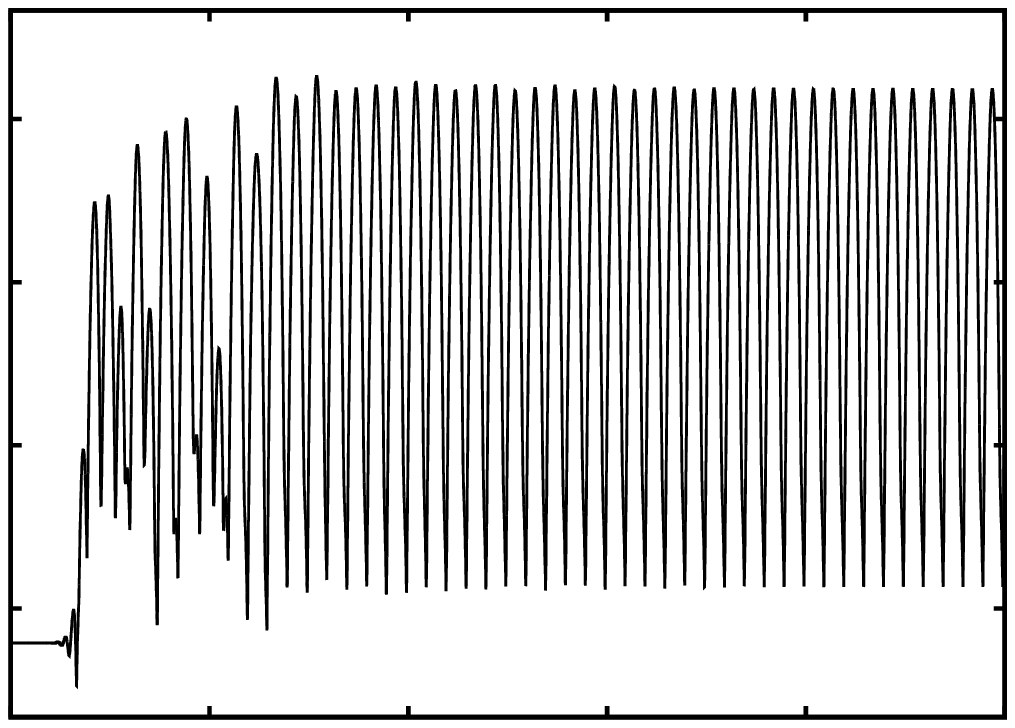} } 
  \scalebox{0.455}{ \input{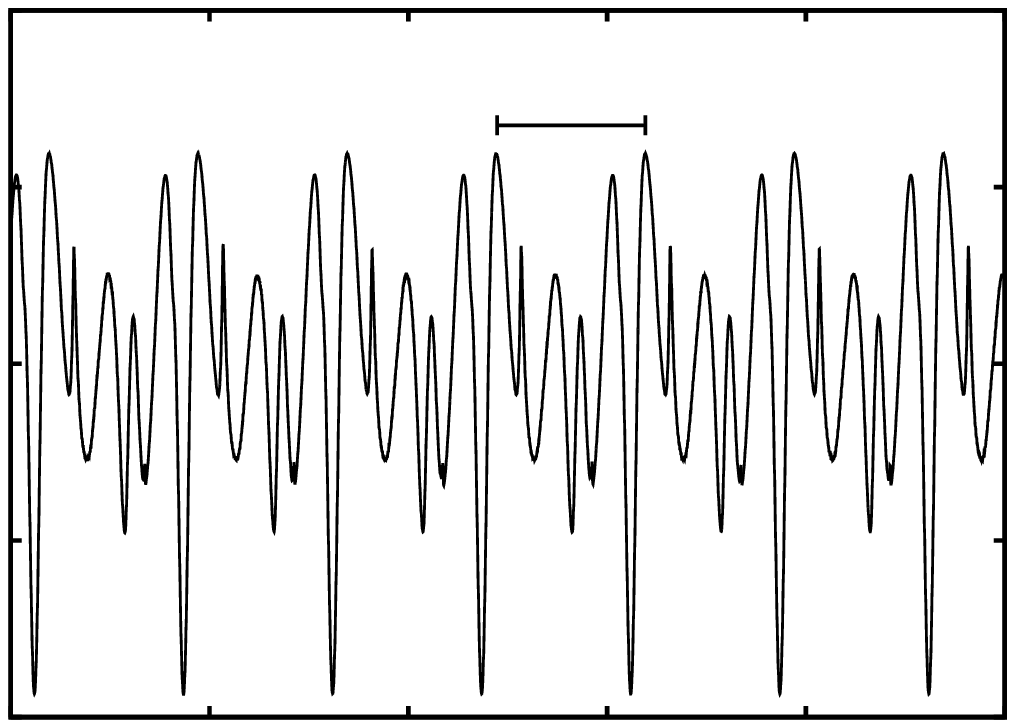} }
   \scalebox{0.455}{ \input{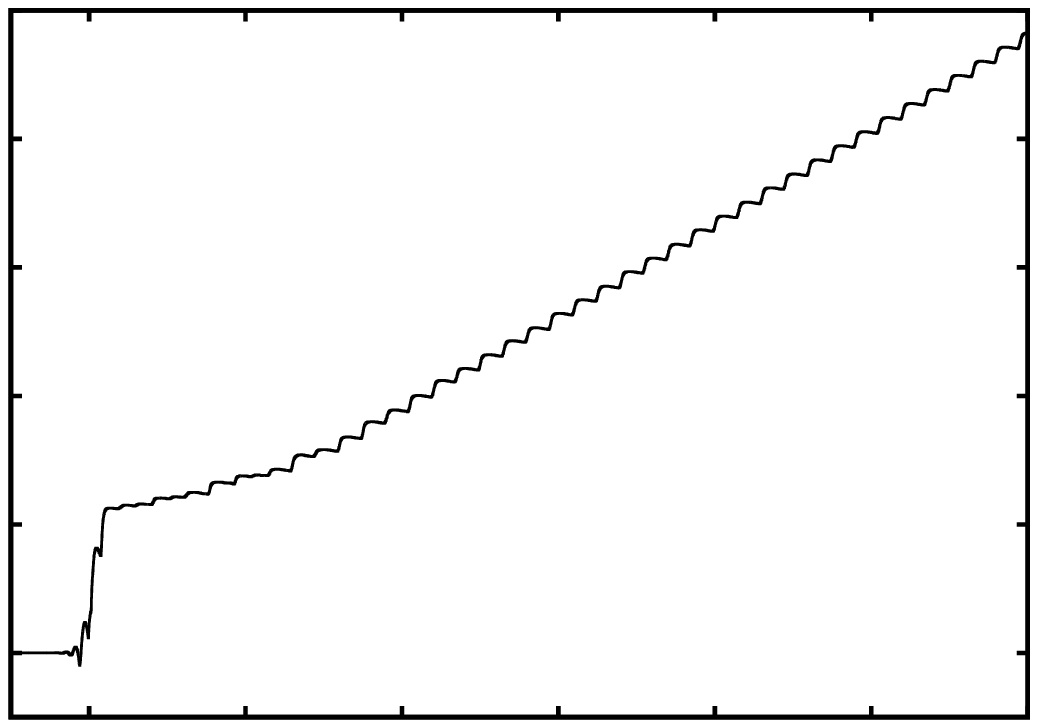} } \\

   \end{array}$
 \caption{Same as Fig. \ref{35m_nonlin} but for M = 27 M$_{\sun}$.}
 \normalsize
 \label{27m_fix_nonlin}
 \end{figure*}

For a model with 35 M$_{\sun}$, the results of the non-linear simulation are shown in Fig. \ref{35m_nonlin}.
As a function of time, the stellar radius, velocity, density and temperature at the 
 outermost grid point and the variation of the bolometric magnitude are given there.
 From the velocity we deduce that the evolution of the instability starts from hydrostatic equilibrium
 with velocity perturbations of the order of 10$^{-6}$ cm s$^{-1}$ 
 (the code picks up the correct unstable modes from numerical noise), undergoes the linear phase of exponential growth 
 and saturates in the non-linear regime with a velocity amplitude of 190 km s$^{-1}$. In this case, 
 the final result of the instabilities are finite amplitude pulsations with a period of 8.4 h (0.35 d). 
A moderate inflation of the star is a consequence of non-linearities. Compared to the hydrostatic value, the mean 
 radius is increased by approximately 10 per cent in the non-linear regime. 
 Fig. \ref{35m_nonlin} also contains the various terms involved
 in the energy balance, where the large hydrostatic values have been subtracted for a meaningful representation. 
 Potential and internal energy with almost identical modulus have opposite sign and nearly cancel each other.
 They are by three and four 
 orders of magnitude
 bigger than the kinetic and the time integrated acoustic energy, respectively, which are of main interest in our study.
 Due to the extreme differences in the order of magnitude of the various energies, the error in the energy balance 
 is particularly important, since we are interested in the smallest energy terms. The error in the energy balance is
 in our normalization given by the sum of all energy terms which is shown in Fig. \ref{35m_nonlin}(i).
 It is smaller than the smallest term in the energy balance by at least four orders of
 magnitude proving the extreme accuracy of our simulation and the reliability of the results. 
 The time integrated acoustic energy (see Fig. \ref{35m_nonlin}h) corresponds to the mechanical energy lost
 from the system by acoustic waves. During a pulsation cycle, there are phases of incoming and outgoing 
 acoustic fluxes. As a consequence, the time integrated acoustic energy is a non-monotonic function. However, integrated 
 over one cycle the outgoing energy exceeds the incoming energy and on average the time integrated acoustic energy
 increases with time. From Fig. \ref{35m_nonlin}(h), we obtain a well defined mean slope of the integrated 
 acoustic energy, which corresponds to a mean mechanical luminosity of the system. Assuming that this mean mechanical 
 luminosity is responsible for mass-loss of the star, we can estimate the mass-loss rate by comparing it to
 the wind kinetic
 luminosity $\frac{1}{2} \, \dot{M} \,v_{\infty} ^{2}$, 
 where $\dot{M}$ and $v_{\infty}$ are the mass-loss rate and terminal wind velocity, respectively 
 \citep[see also][]{grott_2005, yadav_2016, yadav_2017}. The terminal wind velocity is estimated by the escape velocity.
 Thus we obtain 
 from the mean slope of the time integrated acoustic energy (Fig. \ref{35m_nonlin}h), a mass-loss rate 
 of 3.9 $\times$ 10$^{-9}$ M$_{\sun}$ yr$^{-1}$ for the 35 M$_{\sun}$ model.

 The time integrated acoustic energy is made up of the fluxes of the kinetic and compression energies contained 
 in a sequence of hypersonic shock waves with associated overdense shells leaving the outer boundary of the integration domain. 
 In the atmosphere of the star where radiation transport needs to be considered in detail being beyond the scope of the 
 present study, the shock wave together with its overdense shell is then expected to undergo a snowplow phase sweeping up 
 the atmosphere material and transferring its momentum to the expanding shell which leaves the system, thus implying mass-loss. 
 Although this process is certainly not stationary, it provides the motivation to estimate the mass-loss rate by comparing
 the mean acoustic luminosity to the kinetic luminosity of a stationary wind with a terminal wind velocity of the order 
 of the escape speed.

The instabilities considered become stronger with increasing luminosity to mass ratios (see Fig. \ref{23_80_solar_modal}b). 
Therefore, we expect their final result to become more pronounced, i.e., the amplitudes of the finite amplitude 
pulsations to increase with L/M. The results for models having masses of 30 and 27 M$_{\sun}$ 
are displayed in Figs. \ref{30m_fix_nonlin} and \ref{27m_fix_nonlin}, respectively, similar to their counterpart 
in Fig. \ref{35m_nonlin} for 
the 35 M$_{\sun}$ model. The periods and velocity amplitudes of the finite amplitude pulsations amount to 0.94 d and 1.5 d and 
196 km s$^{-1}$ and 281 km s$^{-1}$ for the 
30 and 27 M$_{\sun}$ models, respectively. As expected, also the effect of inflation of the mean radius 
 becomes more pronounced with increasing L/M or strength of the instability, respectively. 
 The increase of the pulsation periods with L/M can then be understood by the period density relation, since 
 both lower masses and inflated radii imply lower densities. 
 The mean mechanical luminosity derived from the mean slope of the time integrated acoustic energy in
 Figs. \ref{30m_fix_nonlin}(c) and \ref{27m_fix_nonlin}(c) provides an estimate for the 
 mass-loss rates of the 30 and 27 M$_{\sun}$ models of 
  8.9 $\times$ 10$^{-9}$ M$_{\sun}$ yr$^{-1}$ and 
7.0 $\times$ 10$^{-8}$ M$_{\sun}$ yr$^{-1}$, respectively.
Again, the increase of the mass-loss rate with L/M supports the idea of a direct relation between the strength of the instability 
and mass-loss. Errors in the energy balance are similar to those shown for the 35 M$_{\sun}$ model.

 \begin{figure*}
\centering $
\LARGE
\begin{array}{ccc}
  \scalebox{0.455}{ \input{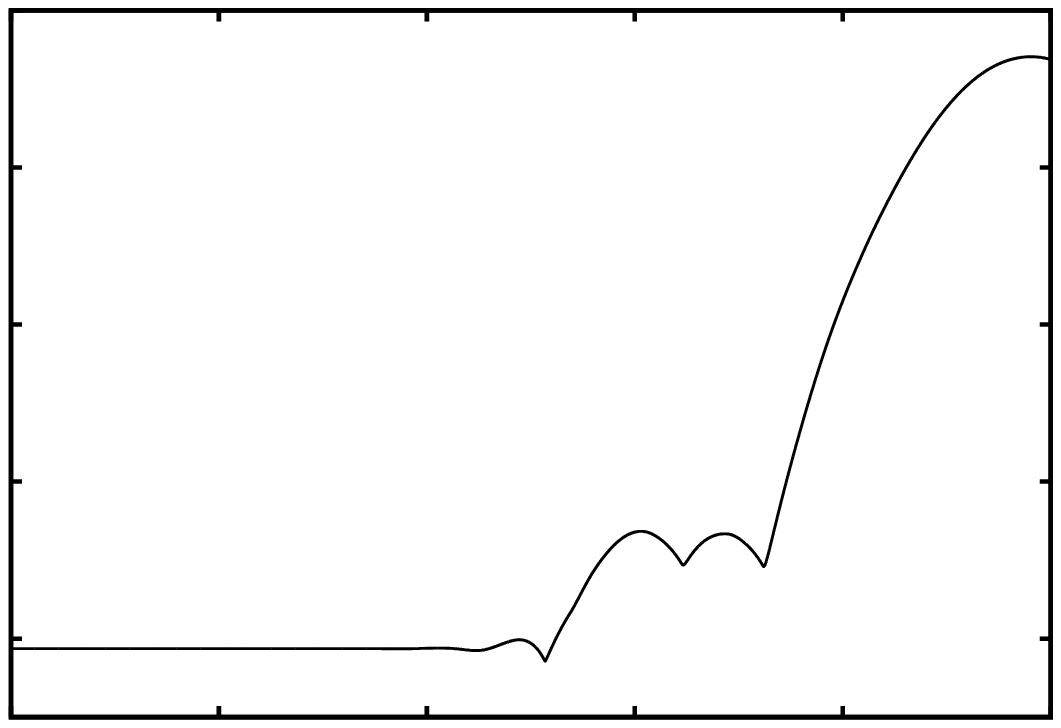} } 
  \scalebox{0.455}{ \input{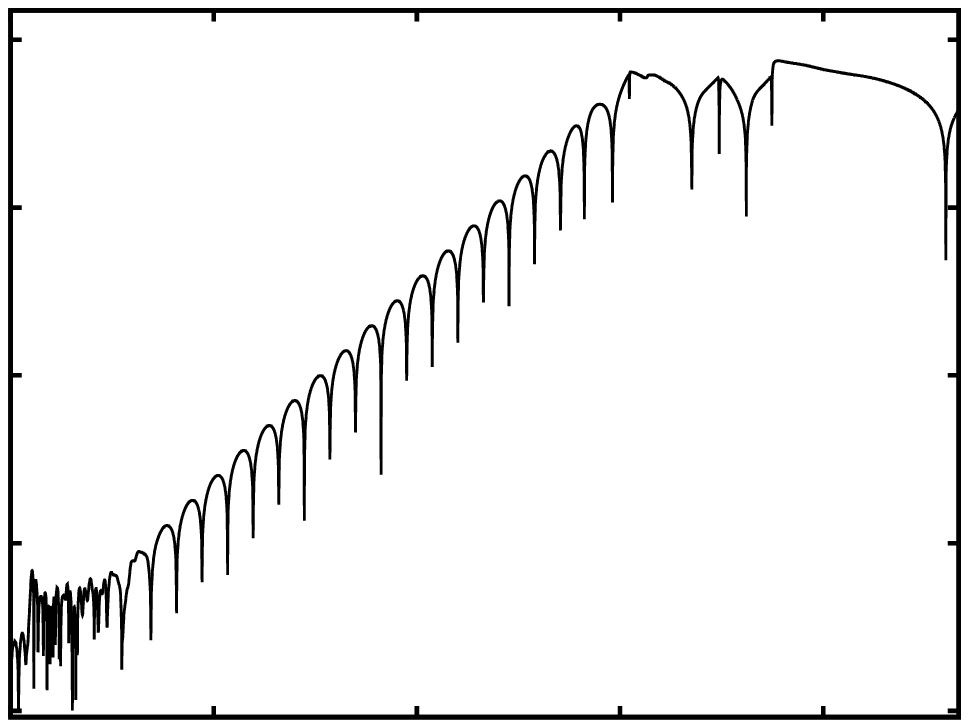} }
   \scalebox{0.455}{ \input{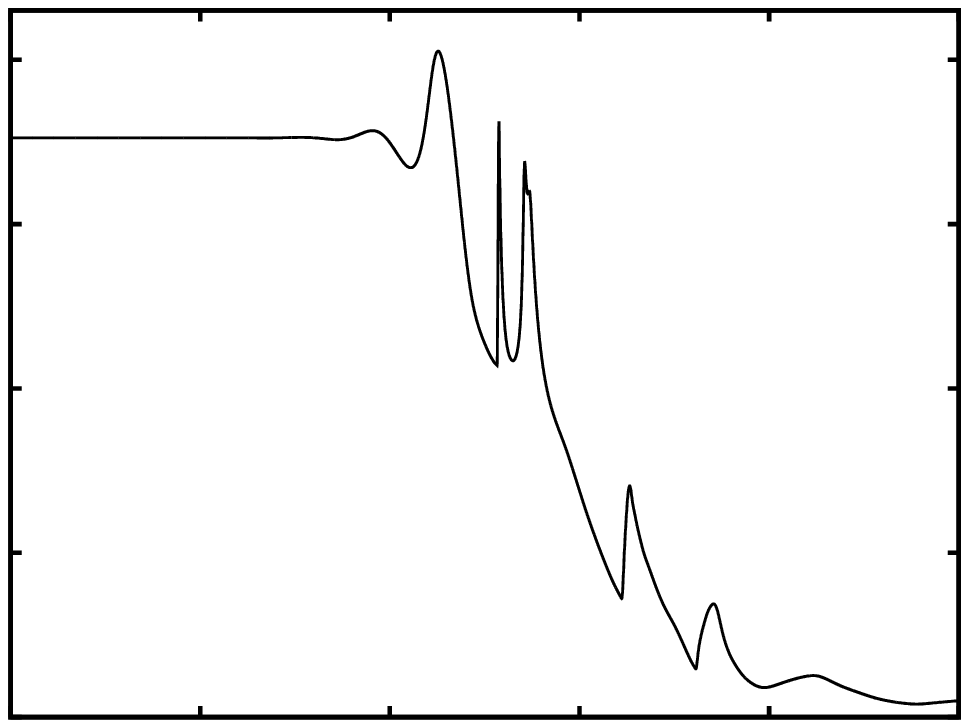} } \\

   \end{array}$
 \caption{Same as Fig. \ref{35m_nonlin} but for M = 23 M$_{\sun}$. 
  For this model the radius is inflated substantially which is associated with a strong temperature decrease.
  Due to the latter, the simulation had to be stopped because opacity data were not available for low temperatures.}
 \normalsize
 \label{23m_nonlin}
 \end{figure*}

The 23 M$_{\sun}$ model having the highest L/M ratio of the models
selected suffers from the most violent instabilities found. 
The results of the simulation of the evolution of these instabilities into the
 non-linear regime are presented in Fig. \ref{23m_nonlin}. The velocity at the outermost grid point shows that 
 the instability starts from hydrostatic equilibrium with perturbations of the order of 10$^{-6}$ cm s$^{-1}$ and saturates
 after the linear phase of exponential growth in the non-linear regime with a maximum velocity of 315 km s$^{-1}$. Associated with 
 these high velocities is an inflation of the stellar radius by a factor of approximately five. As a consequence,
 the temperature at the outermost grid point drops to approximately 5000 K. For even lower temperatures, opacities were not 
 available and we had to stop the simulation. The maximum velocity of 315 km s$^{-1}$ has to be compared with the escape velocity of 
 390 km s$^{-1}$. Even if the evolution could not be followed further, we may take this result as an indication for direct mass-loss. 
 Again, errors in the energy balance are similar to those shown for the 35 M$_{\sun}$ model.

 \subsection{Evolutionary models}
 For selected models on the evolutionary tracks shown in Fig. \ref{hrd}, which have been determined to be unstable 
 by the linear analysis in section 3, we have performed numerical simulations of the evolution of the instabilities 
 into the non-linear regime to determine the final fate of the corresponding objects.

 \subsubsection{Models having log T$_{\rm{eff}}$ = 4.6}
 \label{46}
  
 \begin{figure*}
\centering $
\LARGE
\begin{array}{cccccc}
  \scalebox{0.455}{ \input{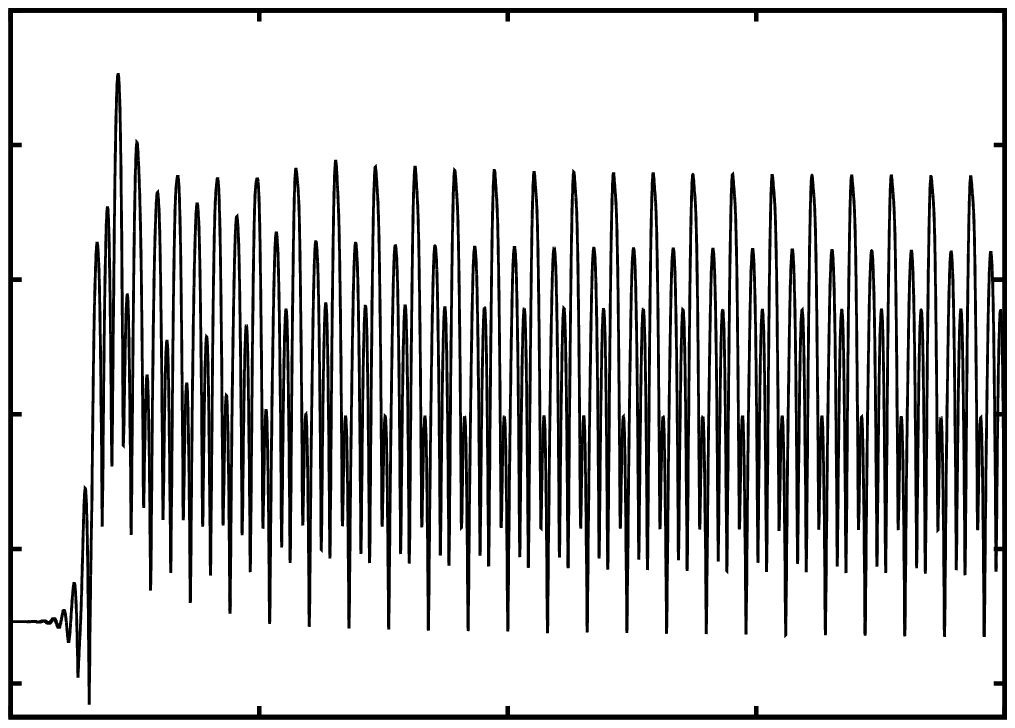} } 
  \scalebox{0.455}{ \input{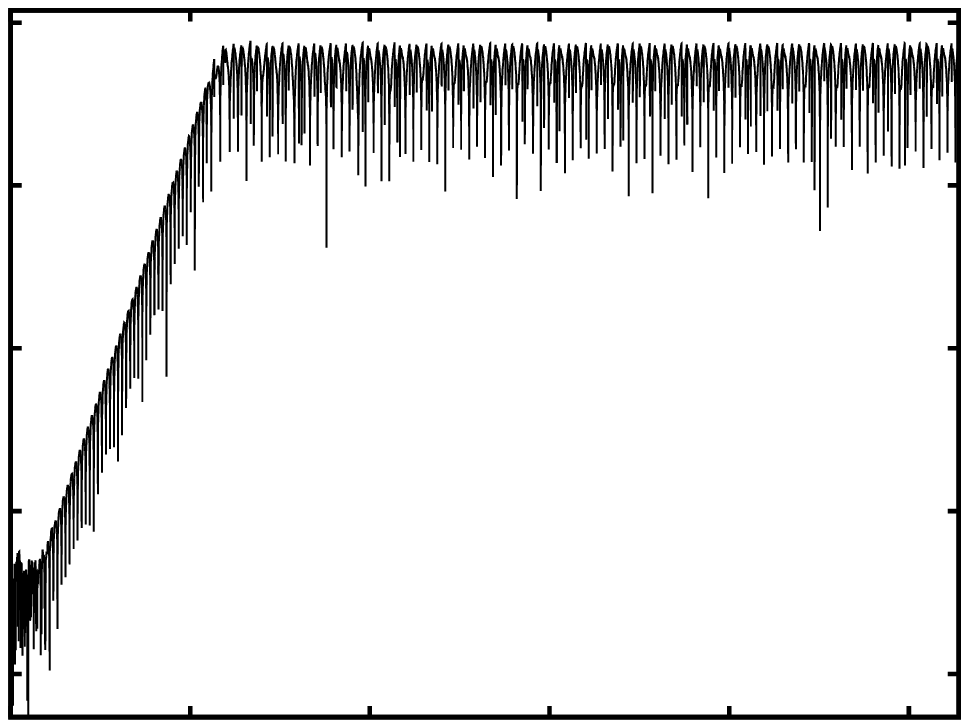} }
   \scalebox{0.455}{ \input{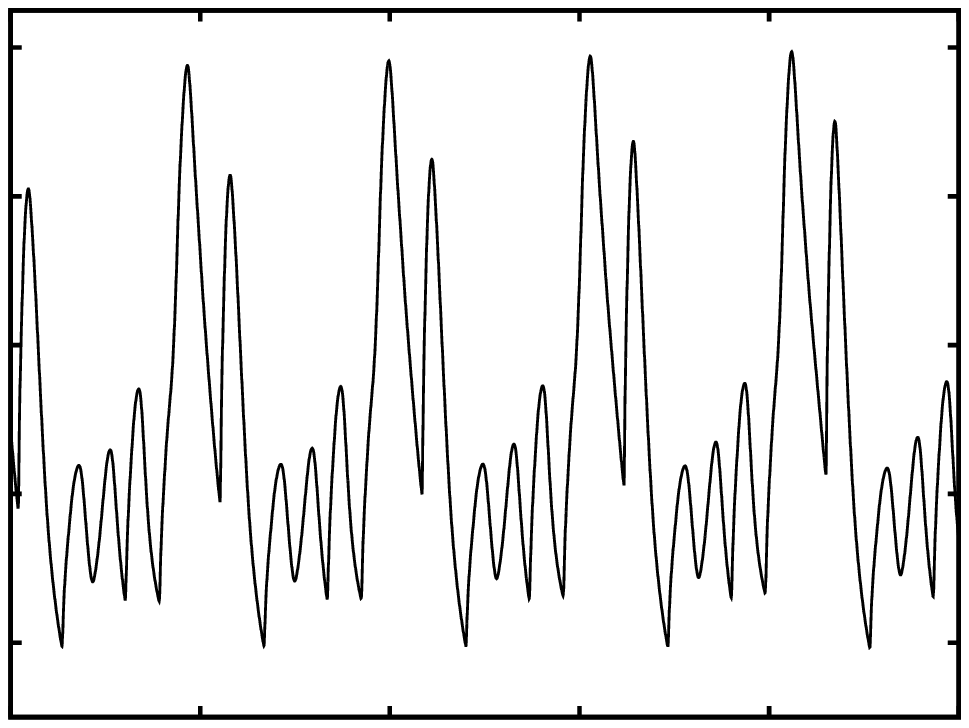} } \\

    \scalebox{0.455}{ \input{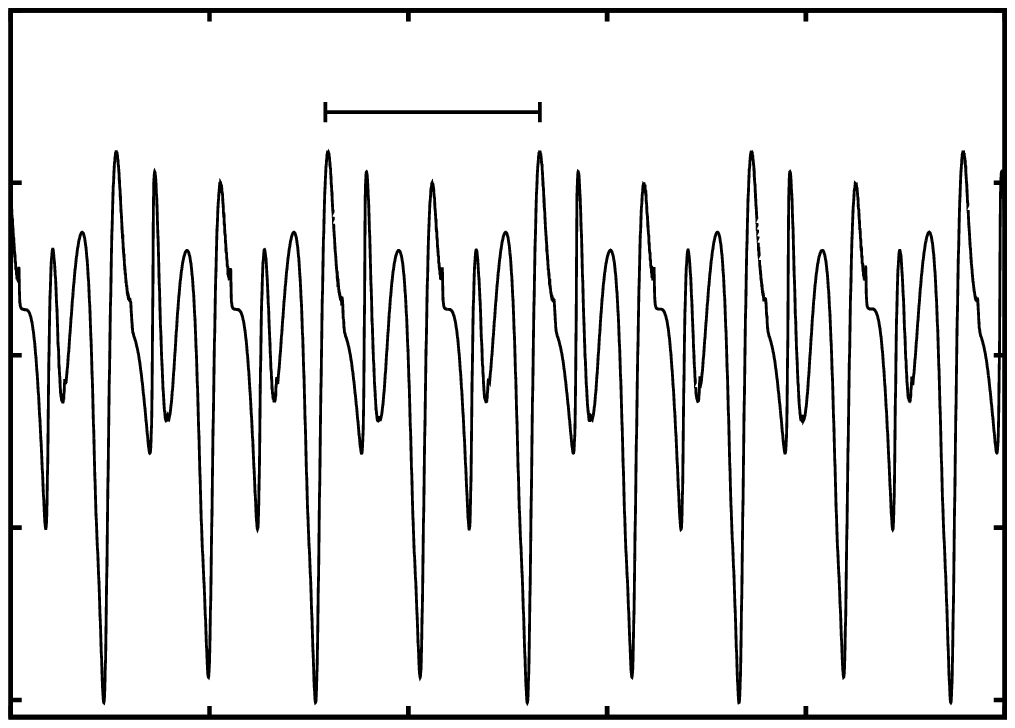} } 
  \scalebox{0.455}{ \input{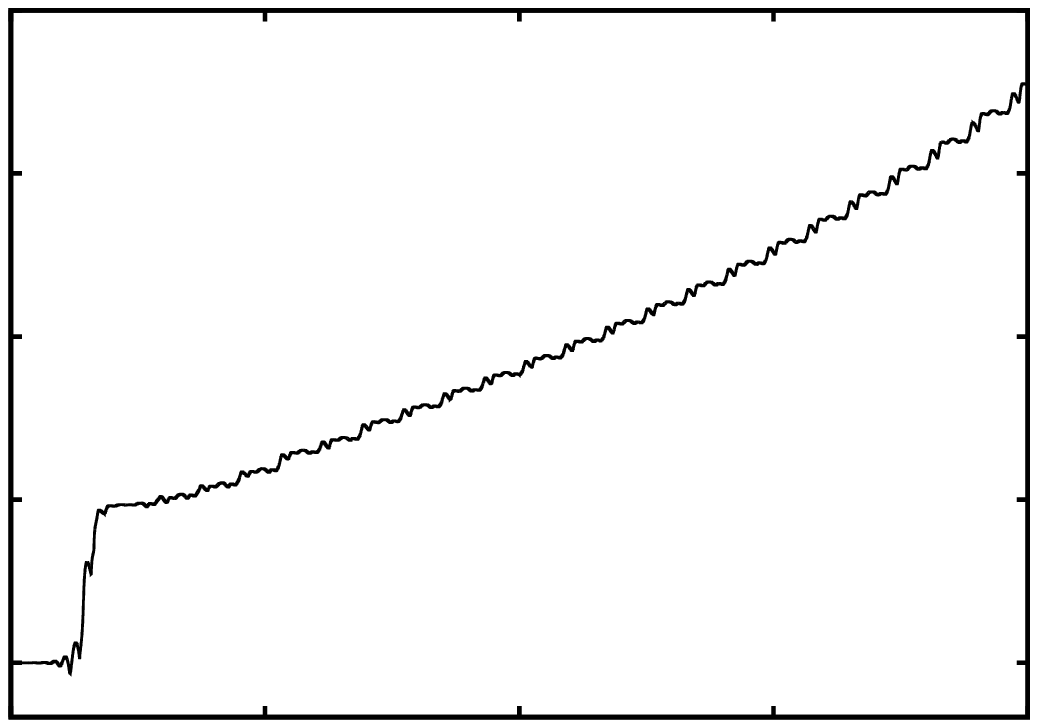} }
   \scalebox{0.455}{ \input{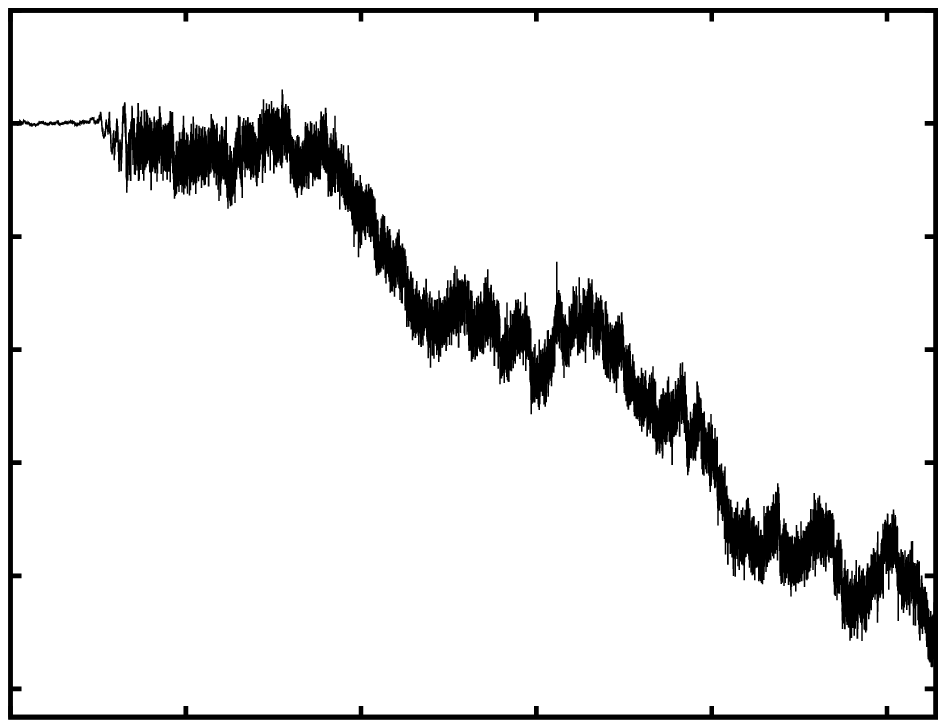} } \\
   \end{array}$
 \caption{Evolution of the instability into the non-linear regime and finite amplitude pulsations 
 for a model having log T$_{\rm{eff}}$ = 4.6 on the evolutionary track of a star with an initial mass of 100 M$_{\sun}$
 (see Fig. \ref{hrd}). As a function of time, the stellar radius, the velocity and temperature at the 
 outermost grid point are shown in (a) - (c), respectively, the variation of the bolometric magnitude is given in (d).
 The velocity amplitude reaches 255 km s$^{-1}$ in the non-linear regime. The time integrated acoustic energy 
 (being the smallest term in the energy balance) and the error of the energy balance are displayed in (e) and (f), respectively.}
 \normalsize
 \label{100m_4p6}
 \end{figure*}
 
\begin{figure*}
\centering $
\LARGE
\begin{array}{cccccc}
  \scalebox{0.455}{ \input{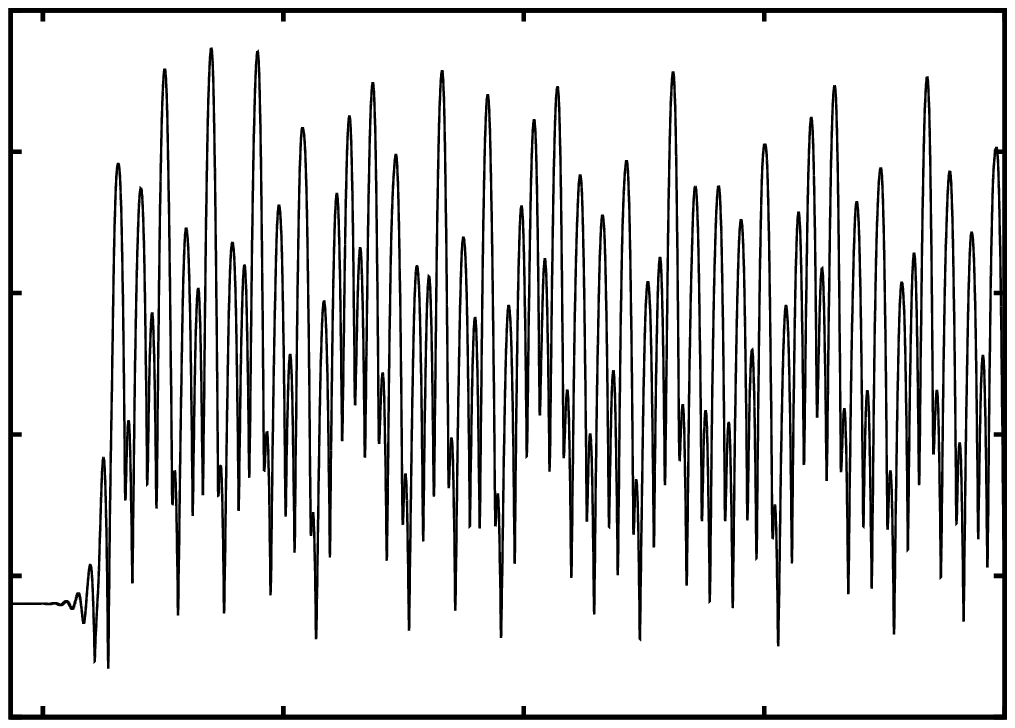} } 
  \scalebox{0.455}{ \input{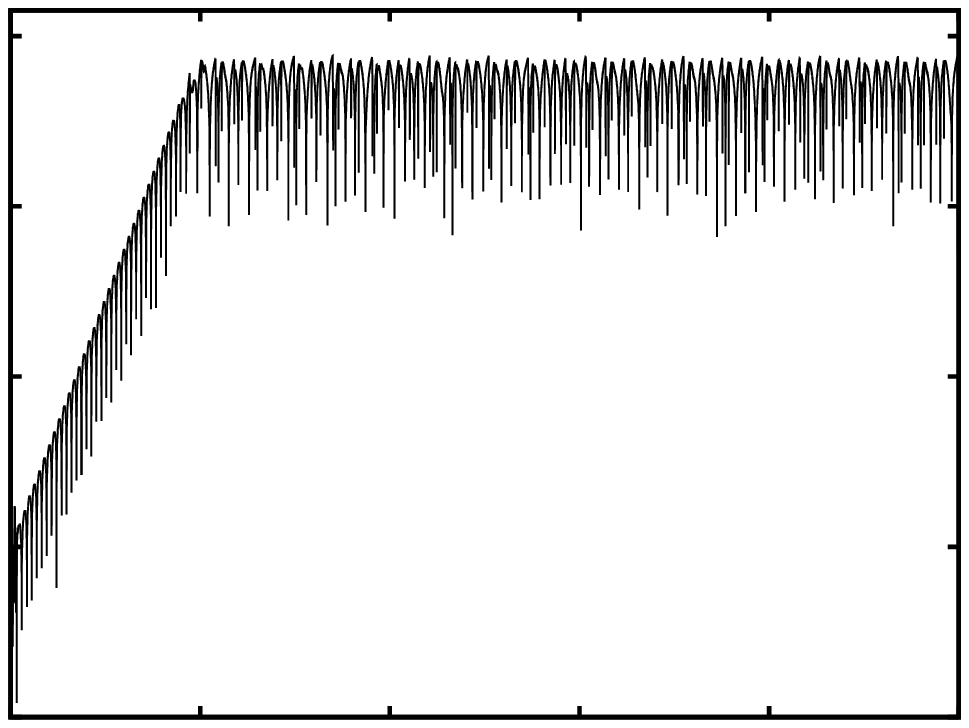} }
   \scalebox{0.455}{ \input{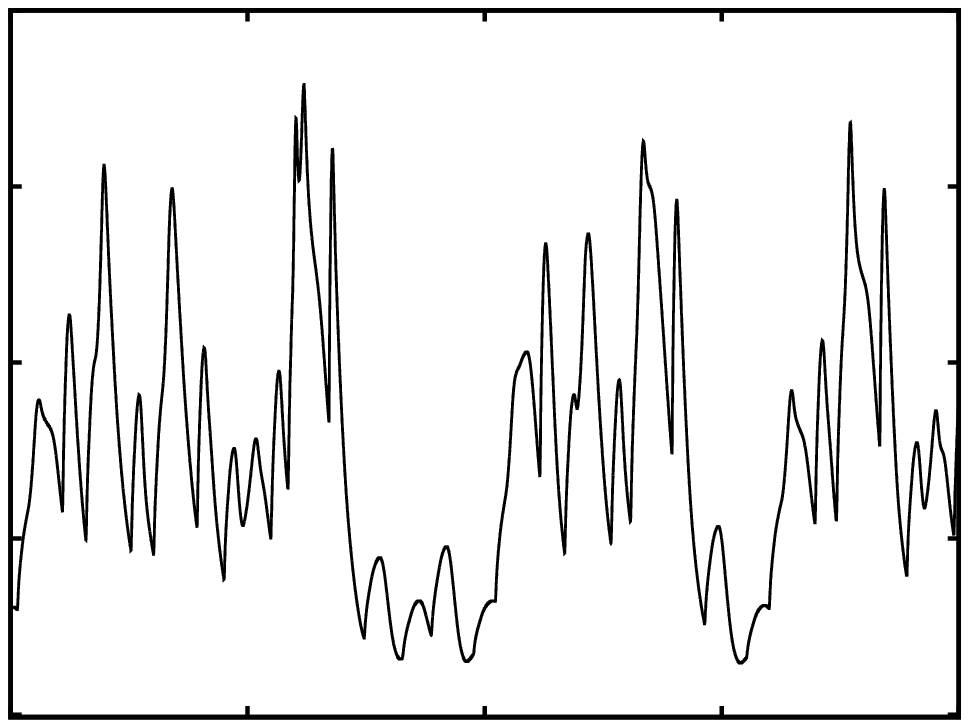} } \\
  \scalebox{0.455}{ \input{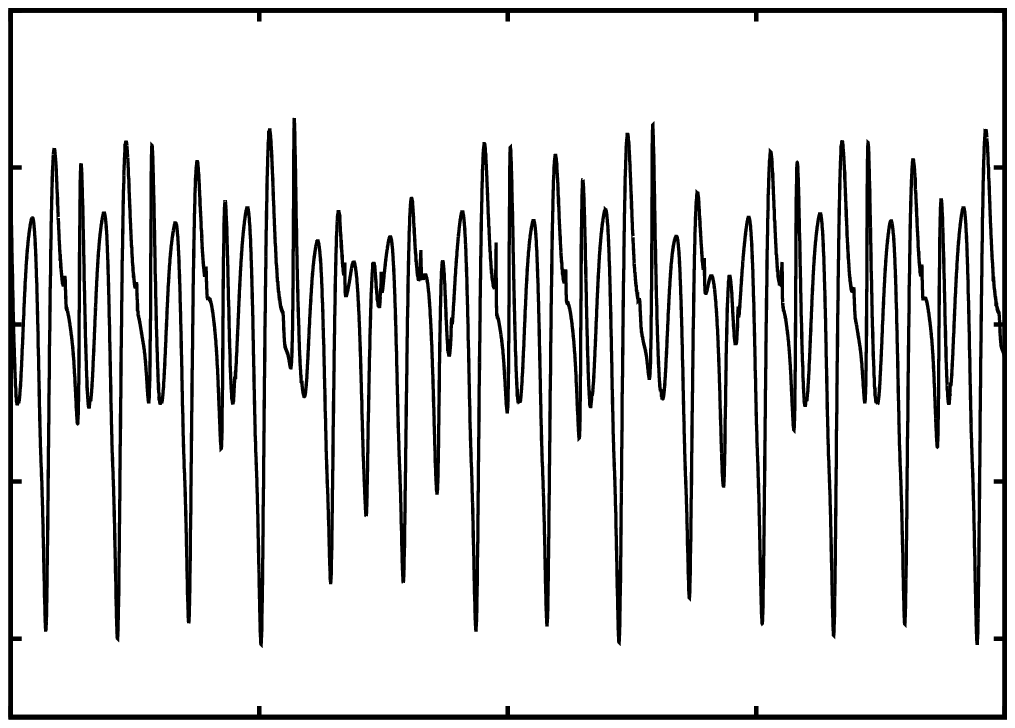} } 
  \scalebox{0.455}{ \input{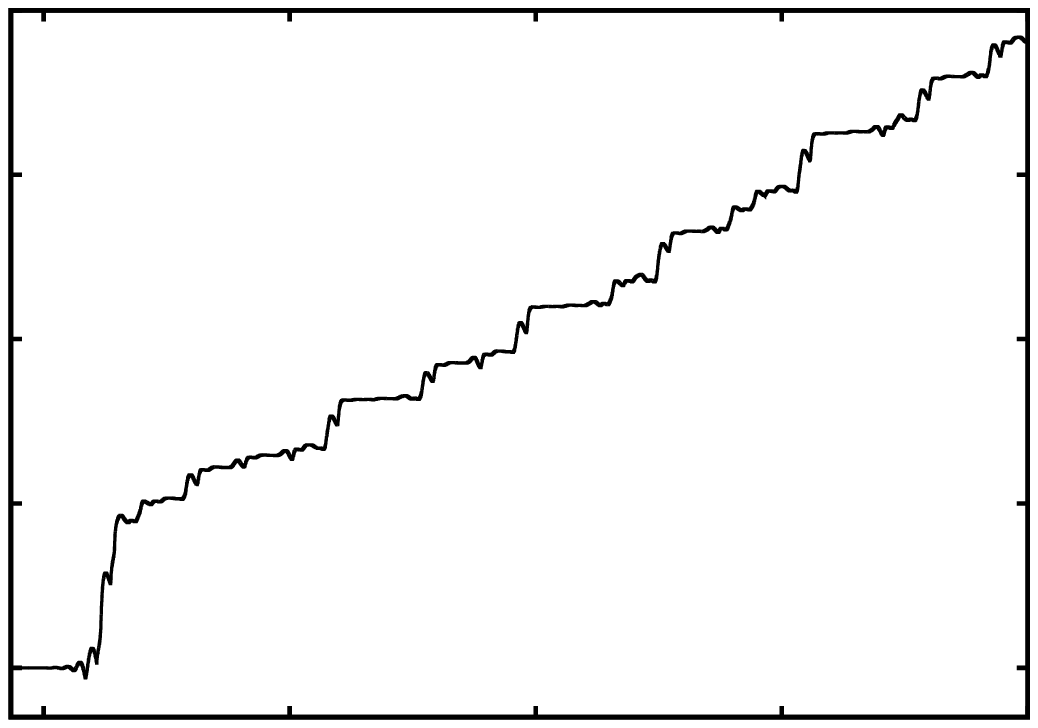} }
   \scalebox{0.455}{ \input{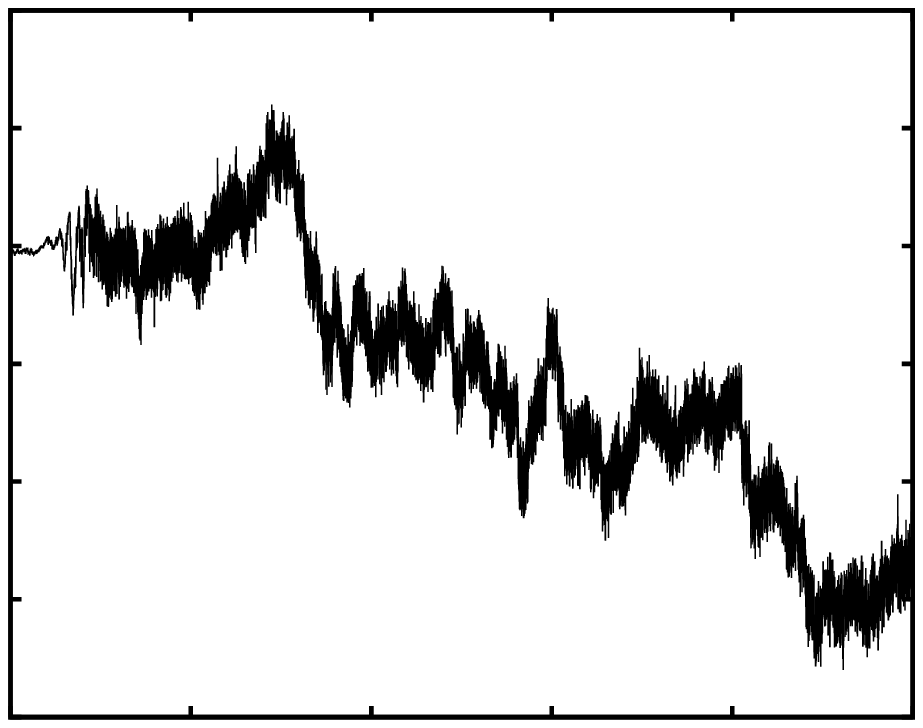} } \\
  \end{array}$
 \caption{Same as Fig. \ref{100m_4p6} but for a star with an effective temperature of log T$_{\rm{eff}}$ = 4.6 and 
 an initial mass of 70 M$_{\sun}$. The finite amplitude pulsation 
 does not exhibit a strictly periodic pattern. The velocity amplitude reaches 262 km s$^{-1}$ in the non-linear regime.}
 \normalsize
 \label{70m_4p6}
 \end{figure*}

 For an effective temperature of log T$_{\rm{eff}}$ = 4.6,
 only models on the evolutionary tracks for initial masses of 100 and 70 M$_{\sun}$ are unstable.  
 The results of our numerical simulations of the evolution of these unstable models 
 are displayed in Figs. \ref{100m_4p6} and \ref{70m_4p6}, respectively. 
 In these figures, the stellar radius (a), the velocity (b) and temperature (c)
at the outermost grid point, the variation of the bolometric magnitude (d), the time integrated acoustic energy (e) and 
 the error in the energy balance (f) are given as a function of time. 
 For both models, the final fate appears to be 
 finite amplitude pulsation where the mean radius is slightly inflated and velocity amplitudes of 255 and 262 km s$^{-1}$ (corresponding
 to $\approx$30 per cent of the escape velocity) are attained. 
  Similar to the previously discussed models, the evolution starts from hydrostatic equilibrium 
  with velocity perturbations of the order of 10$^{-6}$ cm s$^{-1}$ on the numerical noise level, 
   goes through the linear phase of exponential growth and finally saturates in the non-linear regime.
 For the model with an initial mass of 100 M$_{\sun}$ in the non-linear regime, a
 strictly periodic pattern with a period of 3.2 d can be 
 identified, whereas for the 70 M$_{\sun}$ model the pattern is no longer strictly periodic, although at least one 
 dominant period may be deduced from the variation of the bolometric magnitude.

 As in the previous section, the mass-loss rate is estimated by comparing the mean slope of the 
 time integrated acoustic energy
  to the kinetic wind luminosity. The mean slope of the time integrated acoustic energy is well defined 
  for the 70 M$_{\sun}$ model (see Fig. \ref{70m_4p6}) allowing for a unique determination of a mass-loss rate,  whereas 
  the mean slope varies for the 100 M$_{\sun}$ model and only a range of mass-loss rates together with a maximum mass-loss
  rate can be estimated. Accordingly, we obtain a mass-loss rate of  2.1 $\times$ 10$^{-7}$ M$_{\sun}$ yr$^{-1}$ for the 
  70 M$_{\sun}$ model from Fig. \ref{70m_4p6} and a maximum mass-loss rate of  
  1.9 $\times$ 10$^{-7}$ M$_{\sun}$ yr$^{-1}$ for the 100 M$_{\sun}$ model from Fig. \ref{100m_4p6}.  
  From the run of the time integrated acoustic energy (see Fig. \ref{100m_4p6}), we suspect that the latter might be even 
  higher if we would have followed the evolution for more than 100 d. The errors in the energy 
   balance given in Figs. \ref{100m_4p6}(f) and \ref{70m_4p6}(f) are smaller than the smallest term in the 
   energy balance (the time integrated acoustic energy) by more than four orders of magnitude. The other terms in the energy 
   balance exhibit similar orders of magnitude as discussed for the models in the previous section.

 \subsubsection{Models having log T$_{\rm{eff}}$ = 4.45}
\begin{figure*}
\centering $
\LARGE
\begin{array}{cccccc}
  \scalebox{0.455}{ \input{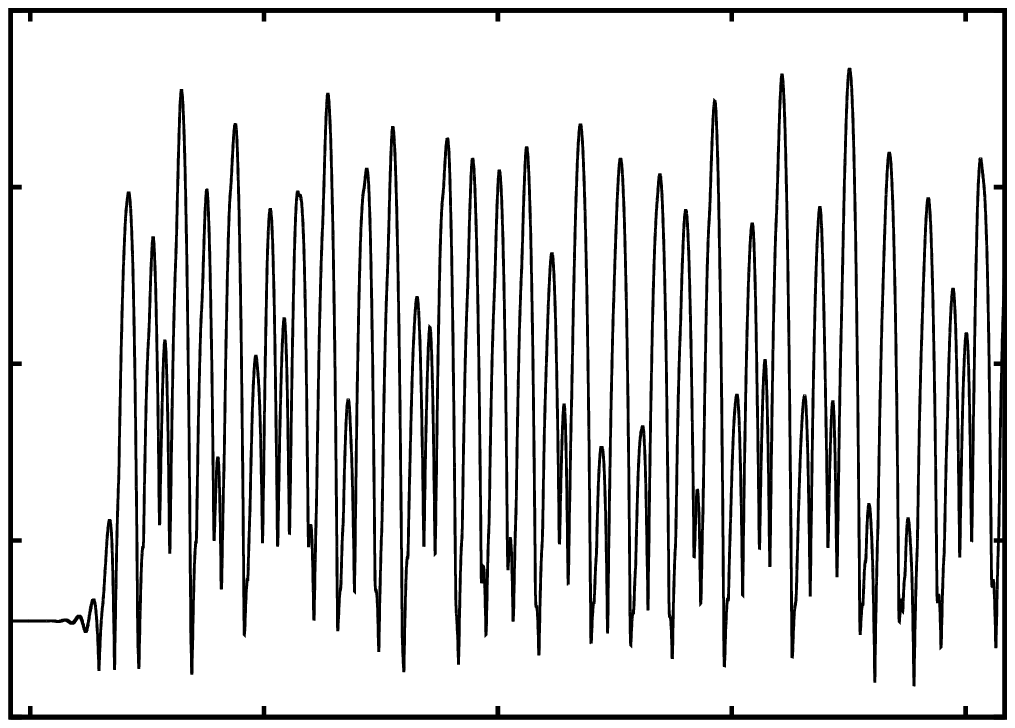} } 
  \scalebox{0.455}{ \input{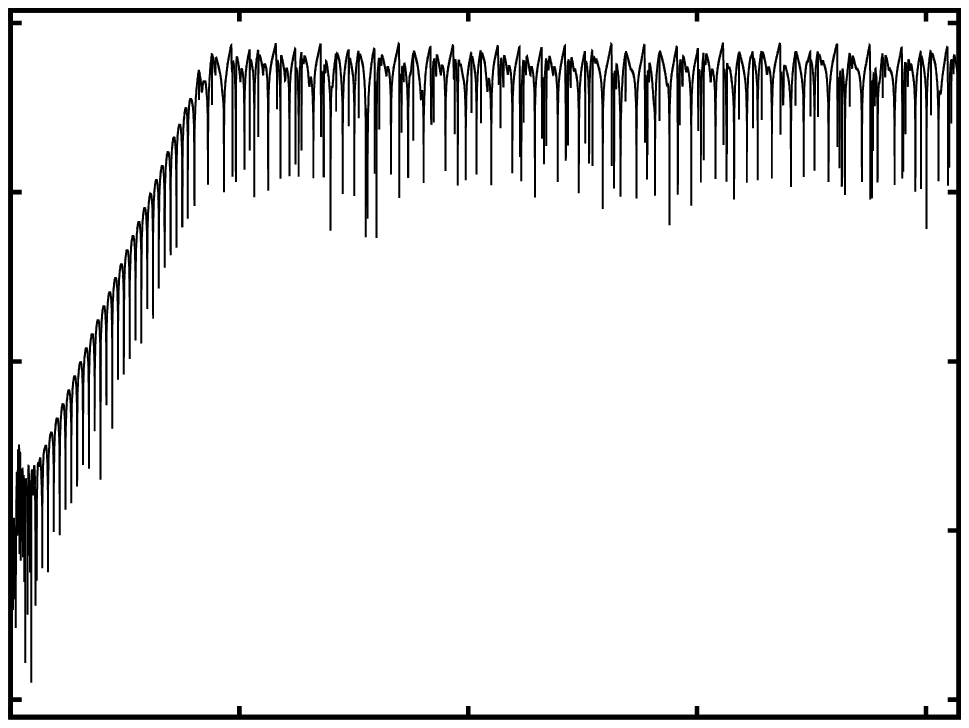} }
   \scalebox{0.455}{ \input{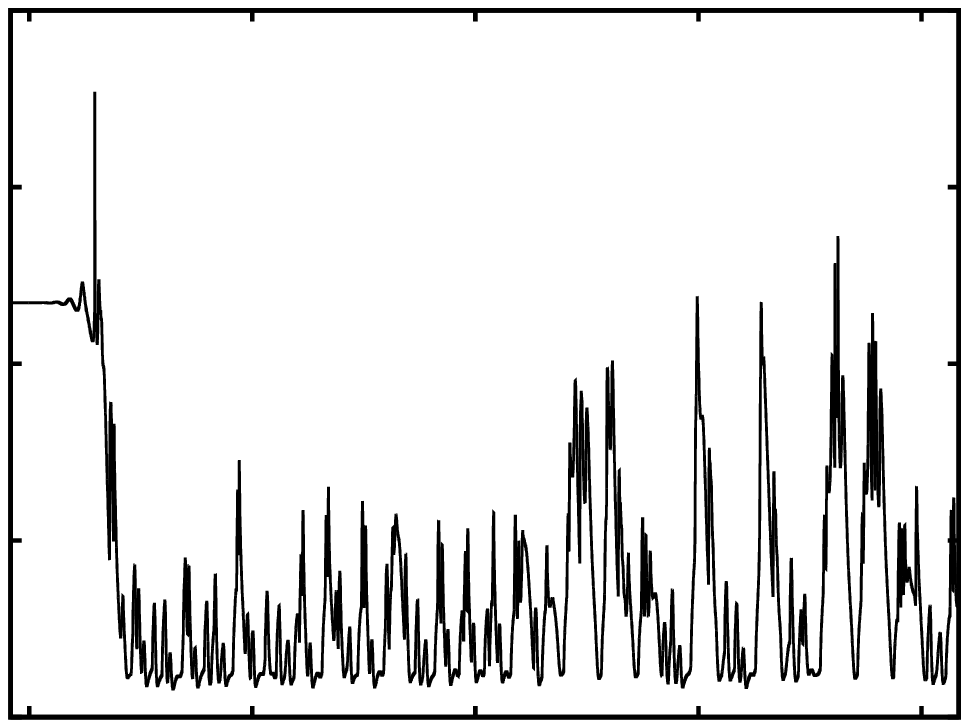} } \\

    \scalebox{0.455}{ \input{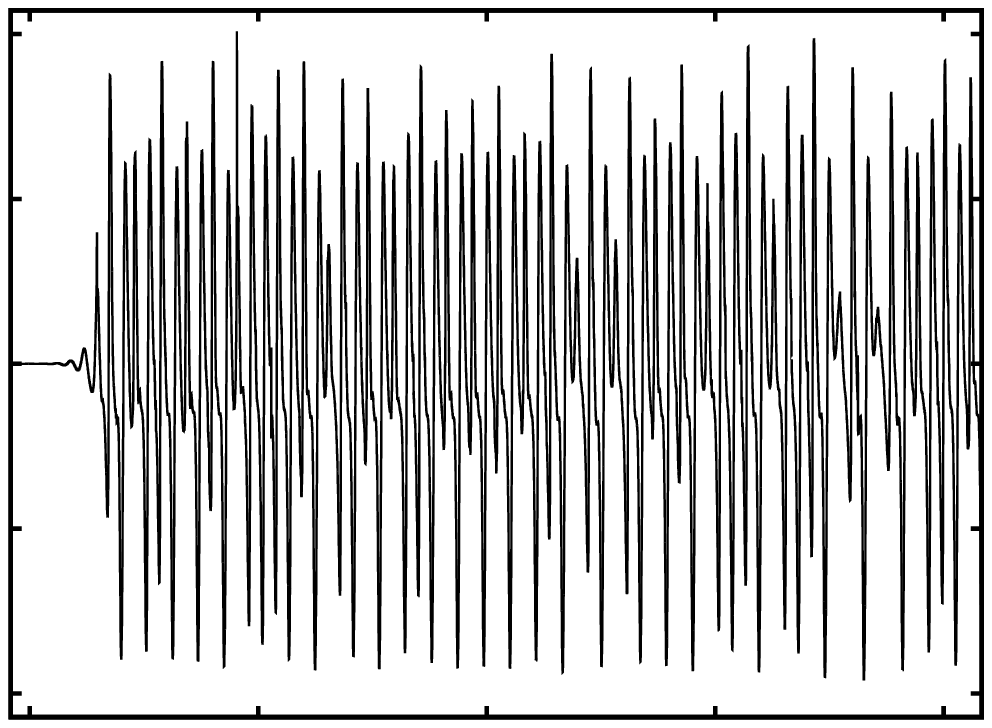} } 
  \scalebox{0.455}{ \input{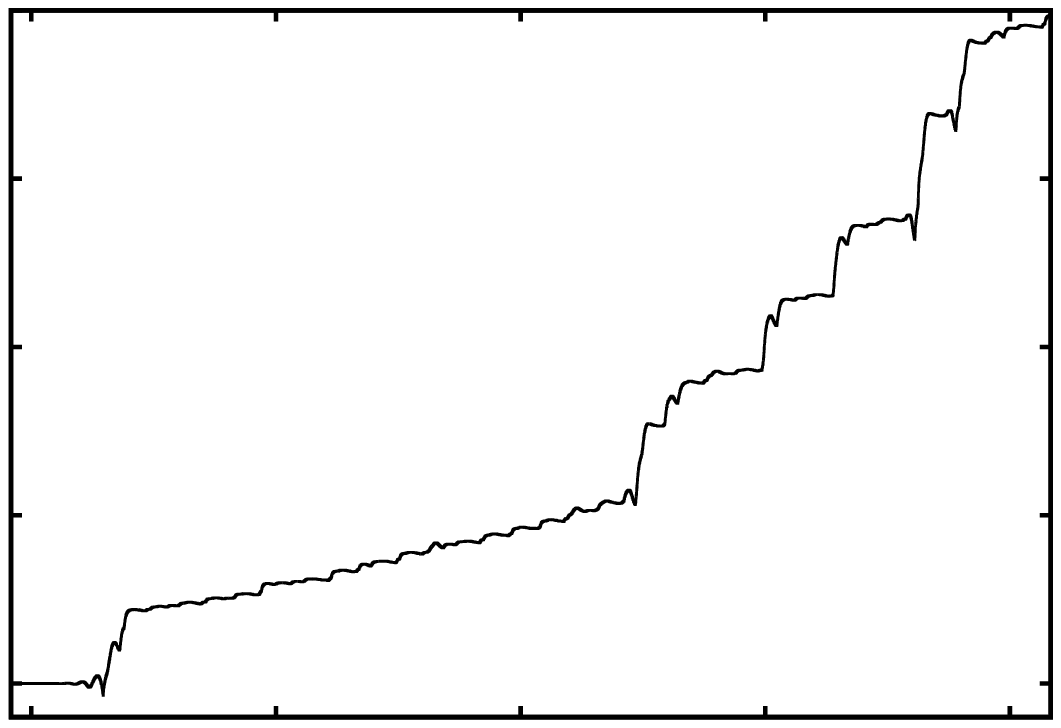} }
   \scalebox{0.455}{ \input{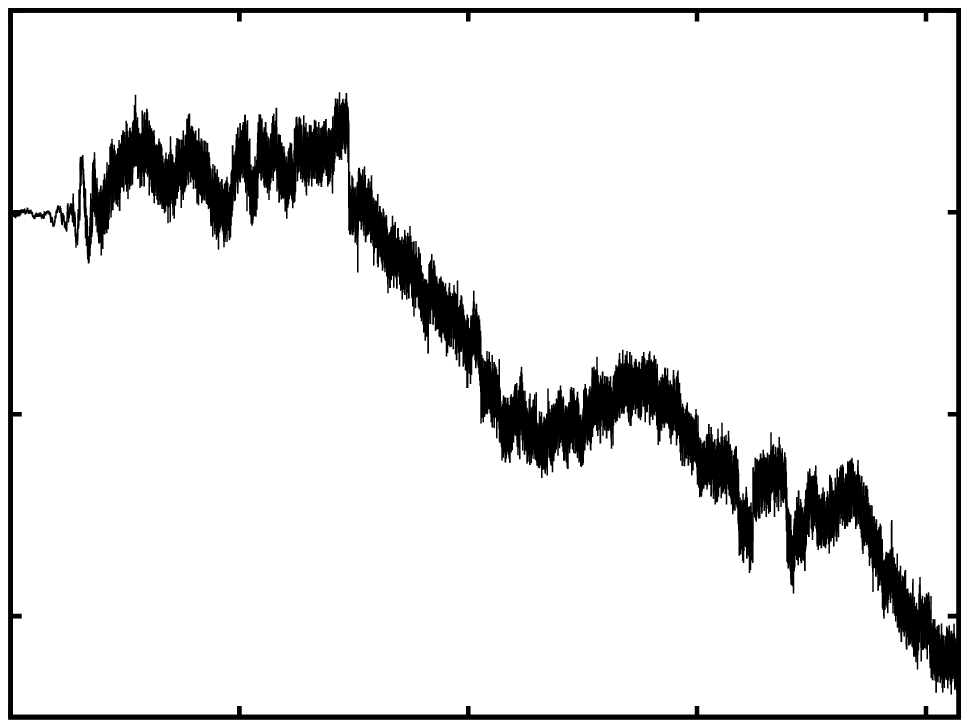} } \\
   \end{array}$
 \caption{Same as Fig. \ref{100m_4p6} but for a model having log T$_{\rm{eff}}$ = 4.45 and an initial mass of 100 M$_{\sun}$. 
 Similar to Fig. \ref{70m_4p6} the finite amplitude pulsation 
 does not exhibit a strictly periodic pattern. The velocity amplitude reaches 215 km s$^{-1}$ in the non-linear regime.}
 \normalsize
 \label{100m_4p45}
 \end{figure*}

\begin{figure*}
\centering $
\LARGE
\begin{array}{cccccc}
  \scalebox{0.455}{ \input{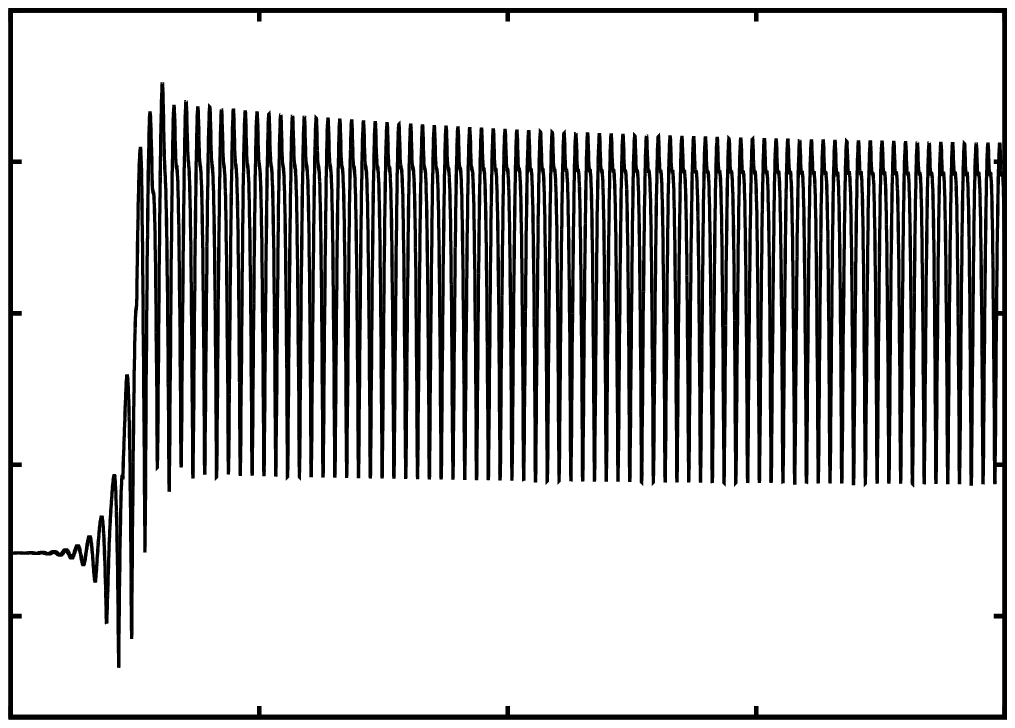} } 
  \scalebox{0.455}{ \input{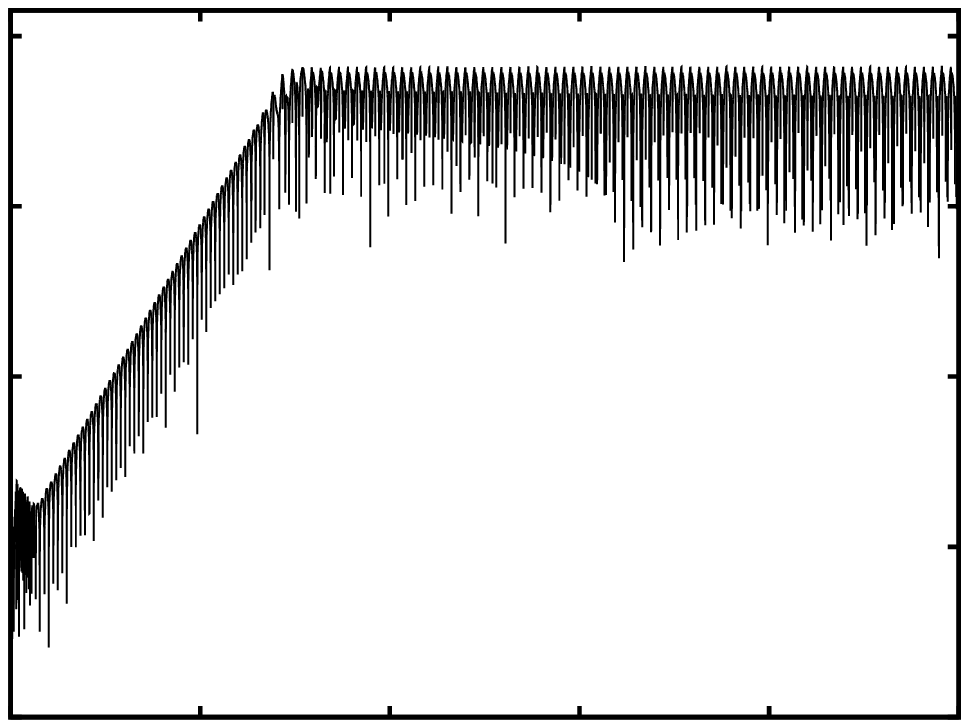} }
    \scalebox{0.455}{ \input{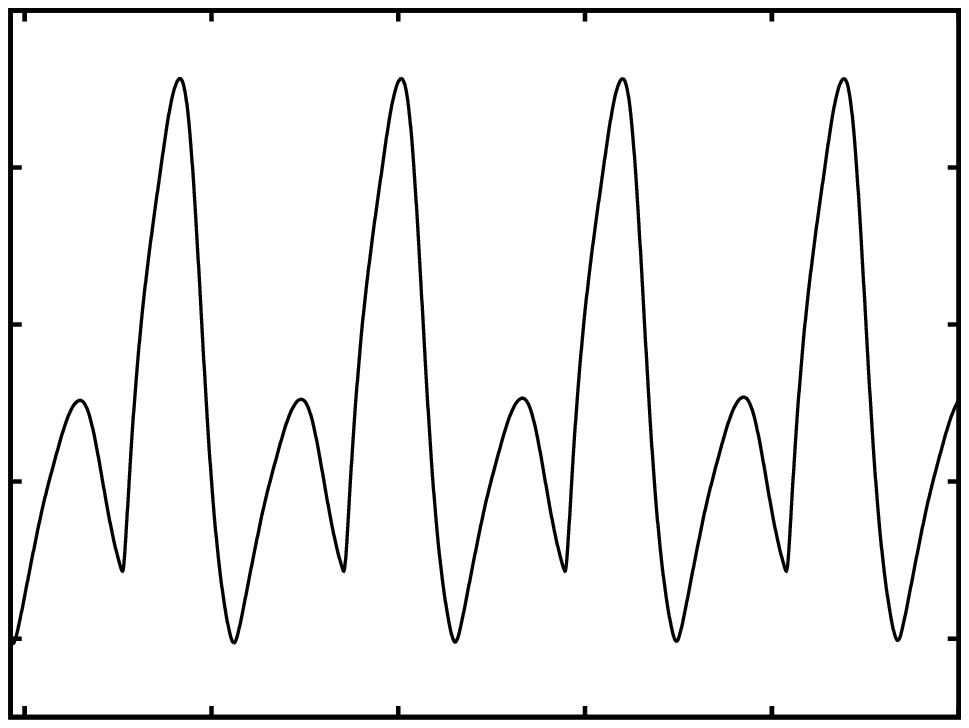} } \\
  \scalebox{0.455}{ \input{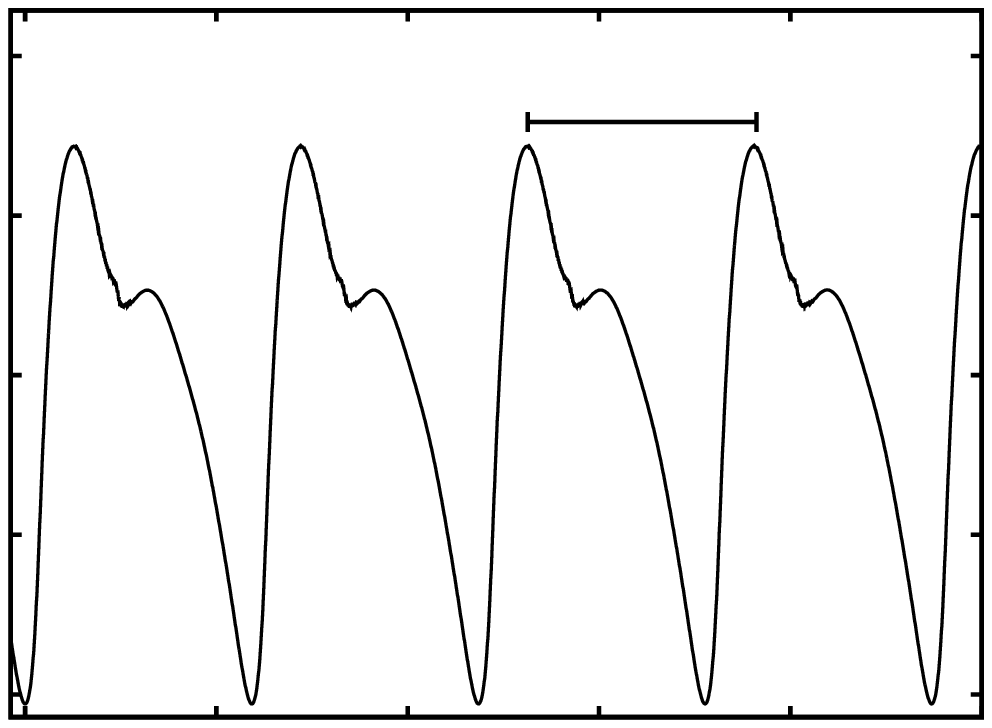} }
  \scalebox{0.455}{ \input{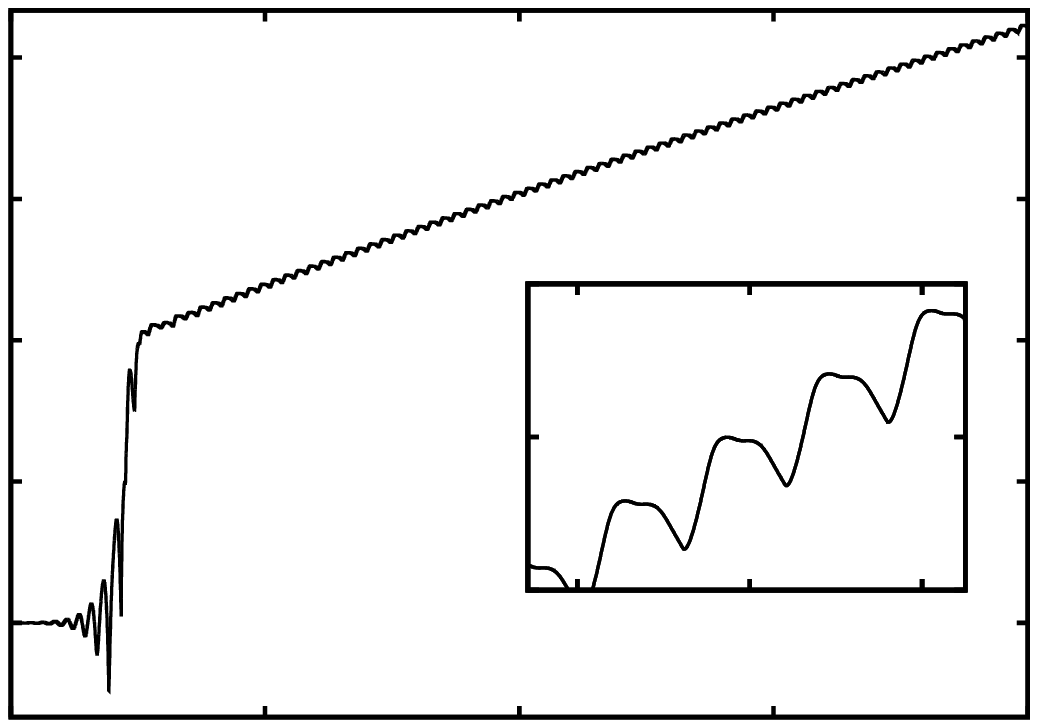} }
   \scalebox{0.455}{ \input{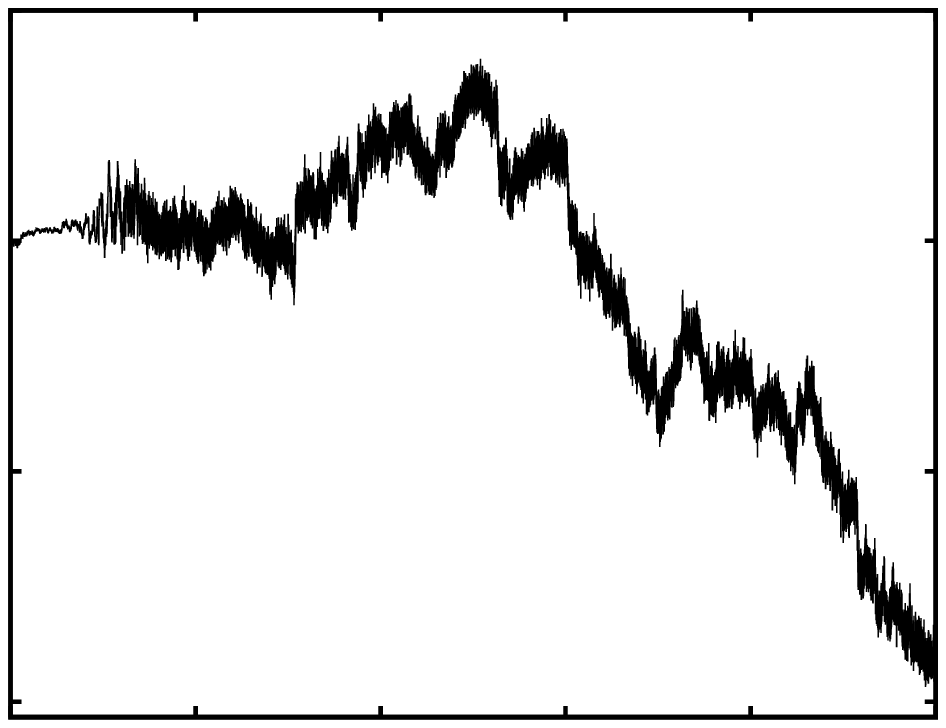} } \\
   \end{array}$
 \caption{Same as Fig. \ref{100m_4p6} but for a star with log T$_{\rm{eff}}$ = 4.45 and an initial mass of 70 M$_{\sun}$.
 The velocity amplitude reaches 131 km s$^{-1}$ in the non-linear regime.}
 \normalsize
 \label{70m_4p45}
 \end{figure*}

\begin{figure*}
\centering $
\LARGE
\begin{array}{cccccc}
  \scalebox{0.455}{ \input{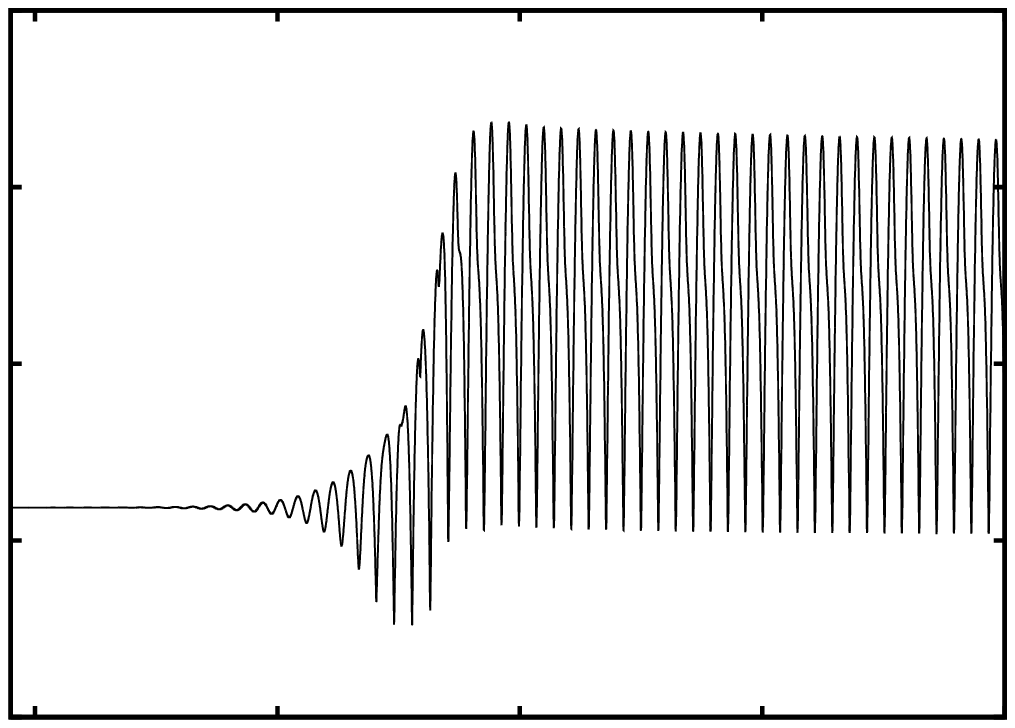} } 
  \scalebox{0.455}{ \input{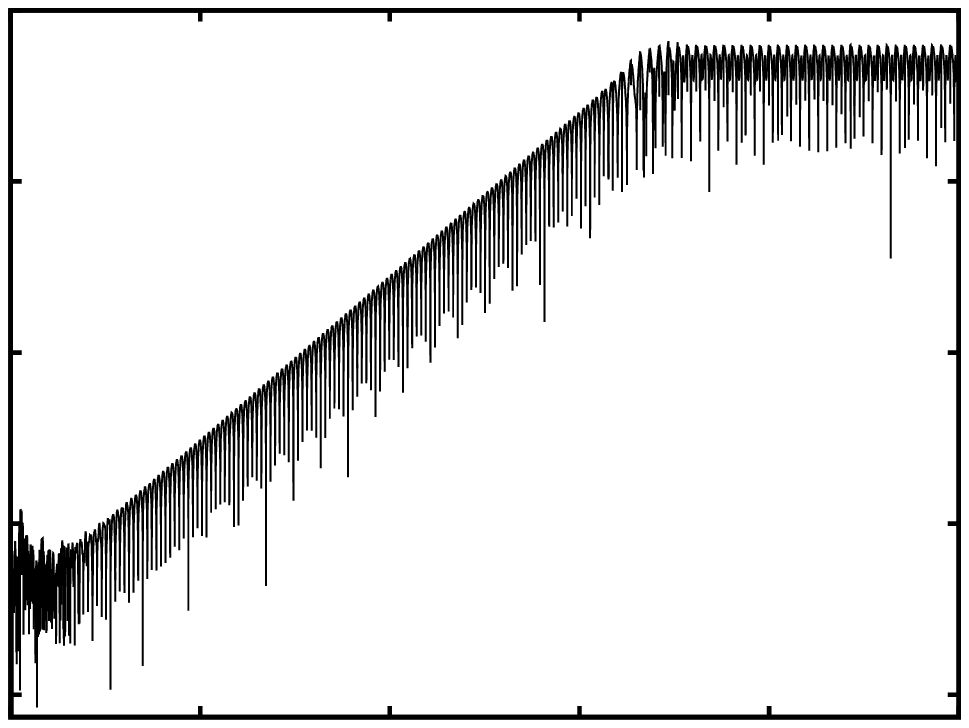} }
    \scalebox{0.455}{ \input{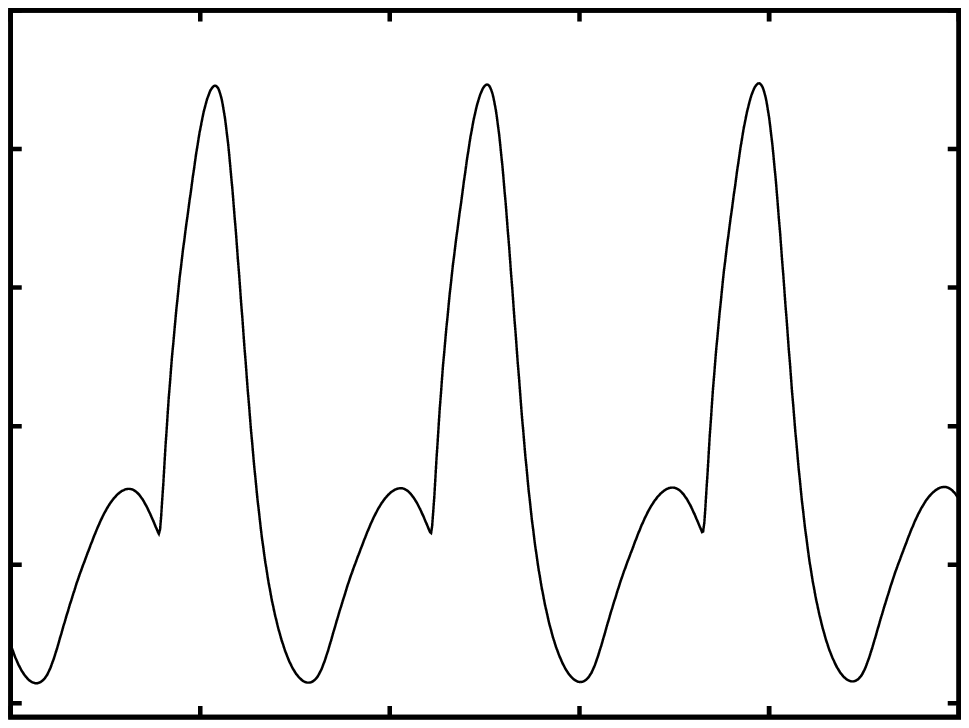} } \\
  \scalebox{0.455}{ \input{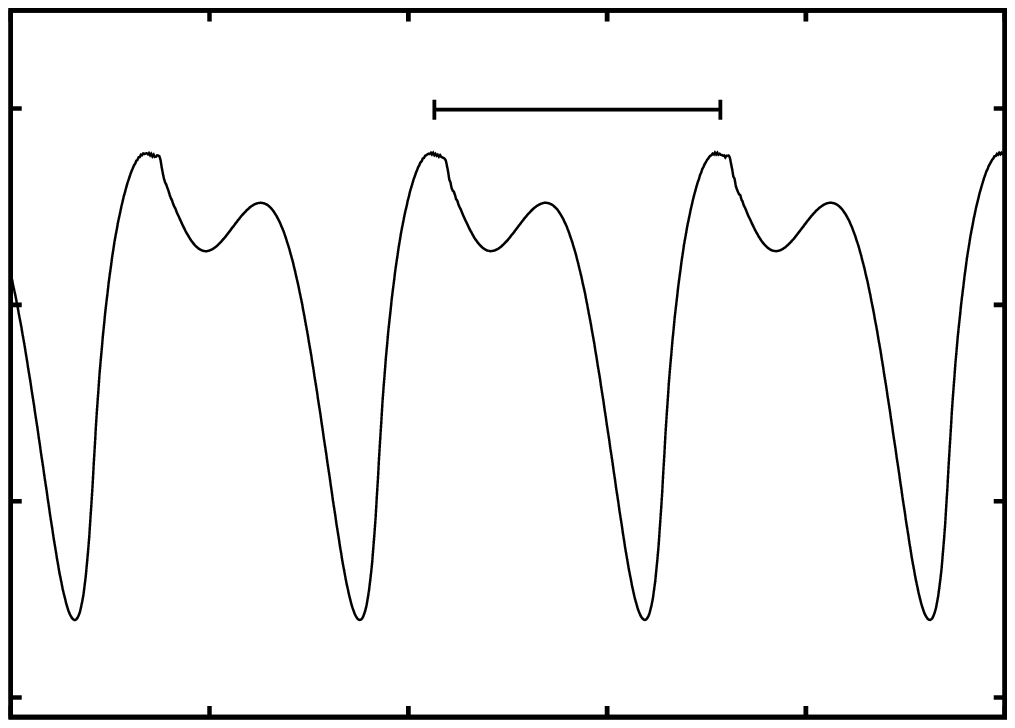} }
  \scalebox{0.455}{ \input{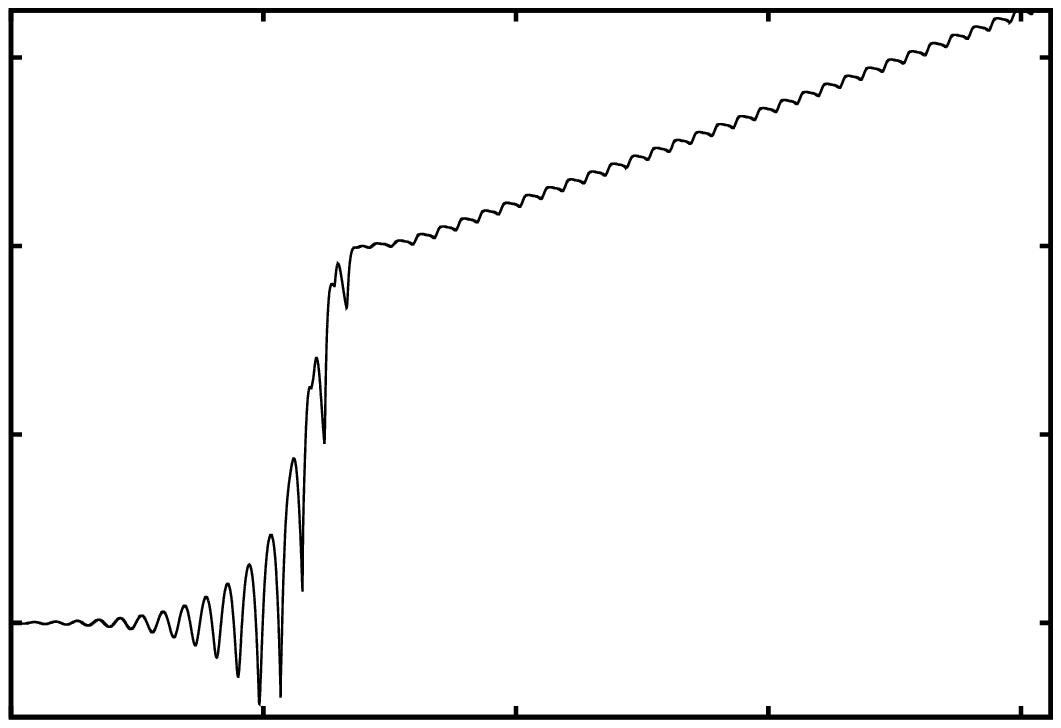} }
   \scalebox{0.455}{ \input{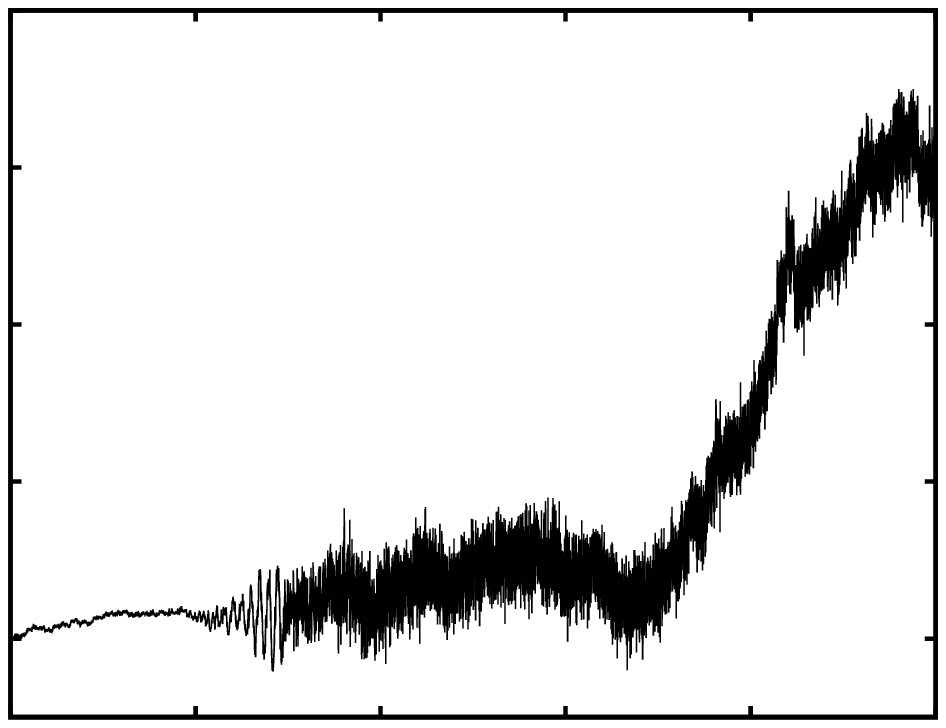} } \\
   \end{array}$
 \caption{Same as Fig. \ref{100m_4p6} but for a star with log T$_{\rm{eff}}$ = 4.45 and an initial mass of 45 M$_{\sun}$.
 The velocity amplitude reaches 139 km s$^{-1}$ in the non-linear regime.}
 \normalsize
 \label{45m_4p45}
 \end{figure*}

\begin{figure*}
\centering $
\LARGE
\begin{array}{cccccc}
  \scalebox{0.455}{ \input{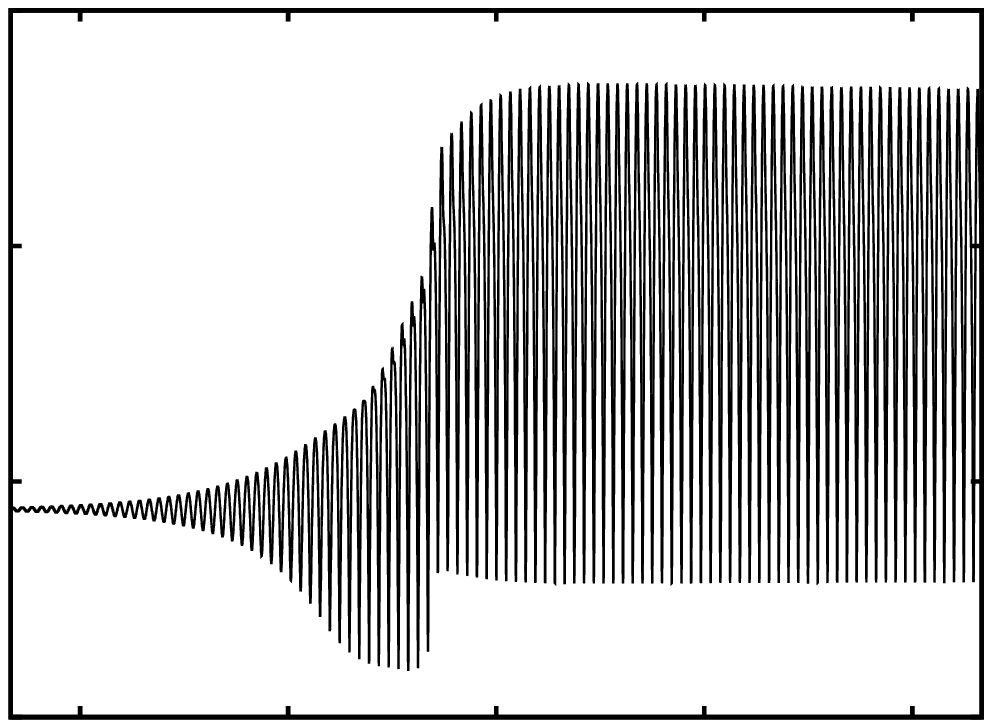} } 
  \scalebox{0.455}{ \input{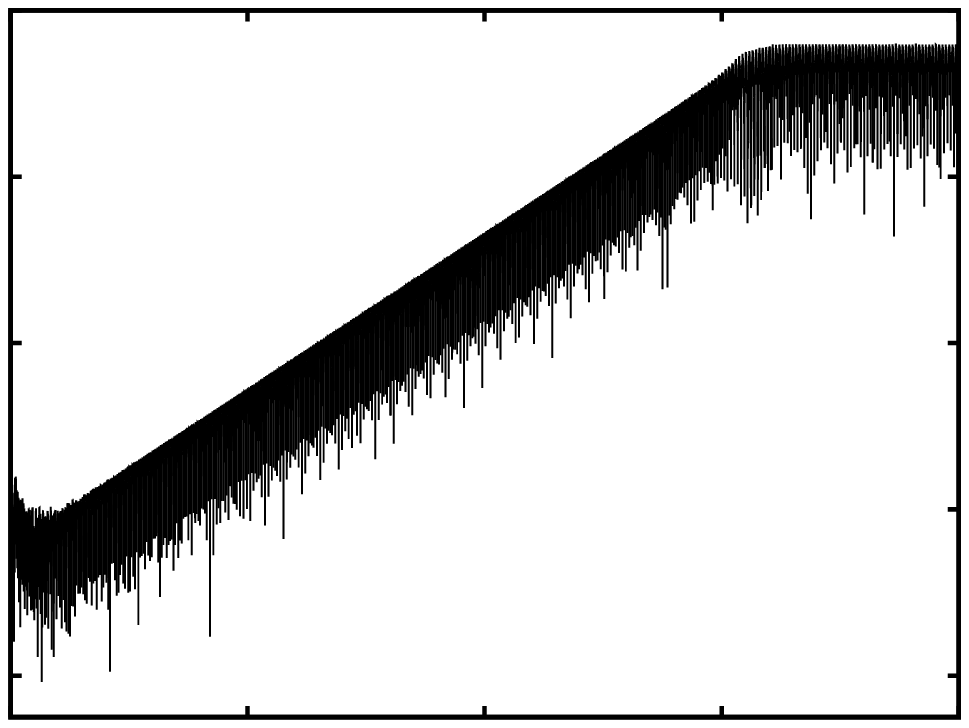} }
    \scalebox{0.455}{ \input{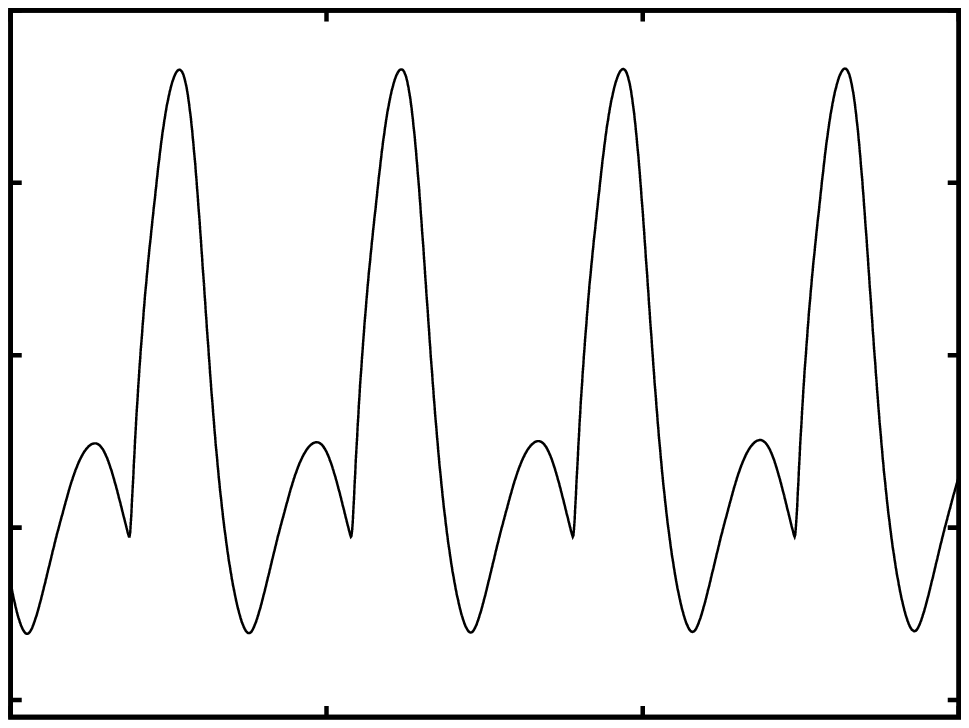} } \\
  \scalebox{0.455}{ \input{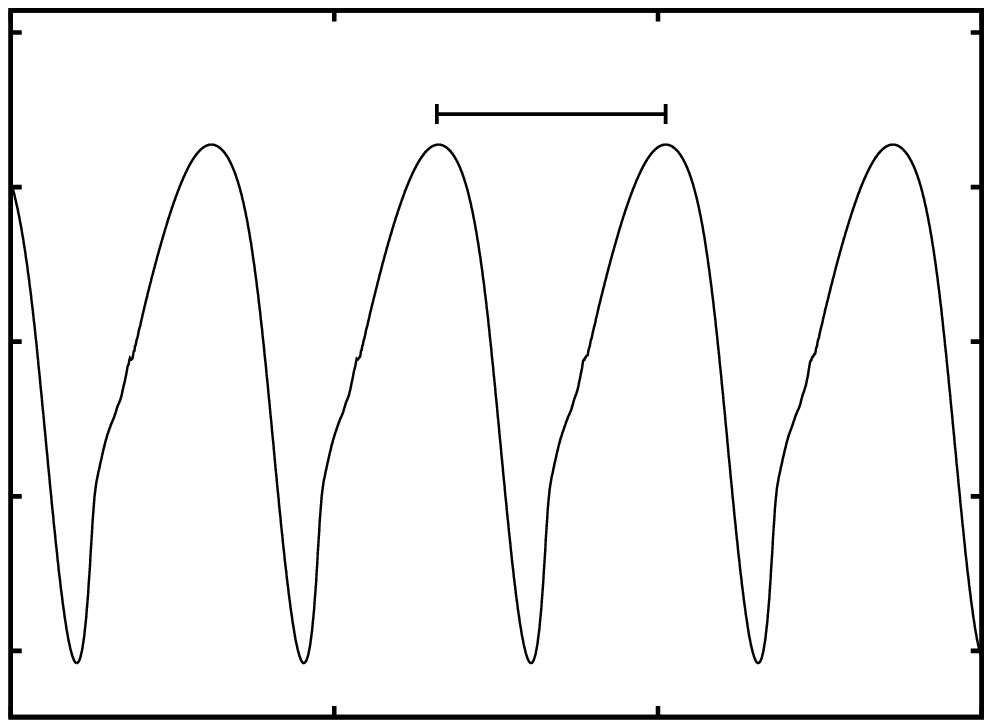} }
  \scalebox{0.455}{ \input{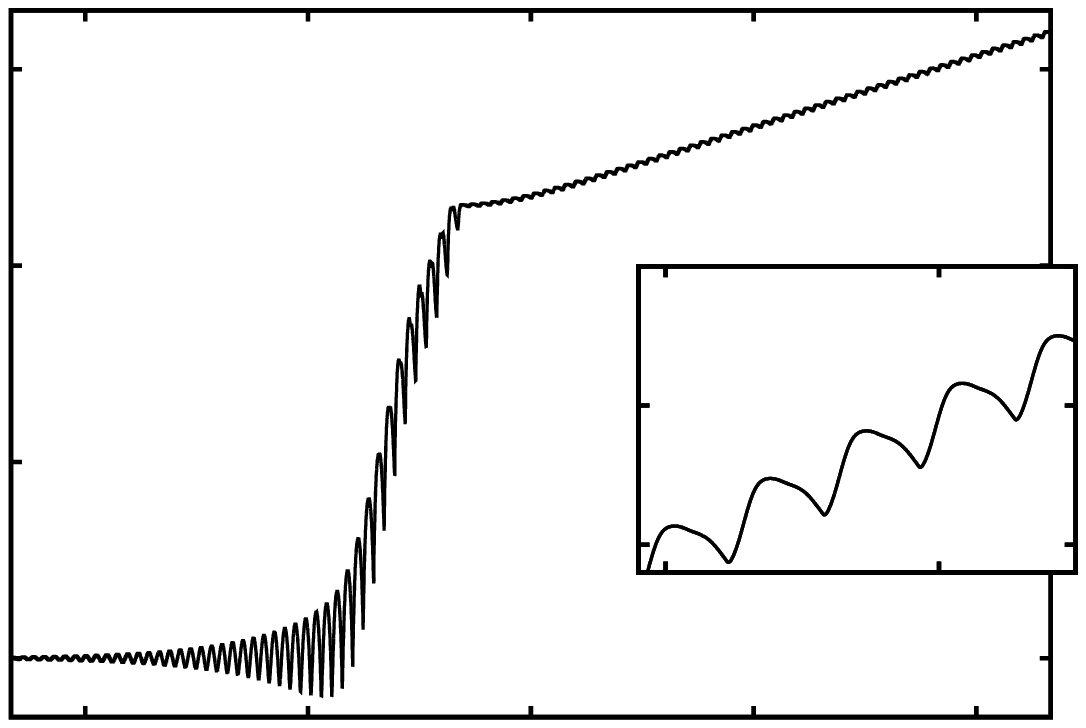} }
   \scalebox{0.455}{ \input{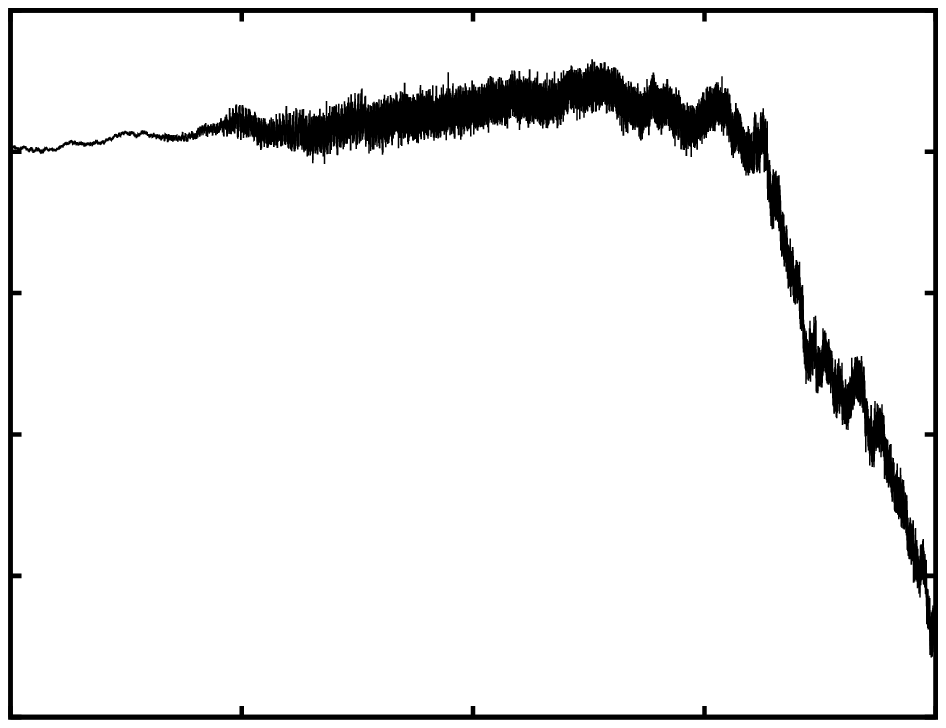} } \\
   \end{array}$
 \caption{Same as Fig. \ref{100m_4p6} but for a star with log T$_{\rm{eff}}$ = 4.45 and an initial mass of 30 M$_{\sun}$.
 The velocity amplitude reaches 130 km s$^{-1}$ in the non-linear regime.}
 \normalsize
 \label{30m_4p45}
 \end{figure*}

For the effective temperature log T$_{\rm{eff}}$ = 4.45, all stellar models on the evolutionary 
tracks for the initial masses of 100, 70, 45 and 30 M$_{\sun}$ are linearly unstable.
Results of the simulations of the evolution of the instabilities into the non-linear regime are shown in 
Figs.  \ref{100m_4p45}, \ref{70m_4p45}, \ref{45m_4p45} and \ref{30m_4p45}, respectively. Essentially the discussion 
of the results given for the models having log T$_{\rm{eff}}$ = 4.6 also holds for those considered here. The velocity
amplitudes attain values between 130 to 215 km s$^{-1}$. Except for the 100 M$_{\sun}$ model, where the finite 
amplitude pulsation is not strictly periodic and a pulsation period cannot be defined, the non-linear pulsation 
periods lie in the range between 0.7 and 2.37 d. 
From the mean slope of the time integrated acoustic energy, we obtain mass-loss rates of 
4.5 $\times$ 10$^{-8}$, 
 4.2 $\times$ 10$^{-8}$ and 1.7 $\times$ 10$^{-8}$ M$_{\sun}$ yr$^{-1}$ for models corresponding to 
 initial masses of 70, 45 and 30 
 M$_{\sun}$, respectively. For the 100 M$_{\sun}$ model, we estimate a maximum mass-loss rate
 of 1.7 $\times$ 10$^{-6}$ M$_{\sun}$ yr$^{-1}$.

\subsubsection{Models having log T$_{\rm{eff}}$ = 4.15}
 \label{415}
\begin{figure*}
\centering $
\LARGE
\begin{array}{cccccc}
  \scalebox{0.455}{ \input{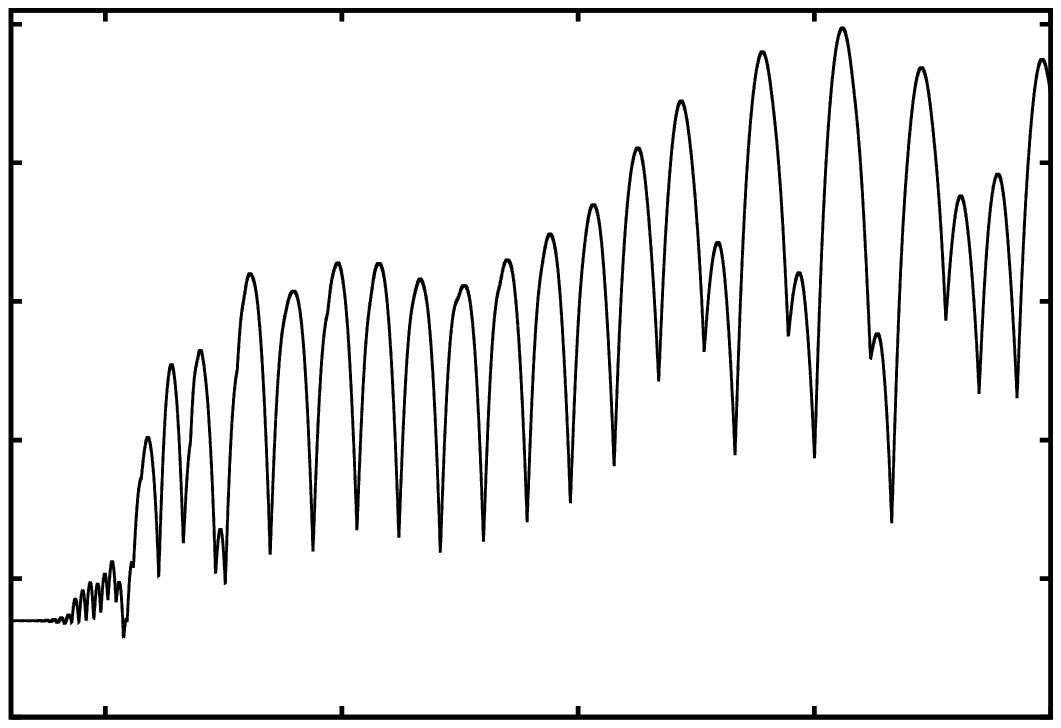} } 
  \scalebox{0.455}{ \input{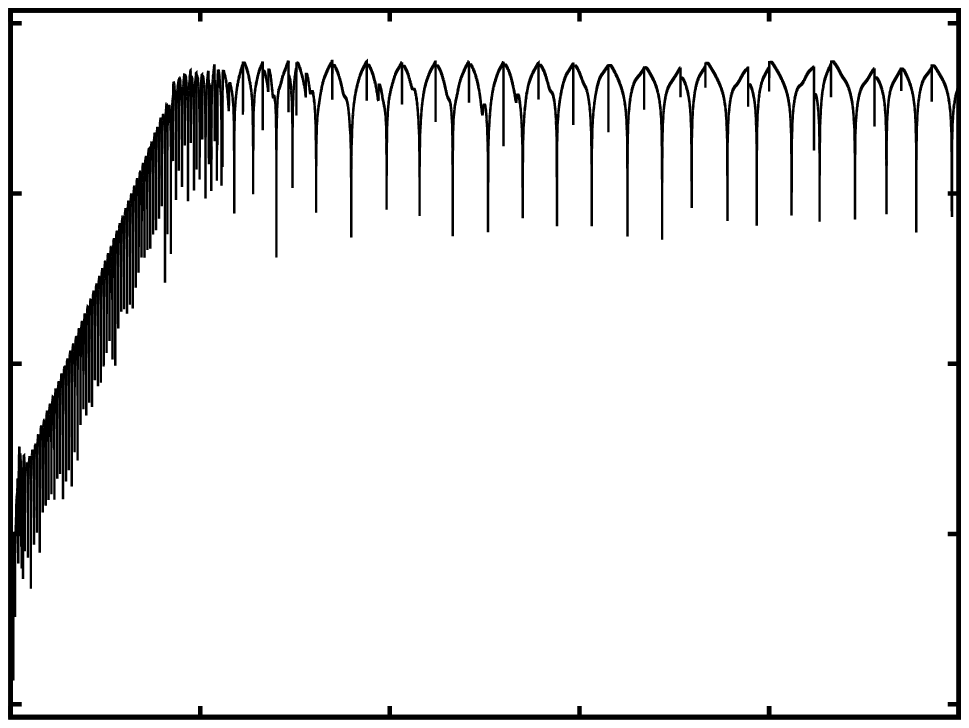} }
   \scalebox{0.455}{ \input{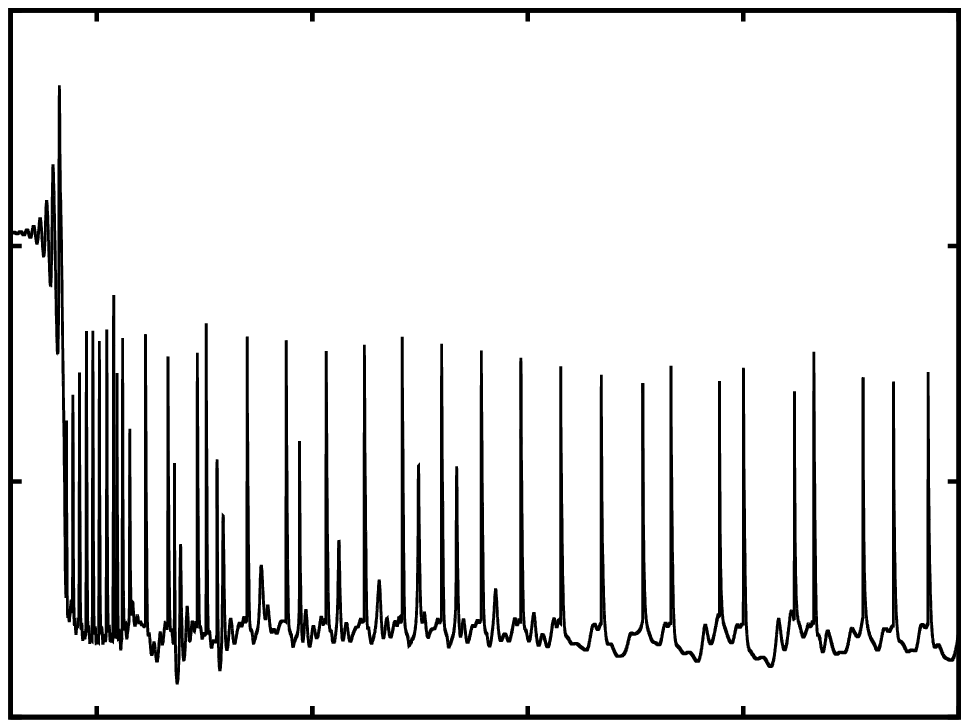} } \\

    \scalebox{0.455}{ \input{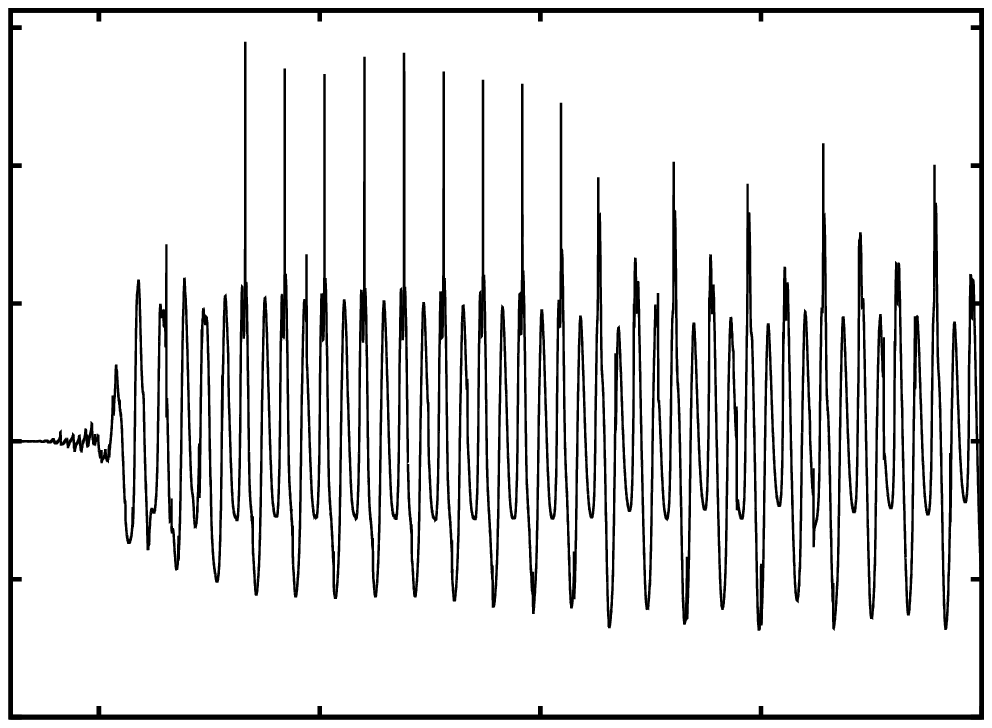} } 
  \scalebox{0.455}{ \input{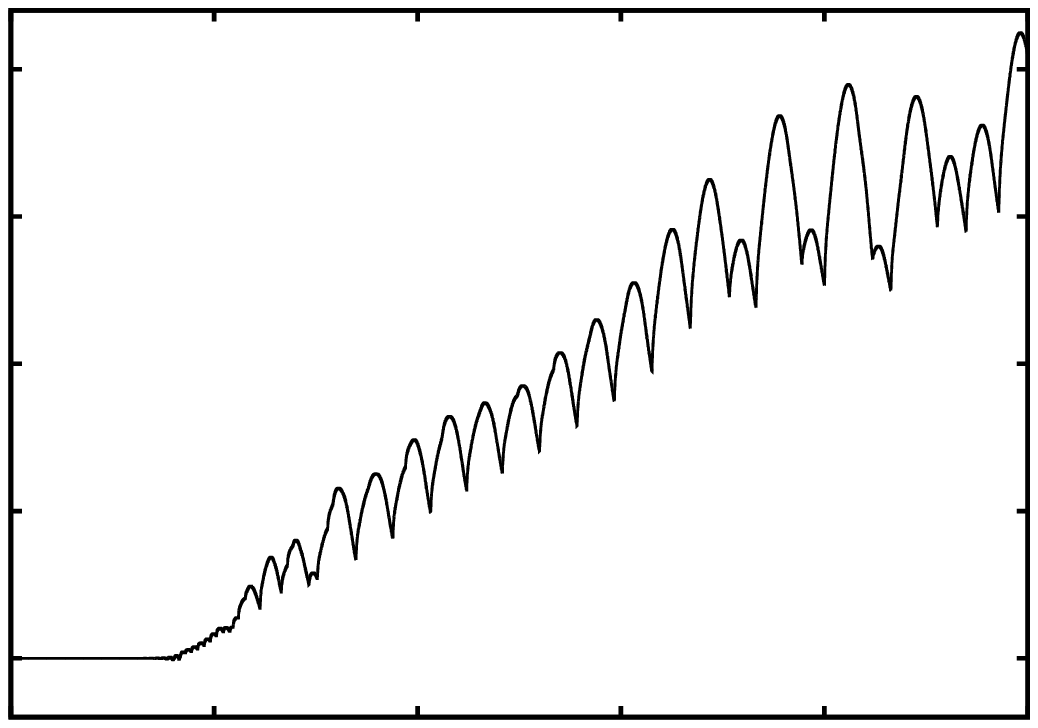} }
   \scalebox{0.455}{ \input{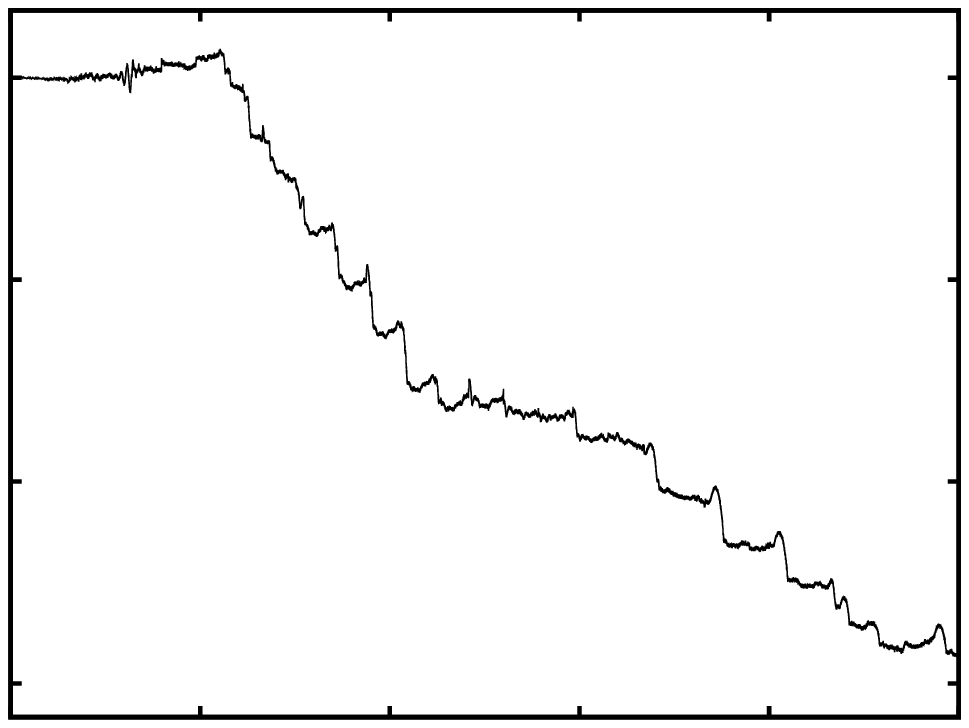} } \\
   \end{array}$
 \caption{Same as Fig. \ref{100m_4p6} but for a model having log T$_{\rm{eff}}$ = 4.15 and an initial mass of 100 M$_{\sun}$.
 Similar to Fig. \ref{100m_4p45} the finite amplitude pulsation 
 does not exhibit a strictly periodic pattern. The radius is considerably inflated and the
 velocity amplitude reaches 131 km s$^{-1}$ 
 in the non-linear regime.}
 \normalsize
 \label{100m_4p15}
 \end{figure*}

\begin{figure*}
\centering $
\LARGE
\begin{array}{ccccccccc}
  \scalebox{0.455}{ \input{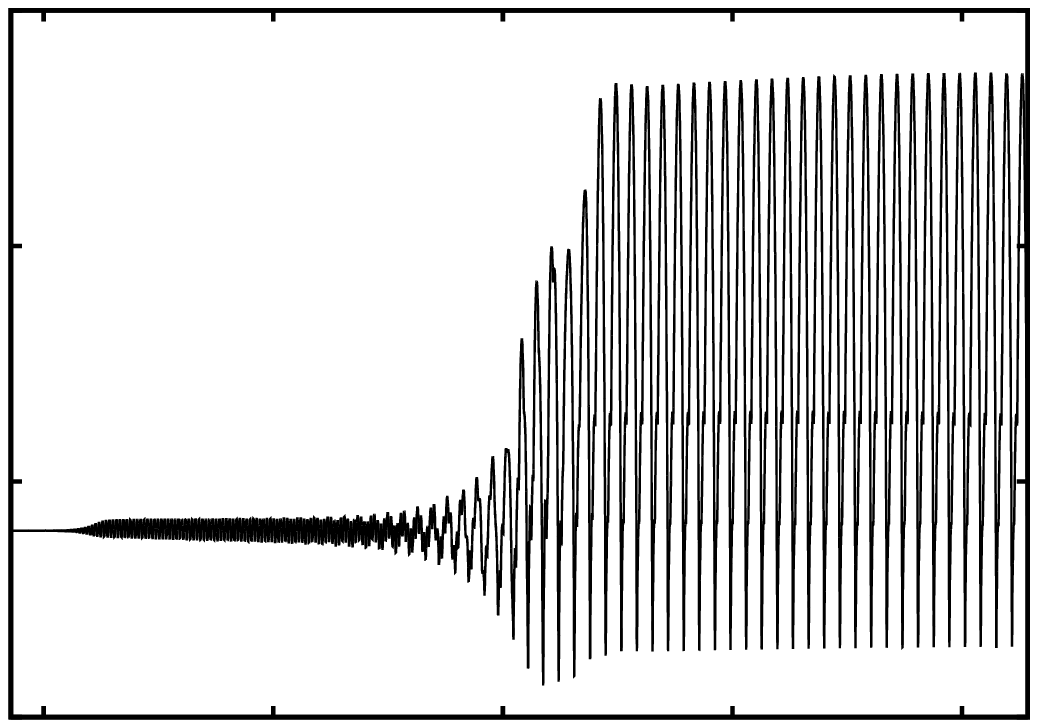} } 
  \scalebox{0.455}{ \input{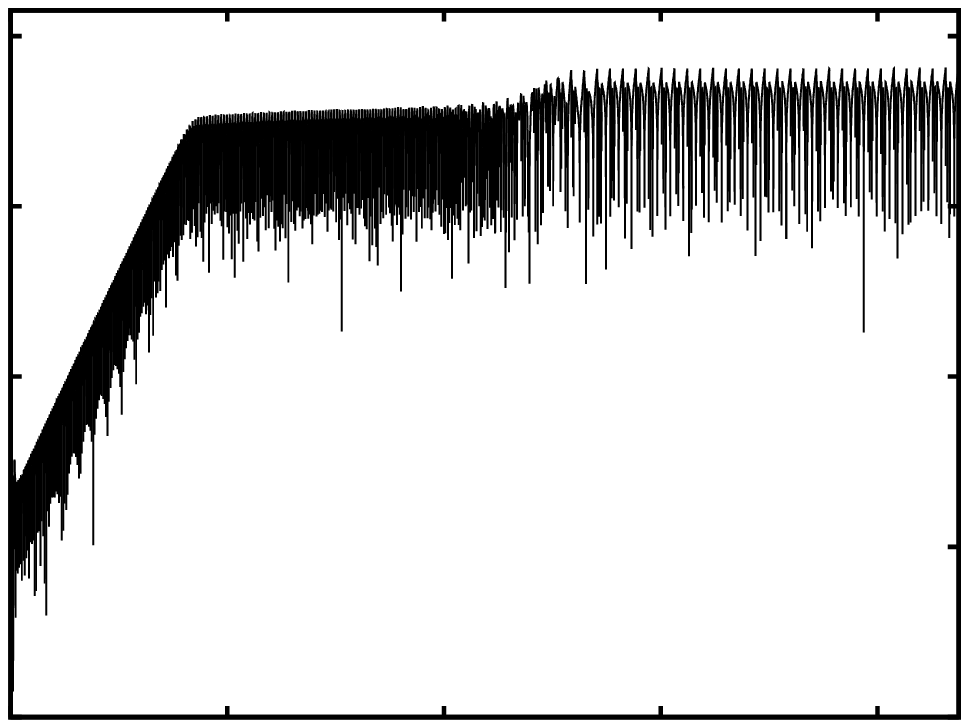} }
   \scalebox{0.455}{ \input{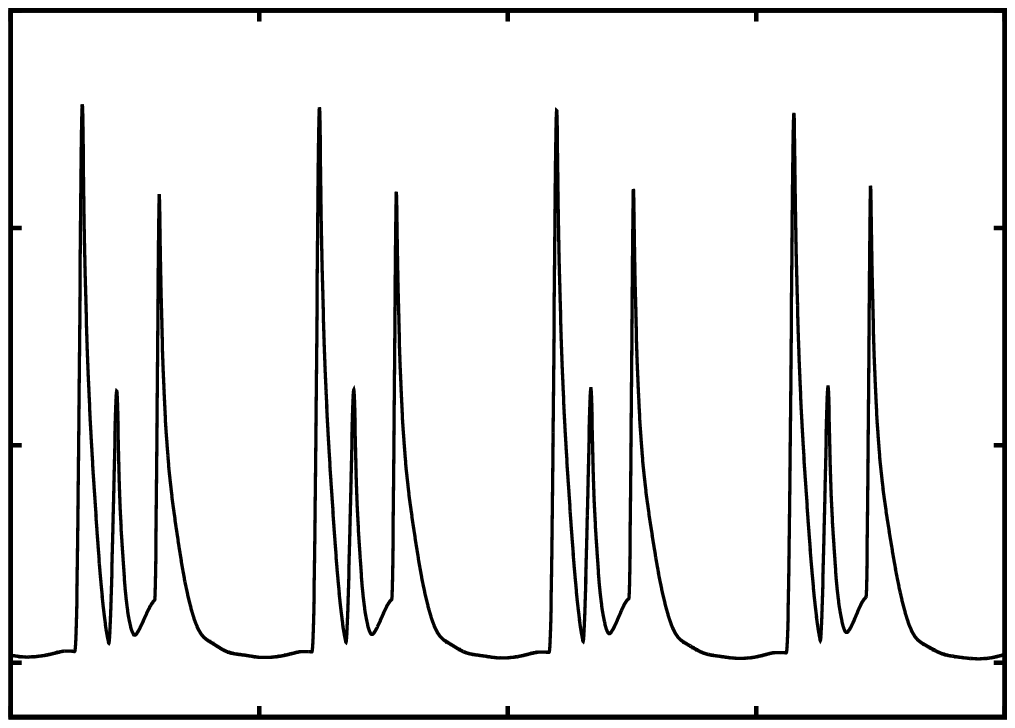} } \\

    \scalebox{0.455}{ \input{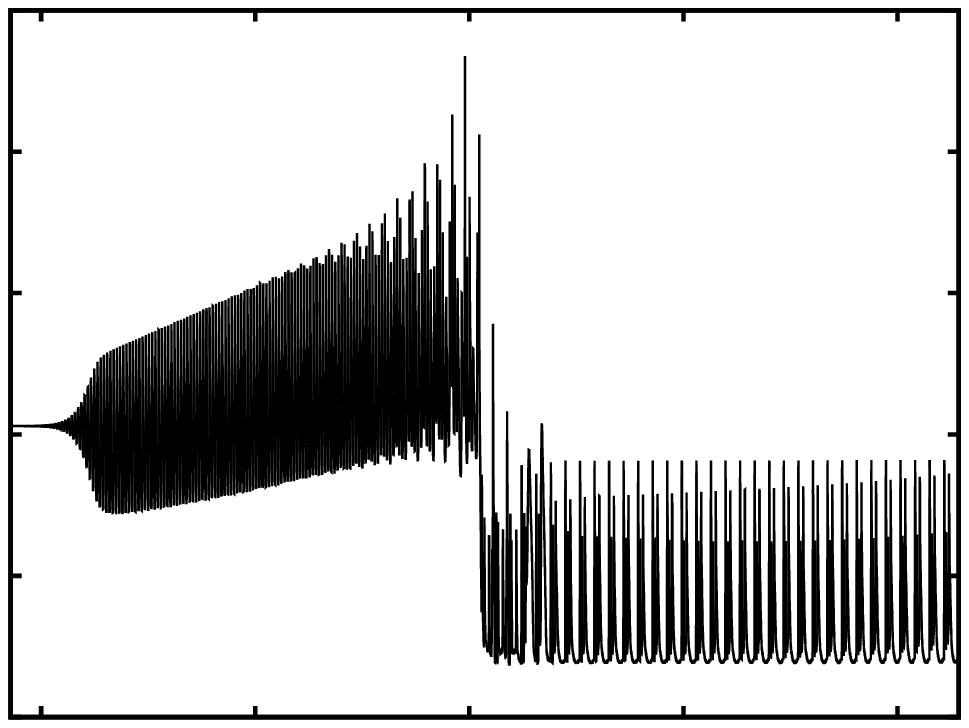} } 
  \scalebox{0.455}{ \input{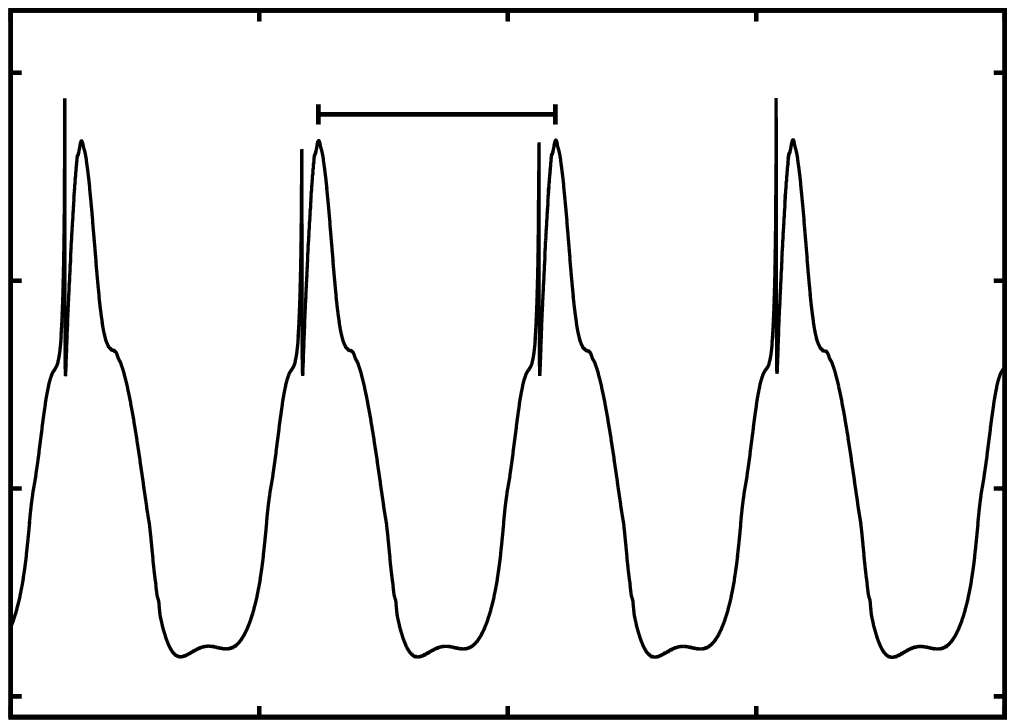} }
   \scalebox{0.455}{ \input{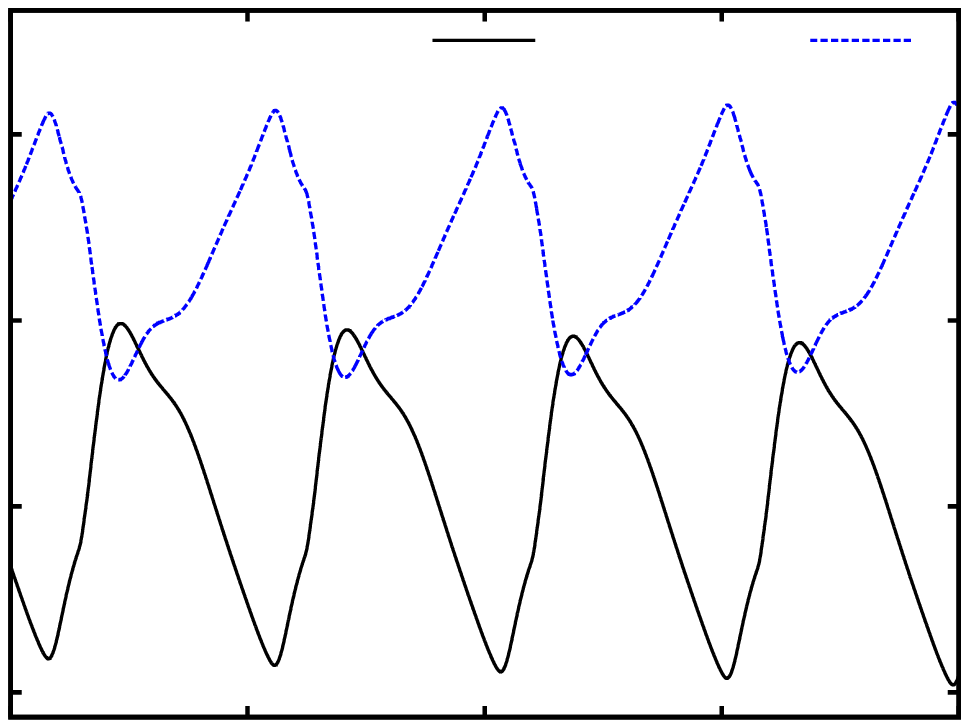} } \\
   
   \scalebox{0.455}{ \input{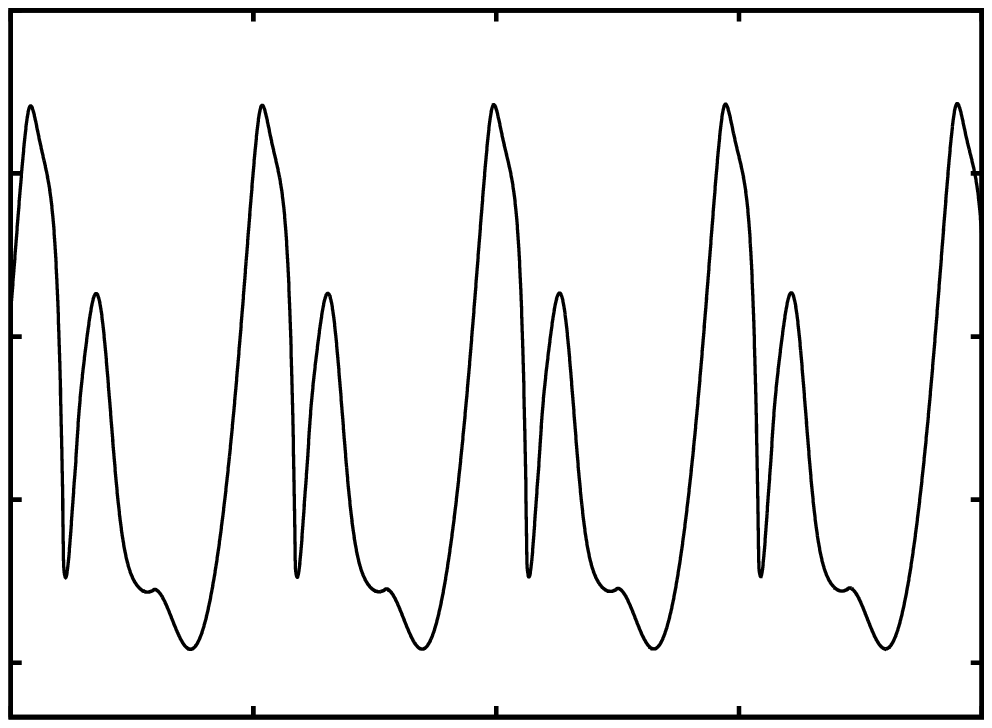} } 
  \scalebox{0.455}{ \input{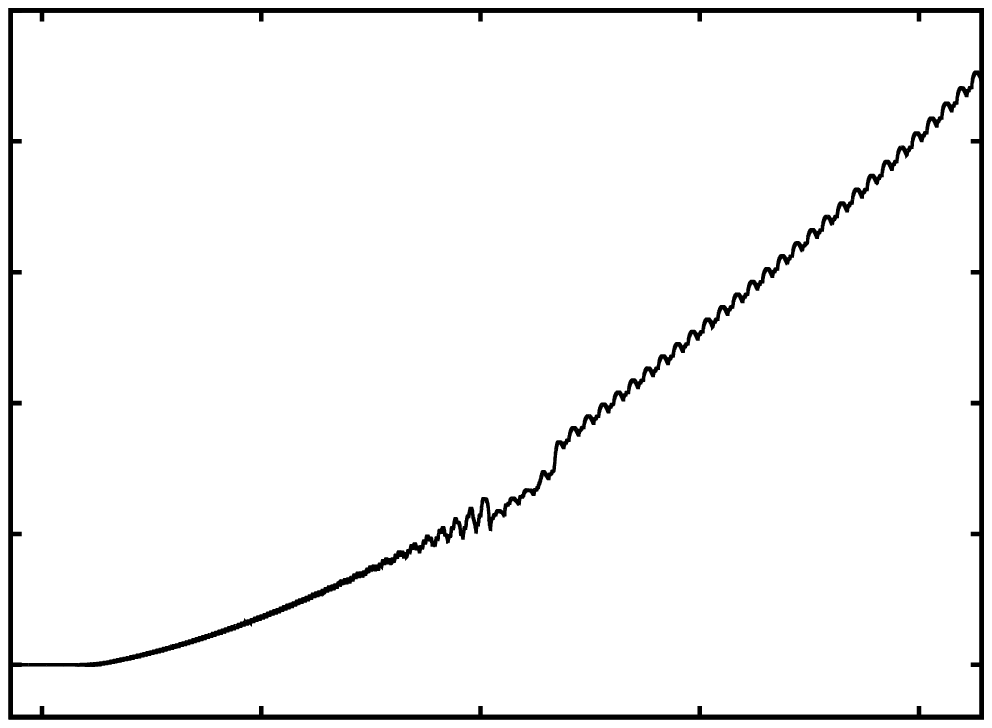} }
   \scalebox{0.455}{ \input{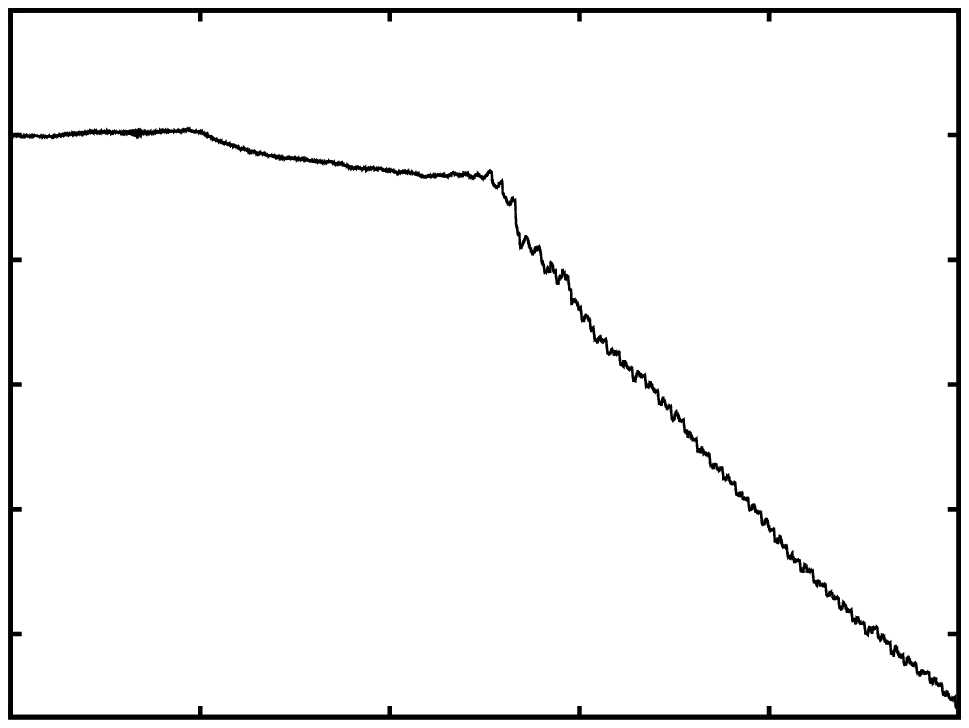} } \\
   \end{array}$
 \caption{Similar to Fig. \ref{100m_4p6} but for a model having log T$_{\rm{eff}}$ = 4.15 and an initial mass of 70 M$_{\sun}$.
 The velocity amplitude reaches 102 km s$^{-1}$ in the non-linear regime. In addition to Fig. \ref{100m_4p6}, the density, 
 the potential, internal and kinetic energies as a function of time are shown in (c), (f) and (g).}
 \normalsize
 \label{70m_4p15}
 \end{figure*}

 For the effective temperature log T$_{\rm{eff}}$ = 4.15, only models having high initial masses of 
 100 and 70 M$_{\sun}$ are linearly unstable. 
 Results of the simulations of the evolution of the instabilities into the non-linear regime are shown in 
Figs.  \ref{100m_4p15} and \ref{70m_4p15}, respectively. Again, essentially the discussion 
of the results given for the models having log T$_{\rm{eff}}$ = 4.6 also holds for those considered here. 
The velocity
amplitudes attain values of 131 and 102 km s$^{-1}$. 
In the 100 M$_{\sun}$ model, the finite 
amplitude pulsation is not strictly periodic and a pulsation period cannot be defined, for the 
70 M$_{\sun}$ model a  non-linear pulsation period of 47.7 d is obtained. 
From the mean slope of the time integrated acoustic energy, we obtain mass-loss rates of 
  2.1 $\times$ 10$^{-6}$ M$_{\sun}$ yr$^{-1}$  and  2.9 $\times$ 10$^{-6}$ M$_{\sun}$ yr$^{-1}$
for models corresponding to initial masses of 100 and 70 M$_{\sun}$, respectively. We note that for the 100  M$_{\sun}$ model,
the mean slope of the time integrated acoustic energy is less well defined than for other models discussed here. Moreover, 
this model exhibits a considerable inflation of the radius. In Fig. \ref{70m_4p15}, we have added plots of the potential, 
internal and kinetic energy terms as a function of time for the 70 M$_{\sun}$ model, since in this case they are higher 
by one order of magnitude compared to the values obtained for the 35 M$_{\sun}$ model discussed in 
Fig. \ref{35m_nonlin}.

 \subsection{ Long term simulations for the determination of the mean acoustic luminosity }
\begin{figure}
\centering $
\Large
\begin{array}{c}
  \scalebox{0.62}{ \input{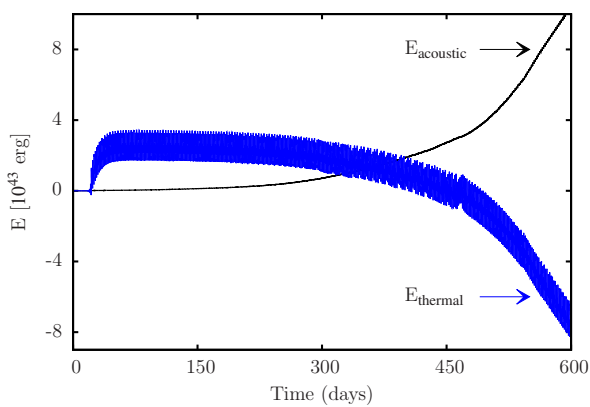} } \\
 \end{array}$
 \caption{ Time integrated acoustic and thermal energies as a function of time 
 for an evolutionary model having an initial mass of 100 M$_{\sun}$ and log T$_{\rm{eff}}$ = 4.6.}
 \normalsize
 \label{long_run10046}
 \end{figure}

 For some unstable models, the mean slope of the time integrated acoustic energy does not yet reach a constant 
 value even for sufficiently advanced times in the non-linear regime, where the finite amplitude pulsations have reached 
 a quasistationary state otherwise (see subsections \ref{46} and \ref{415}).
 For the evolutionary model having an initial mass of 100 M$_{\sun}$ and log T$_{\rm{eff}}$ = 4.6 
 (see also Fig. \ref{100m_4p6} and its discussion), we have 
 therefore performed long term simulations covering more than 200 pulsation periods. The results, i.e., the time 
 integrated acoustic and thermal energies as a function of time are shown in Fig. \ref{long_run10046}. After 500 d, 
 the mean slope both of the time integrated acoustic and thermal energy appears to reach a constant value. For 
 the time integrated acoustic energy, it corresponds to a mass-loss rate of  
 1.8 $\times$ 10$^{-5}$ M$_{\sun}$ yr$^{-1}$. 
 For a 100 M$_{\sun}$ star with a typical life time of around 10$^{6}$ yr, a mass-loss rate of this order
 can affect its evolution.    
From Fig. \ref{long_run10046}, we deduce that the mean slopes of the time integrated acoustic and thermal energies 
have approximately the same modulus but opposite sign. Accordingly, the associated acoustic and thermal luminosities 
(the latter being negative)
cancel. This is interpreted as a transformation of thermal flux into acoustic flux on the basis of the pulsations
acting as a Carnot type heat engine. Note that the total thermal luminosity being the sum of the stationary thermal 
luminosity and the variable contribution are always positive. In our discussions, we have always subtracted stationary 
contributions from the hydrostatic background model. Similar to the model having an initial mass of 100 M$_{\sun}$ and log T$_{\rm{eff}}$ = 4.6, 
we have performed a long term simulation for the model having 
 100 M$_{\sun}$ and log T$_{\rm{eff}}$ = 4.45 (see Fig. \ref{100m_4p45}). In this case, the mass-loss rate is found to increase from 
 1.7 $\times$ 10$^{-6}$ M$_{\sun}$ yr$^{-1}$ to 1.4 $\times$ 10$^{-4}$ M$_{\sun}$ yr$^{-1}$.

  \begin{figure*}
\centering $
\Large
\begin{array}{cccccc}
  \scalebox{0.6}{ \input{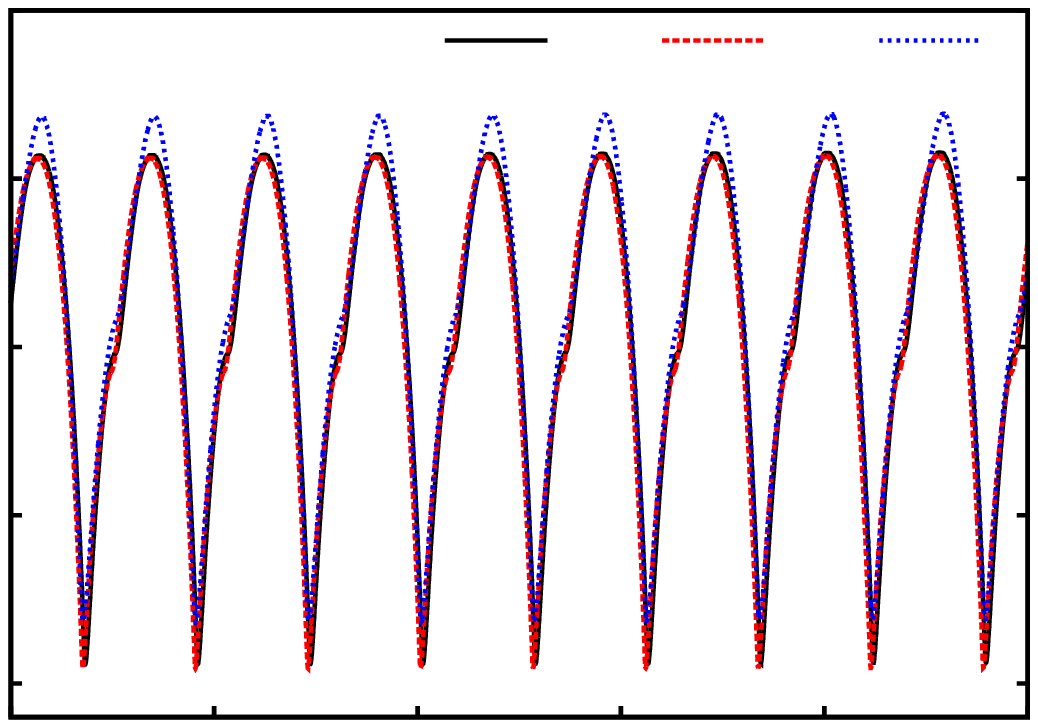} } 
  \scalebox{0.6}{ \input{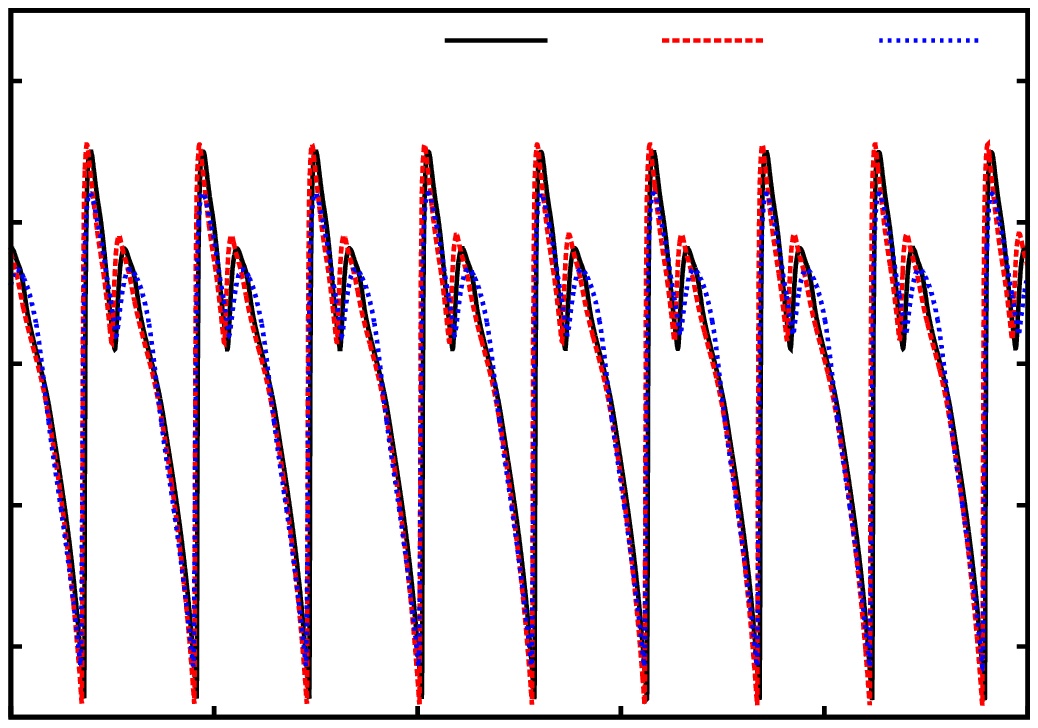} } \\
    \scalebox{0.6}{ \input{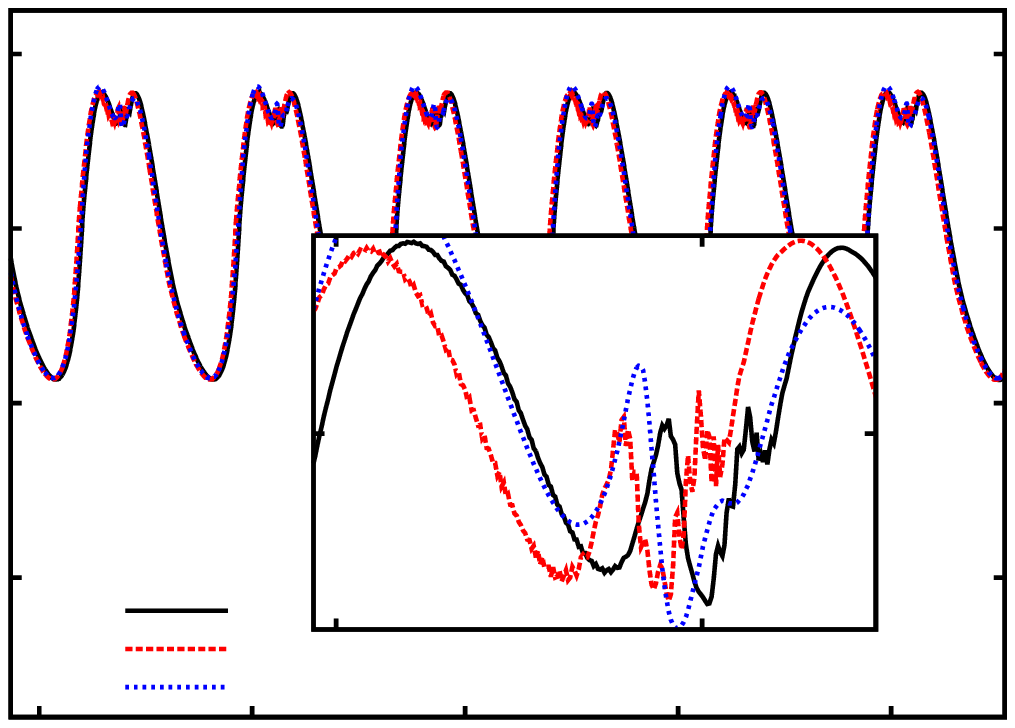} }
  \scalebox{0.6}{ \input{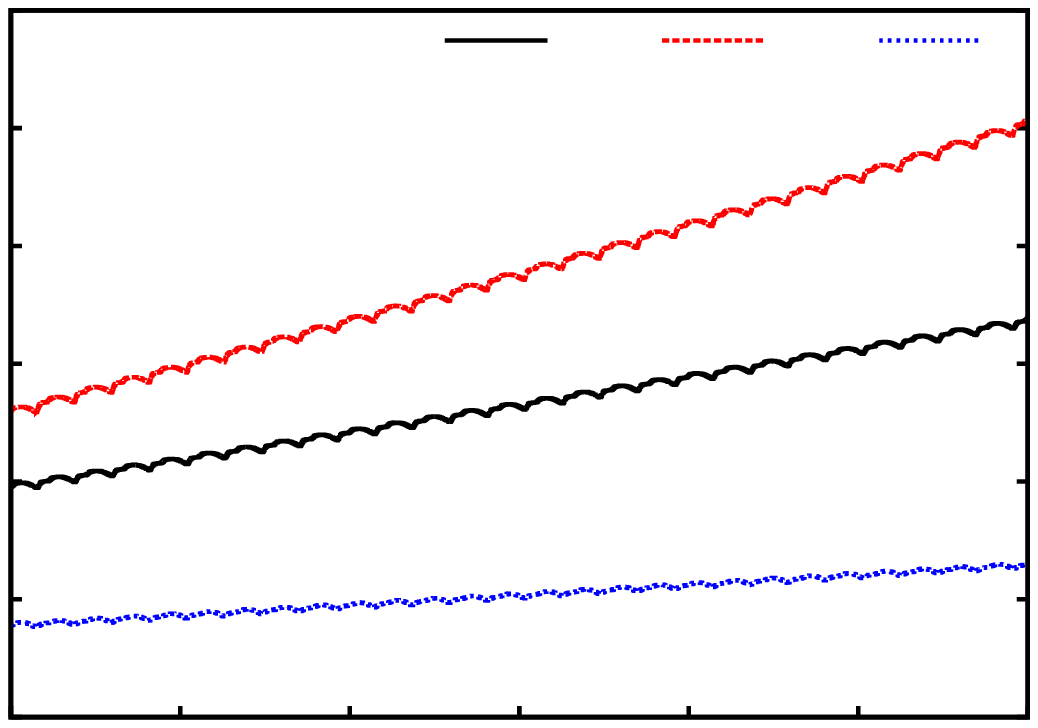} } \\
  \scalebox{0.6}{ \input{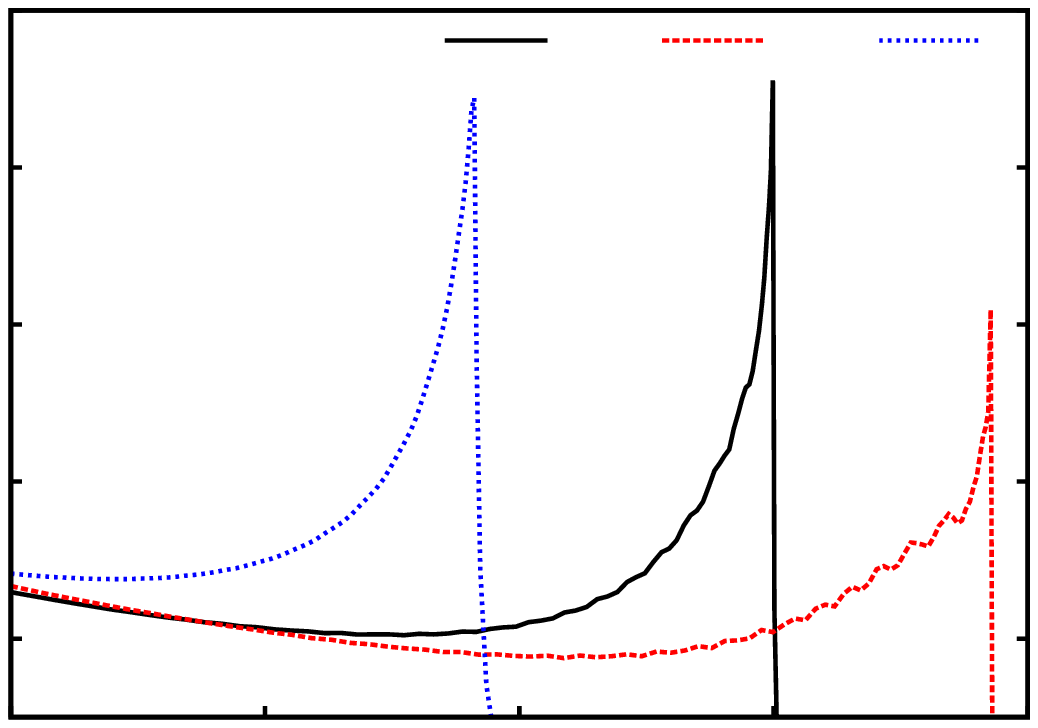} }
  \scalebox{0.6}{ \input{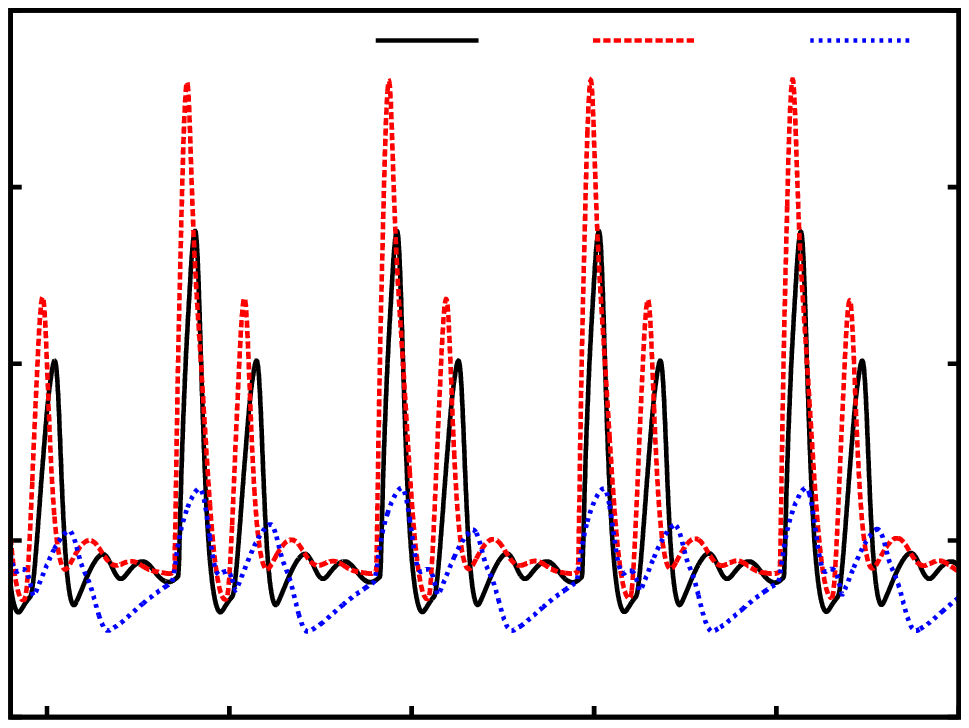} } \\
  
   \end{array}$
 \caption{ The influence of the artificial viscosity parameter ($\nu_{0}$ = 5, 10, 100)
 on the finite amplitude pulsations of an evolutionary model
 with log T$_{\rm{eff}}$ = 4.3 and an initial mass of 70 M$_{\sun}$. Radius, velocity at the outermost grid 
 point, the variation of the bolometric magnitude, 
 the time integrated acoustic energy and the temperature at the outermost grid point are given as a function of
 time in (a), (b), (c), (d) and
 (f) respectively, the density stratification is given in (e) as a function of radius at 
 comparable instances of time.}
 \normalsize
 \label{2arti}
 \end{figure*}

 \subsection{Role of artificial viscosity in non-linear regime}
 
 The instabilities identified here have the character of sound wave. 
 In the course of evolution, any sound wave will steepen and finally form a shock wave. 
 Thus, the appearance of shock waves is expected, when the instabilities are followed into the non-linear regime.
 Any numerical technique faces difficulties in representing discontinuous functions such as shock waves.
 An approach to handle shock waves without violating the requirement of the conservativity of the numerical 
 scheme consists of the introduction of artificial viscosity \cite[see,][]{neumann_1950}. Meanwhile, many 
 forms of the artificial viscosity have been proposed \citep[see,e.g.,][]{tscharnuter_1979}.  
 In this study, we adopt the artificial viscosity as given by \citet{grott_2005}. Among others, \citet{noh_1987} has pointed 
 out that the artificial viscosity can imply artefacts and may lead to erroneous numerical results.
 Therefore, the influence of the artificial viscosity on the numerical solutions needs to be studied. 
 In particular, the dependence on the viscosity parameter  $\nu_{0}$ introduced by \citet{grott_2005} has to be investigated. 
 Its appropriate choice is expected to depend on the stellar model considered.

 In the linear regime of exponential growth of the instabilities, the influence of the artificial viscosity is negligible since 
 the compression rates defining the artificial viscosity remain sufficiently small. In any case, artificial viscosity
 is important and implies consequences only in the vicinity of shocks. Numerical oscillations of, e.g., the density 
 stratification in the vicinity of a shock (Gibbs phenomenon) indicate that the numerical viscosity chosen is too small.
 In general, the viscosity parameter is chosen as small as possible but high enough such that these oscillations 
 are marginally suppressed. A value of $\nu_{0}$ = 10 turned out to be appropriate for many stellar models. The 
 influence of $\nu_{0}$ on the results is illustrated in Fig. \ref{2arti} for an evolutionary model with 
  log T$_{\rm{eff}}$ = 4.3 and an initial mass of 70 M$_{\sun}$. 
 From Fig. \ref{2arti}, we deduce that radius, velocity, variations of the bolometric magnitude and the density stratification
 even in the vicinity of a shock are not substantially affected by varying the artificial viscosity even by two 
 orders of magnitude. Moreover, a value of  $\nu_{0}$ = 10 is apparently sufficient to suppress the Gibbs phenomenon. The latter 
 seems to be more pronounced in the variation of the bolometric magnitude when considering it on a smaller scale. 
 In the most extreme case, i.e., at the outermost grid point, the variation of the temperature in the passing shock is 
 considerably reduced by artificial viscosity. Thus conclusions concerning temperature variations should be drawn with 
 caution. Substantially affected by the artificial viscosity is the mean slope of the time integrated acoustic energy used for 
 the estimate of the mass-loss rate. It increases with decreasing $\nu_{0}$ implying an (expected) 
 enhanced redistribution of mechanical into thermal energy with increasing artificial viscosity. Therefore the mass-loss rates 
 based on the mean slope of the time integrated acoustic energy derived here should be considered as lower limits to the 
 actual pulsationally driven mass-loss rates. From the mean slope of the time integrated acoustic energy (Fig. \ref{2arti}d), 
 we obtain for the viscosity parameter $\nu_{0}$ = 5, 10 and 100 a mass-loss rate of  
 5.1 $\times$ 10$^{-7}$ M$_{\sun}$ yr$^{-1}$, 3.0 $\times$ 10$^{-7}$ M$_{\sun}$ yr$^{-1}$ and 
 1.1 $\times$ 10$^{-7}$ M$_{\sun}$ yr$^{-1}$, respectively.

\begin{figure}
\centering $
\Large
\begin{array}{cc}
  \scalebox{0.62}{ \input{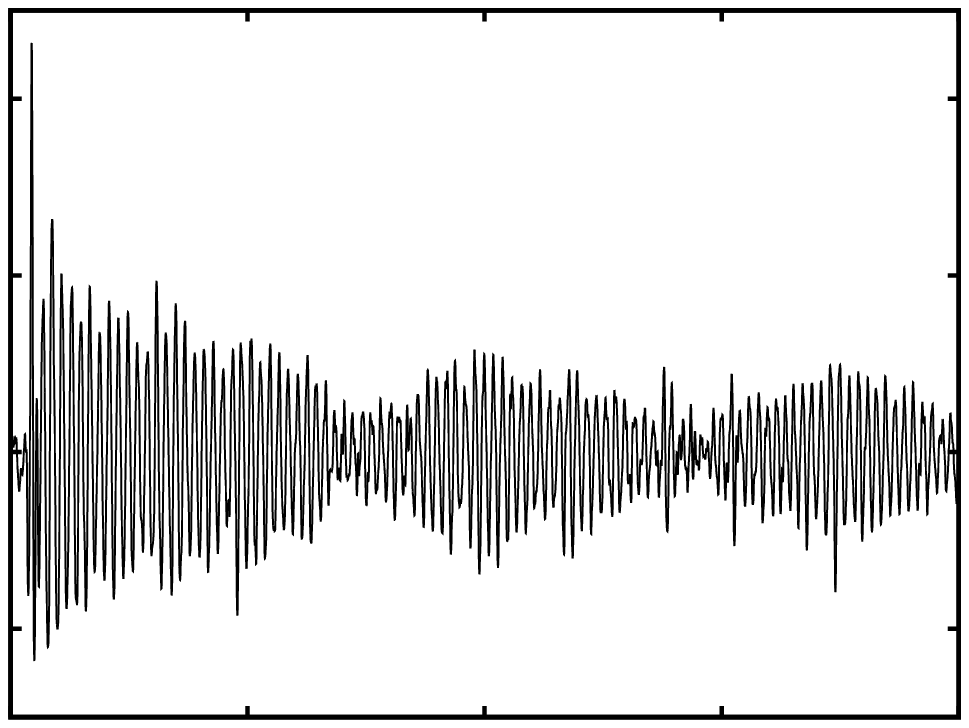} } \\
   \scalebox{0.62}{ \input{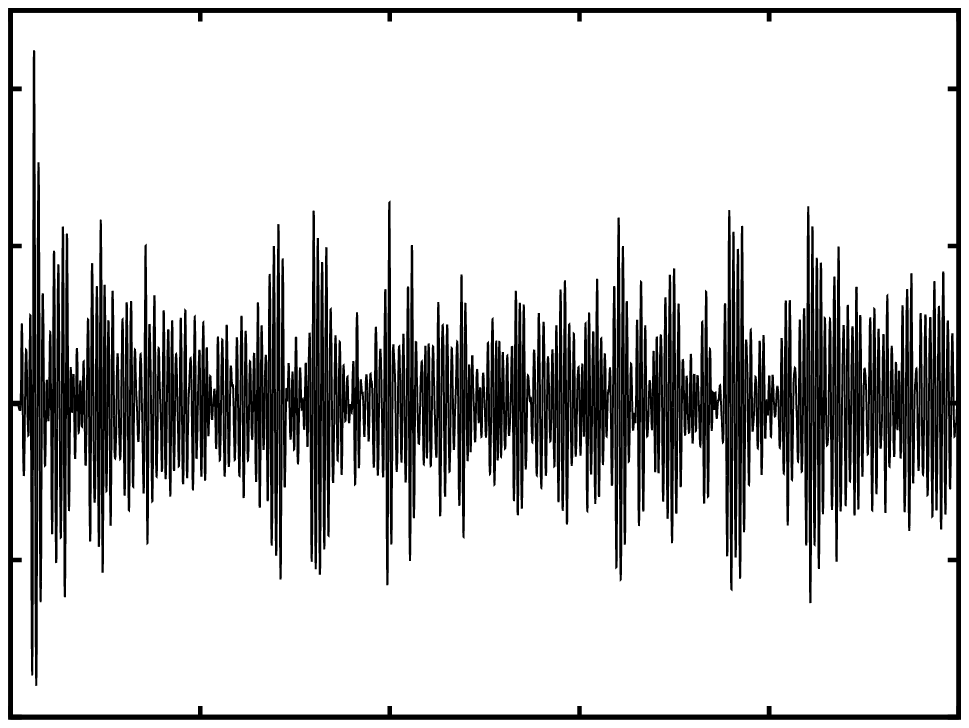} } \\
 \end{array}$
 \caption{ The velocity at the outermost grid point 
 as a function of time for two linearly stable evolutionary models with 
 an effective temperature of log T$_{\rm{eff}}$ = 4.15 and an initial mass of 45 M$_{\sun}$ (a), 
 and an effective temperature of log T$_{\rm{eff}}$ = 4.3 and an initial mass of 30 M$_{\sun}$ (b). 
 The velocity remains on the numerical noise level.}
 \normalsize
 \label{stable_v}
 \end{figure}
 
 \subsection{Linearly stable models}
 
For unstable models, we have seen that the velocity amplitude grows from numerical noise ($10^{-4}$ cm s$^{-1}$) without 
any external perturbation and saturates in the non-linear regime with maximum amplitudes of the order of 
100 km s$^{-1}$. To test and validate the code, we have also performed simulations of the evolution of  
linearly stable models \citep[see also,][]{yadav_2017}. In this case, the code 
 should not pick up any instability and, as a result, the velocity amplitude should always 
 remain on the numeral noise level of the scheme. 
 The latter is controlled by the prescribed relaxation accuracy of the initial hydrostatic model. 
 As an example, velocities at the
 outermost grid point are given as a function of time in Fig. \ref{stable_v} for two linearly stable models. 
 For both models, the velocity amplitudes remain on the numerical noise level which proofs that the final 
 results of the simulations are due to physical instabilities (if present) rather than to numerical instabilities and artefacts.

 \begin{figure}
\centering $
\Large
\begin{array}{c}
  \scalebox{0.63}{ \input{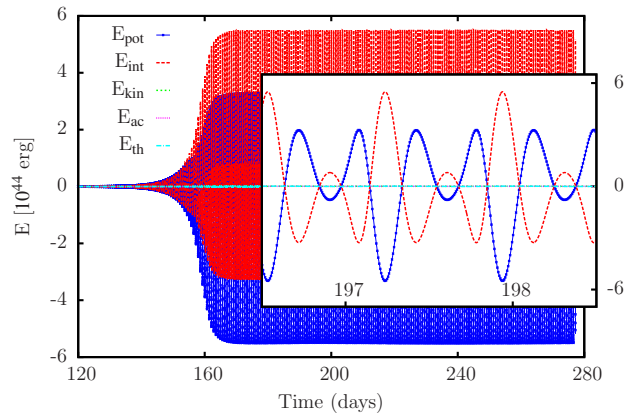} } \\
 \end{array}$
 \caption{Potential, internal, kinetic, time integrated acoustic and time integrated thermal energy 
 as a function of time for an evolutionary model with log T$_{\rm{eff}}$ = 4.45 having an initial mass
 of 30 M$_{\sun}$.}
 \normalsize
 \label{30m_all_energy}
 \end{figure}

 \begin{figure}
\centering $
\Large
\begin{array}{c}
  \scalebox{0.63}{ \input{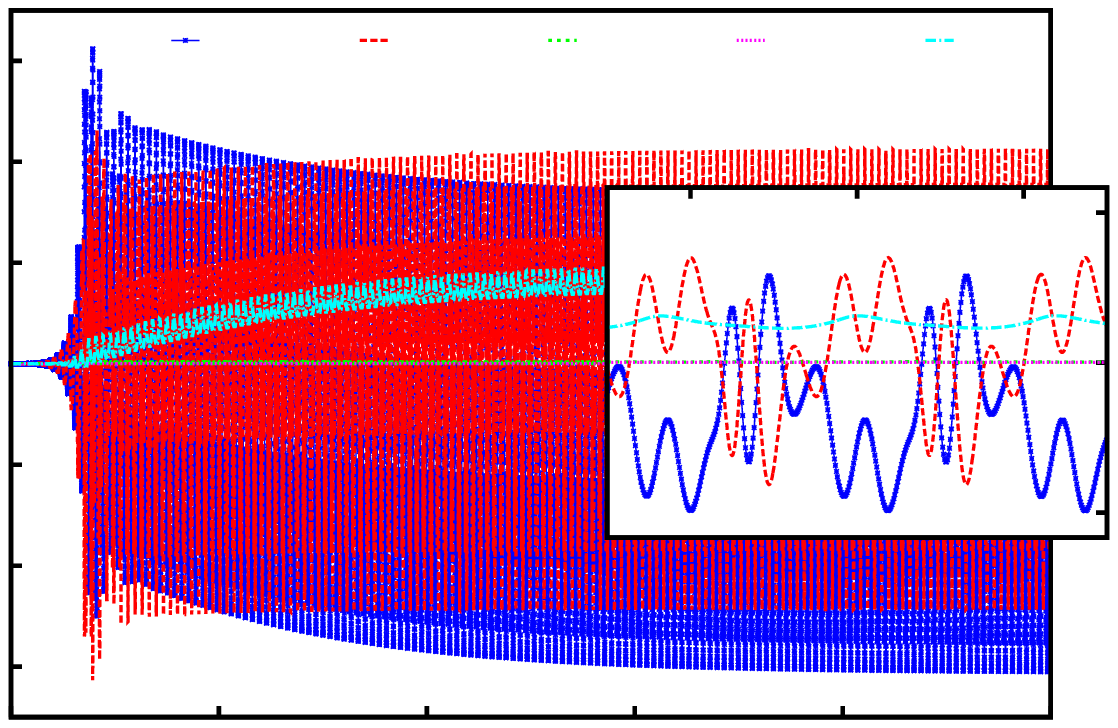} } \\
 \end{array}$
 \caption{Same as Fig. \ref{30m_all_energy} but for an initial mass of 70 M$_{\sun}$.}
 \normalsize
 \label{70m_all_energy}
 \end{figure}

 \begin{figure}
\centering $
\Large
\begin{array}{c}
  \scalebox{0.63}{ \input{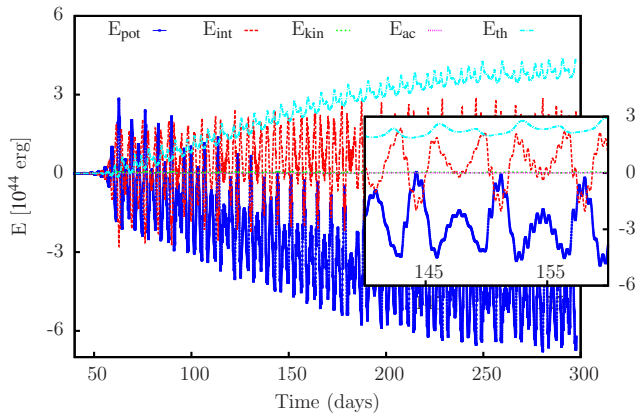} } \\
 \end{array}$
 \caption{Same as Fig. \ref{30m_all_energy} but for an initial mass of 100 M$_{\sun}$.}
 \normalsize
 \label{100m_all_energy}
 \end{figure}


\section{Discussion and conclusions}
\label{dac}
The linear instability of massive stars in the upper domain of the HR diagram has been known now for more than 
twenty years \citep[see e.g.,][]{glatzel_1993, kiriakidis_1993}. Instabilities with growth rates 
in the dynamical range are associated with the appearance of strange modes and rely on a non-classical mechanism
described by \citet{glatzel_1994}. They fall into three groups related to the three opacity maxima due to the contribution
of heavy elements, helium and hydrogen ionization, respectively. In the first part of this paper, we have redone the 
linear stability analysis of massive stars on the basis of the latest available opacity and equation of state tables. As 
a result, the previous investigations have been confirmed qualitatively. In particular, for the range of models 
considered, the maxima in the growth rates related to opacity maxima due to heavy elements and the first helium
ionization have been established. Contrary to the linear stability analysis, non-linear investigations concerning the final  
result of the instabilities are rare \cite[see e.g.,][]{glatzel_1999, grott_2005, yadav_2016, yadav_2017}. 
In particular, a systematic non-linear study of massive stars in the upper domain of the HRD is not yet available. 
Accordingly, the latter was the motivation and main subject of this paper.

The lack of numerical studies of the evolution of stellar instabilities into the non-linear regime is due to 
extreme numerical difficulties caused by huge differences in the order of magnitude of the various forms of 
the energy involved in a stellar pulsation. Typically gravitational and internal energies exceed the kinetic energy 
of interest by four orders of magnitude. Thus, for meaningful results,
the numerical scheme has to represent the energy balance correctly with a relative accuracy better than 10$^{-5}$. 
This condition requires the numerical scheme to be intrinsically fully conservative with respect to energy.
For non conservative schemes, numerical errors lead to kinetic energies comparable to gravitational and 
internal energies and thus to erroneously high velocity amplitudes and mass-loss 
\citep[see, e.g.,][]{appenzeller_1970, moriya_2015}. For the present study, we have adopted the fully conservative 
scheme described by \cite{grott_2005}. 

In any case, the result of the instabilities considered are finite amplitude pulsations where the velocity amplitudes 
(of the order of 100 km s$^{-1}$) attain a significant fraction of the escape velocity. Caused by sequences of shock waves the 
envelope becomes considerably inflated in the non-linear regime for many models. As a consequence, the final finite 
amplitude pulsation period is bigger than the linearly determined period of the unstable mode. Therefore, for 
a comparison with observed pulsation periods, the finite amplitude pulsation periods should be used rather than periods 
determined by a linear stability analysis. The inflation of the envelope is also associated with a decrease in 
temperature which occasionally was so strong that for the temperatures reached, opacity data were no longer available. 
Unfortunately, in these cases we had to stop the simulation. In the non-linear regime, the mean slope of the
time integrated acoustic energy was always positive indicating an outward directed mechanical acoustic luminosity. 
Comparing this acoustic luminosity with a kinetic wind luminosity, we were able to estimate pulsationally driven 
mass-loss rates. Depending on the stellar model mass-loss rates in the range between 
10$^{-9}$ M$_{\sun}$ yr$^{-1}$ and 10$^{-4}$ M$_{\sun}$ yr$^{-1}$ have been derived.

Compared to unstable ZAMS models, the pulsation periods of the evolved models considered here having substantially
larger radii are significantly longer as a consequence of the period density relation. Similarly, the maximum mass-loss rate
of 10$^{-4}$ M$_{\sun}$ yr$^{-1}$ for the evolved models compared to 10$^{-7}$ M$_{\sun}$ yr$^{-1}$ for ZAMS objects 
 is due to larger radii and less 
bound envelopes, where the velocity amplitudes are of the order of 100 km s$^{-1}$ in both cases. For the extended evolved models, 
the velocity amplitudes amount to a larger fraction of the escape speed.

For the representation of shock waves occurring in the course of the evolution of the instabilities, artificial
viscosity had to be introduced. Whether and how it influences the solution had to be studied. Quantities as pulsation 
periods, velocities (including amplitudes), bolometric magnitudes and radius variations are almost unaffected even 
for a large variation of the viscosity parameter. However, the time integrated acoustic energy and its mean slope 
sensitively depends on the artificial viscosity, where the mean slope increases with decreasing artificial viscosity. 
Thus, the estimated mass-loss rates have to be regarded as lower limits to the actual values.

The dependence on the stellar model of the run of the various forms of the energy  
is noteworthy. As an example, 
in Figs. \ref{30m_all_energy}, \ref{70m_all_energy} and \ref{100m_all_energy}
the energies 
are given as a function of time for three evolutionary models with log T$_{\rm{eff}}$ = 4.45
having initial masses of 30, 70 and 100 M$_{\sun}$,  respectively. In any case, 
the relative error in the energy balance lies below 10$^{-7}$ and 
the time integrated acoustic energy as well as the kinetic energy is much smaller than the other terms.
For the 30 M$_{\sun}$ model, even the time integrated thermal flux is negligible compared to the internal and the
gravitational potential energy (see Fig. \ref{30m_all_energy}). 
The latter form a symmetric pair and almost cancel each other. For the 70 M$_{\sun}$
model, this symmetry is lifted, since the time integrated thermal flux becomes comparable to the internal and 
the potential energies. Now the sum of the time integrated thermal flux, the potential and the internal energy 
approximately cancel (see Fig. \ref{70m_all_energy}). 
This situation becomes even more pronounced for the 100 M$_{\sun}$ model, where the time integrated thermal flux 
even exceeds the internal energy (see Fig. \ref{100m_all_energy}).

The importance of the time integrated thermal flux has been realized for stellar models, for which  
we were able to perform long term simulations of the evolution of the finite amplitude pulsations.
In these cases, the mean slope of the time integrated acoustic energy (and the associated mass-loss rate) was found 
to increase considerably in the course of long term evolution. Simultaneously, the mean slopes of 
the time integrated acoustic and (significant) thermal energies have approximately the same modulus but opposite sign. 
This means that acoustic and thermal luminosities cancel, which is interpreted as a transformation of thermal flux 
into acoustic flux. The observation that in the course of long term evolution, the mean acoustic luminosity 
(and with it the mass-loss rate) may be supported by and at the cost of the thermal luminosity, thus being considerably
enhanced, indicates that the mass-loss rates estimated here have to be regarded as lower limits, since long term 
simulations could be performed only for selected models.

This study was restricted to the consideration of spherically symmetric perturbations and pulsations. 
Both from observations \citep[][]{kraus_2015} and linear non-radial stability analyses \citep[][]{glatzel_1996}, 
we know that the models considered here do also suffer from non-radial strange mode instabilities. Therefore, 
an investigation of the evolution of non-radial instabilities into the non-linear regime is necessary. However,
corresponding numerical tools satisfying the conservativity requirement are not yet available. 
Thus, the development of appropriate schemes is highly recommended.

\section*{Acknowledgements}

APY gratefully acknowledges financial support by a SmartLink - Erasmus Mundus post-doc fellowship.




\bibliographystyle{mnras}
\bibliography{first} 








\bsp	
\label{lastpage}
\end{document}